\documentclass[final,authoryear,5p,times,twocolumn]{elsarticle}
\usepackage{amsmath}
\usepackage[usenames,dvipsnames]{color}
\usepackage{booktabs}
\usepackage{lineno,hyperref}
\modulolinenumbers[5]

\journal{Journal of Quantitative Spectroscopy \& Radiative Transfer}

\newcommand{\tripoli}{{\sc Tripoli-4}\textsuperscript{ \textregistered}}

\begin{document}

\begin{frontmatter}

\title{Benchmark solutions for transport in $d$-dimensional Markov binary mixtures}


\author[label1]{Coline Larmier}
\address[label1]{Den-Service d'Etudes des R\'eacteurs et de Math\'ematiques Appliqu\'ees (SERMA), CEA, Universit\'e Paris-Saclay, 91191 Gif-sur-Yvette, FRANCE.}
\author[label1]{Fran\c{c}ois-Xavier Hugot}
\author[label1]{Fausto Malvagi}
\author[label1]{Alain Mazzolo}
\author[label1]{Andrea Zoia\corref{cor1}}
\cortext[cor1]{Corresponding author. Tel. +33 (0)1 69 08 95 44}
\ead{andrea.zoia@cea.fr}





\begin{abstract}
Linear particle transport in stochastic media is key to such relevant applications as neutron diffusion in randomly mixed immiscible materials, light propagation through engineered optical materials, and inertial confinement fusion, only to name a few. We extend the pioneering work by Adams, Larsen and Pomraning~\cite{benchmark_adams} (recently revisited by Brantley~\cite{brantley_benchmark}) by considering a series of benchmark configurations for mono-energetic and isotropic transport through Markov binary mixtures in dimension $d$. The stochastic media are generated by resorting to Poisson random tessellations in $1d$ slab, $2d$ extruded, and full $3d$ geometry. For each realization, particle transport is performed by resorting to the Monte Carlo simulation. The distributions of the transmission and reflection coefficients on the free surfaces of the geometry are subsequently estimated, and the average values over the ensemble of realizations are computed. Reference solutions for the benchmark have never been provided before for two- and three-dimensional Poisson tessellations, and the results presented in this paper might thus be useful in order to validate fast but approximated models for particle transport in Markov stochastic media, such as the celebrated Chord Length Sampling algorithm.
\end{abstract}


\end{frontmatter}


\section{Introduction}

Linear transport through heterogeneous and disordered media emerges in several applications in nuclear science and engineering. Examples are widespread and concern for instance neutron diffusion in pebble-bed reactors~\cite{larsen} or randomly mixed immiscible materials~\cite{pomraning, renewal}, and inertial confinement fusion~\cite{zimmerman, zimmerman_adams, haran}. Besides, the spectrum of applications is fairly broad and far reaching~\cite{kendall, torquato}, and concerns also light propagation through engineered optical materials~\cite{NatureOptical, PREOptical, PREQuenched} or turbid media~\cite{davis, kostinski, clouds}, tracer diffusion in biological tissues~\cite{tuchin}, and radiation trapping in hot atomic vapours~\cite{NatureVapours}, only to name a few. The key goal of particle transport theory in stochastic media consists in deriving a formalism for the description of the ensemble-averaged angular particle flux $\langle \varphi({\bf r}, {\boldsymbol \omega}) \rangle$, where $\varphi({\bf r}, {\boldsymbol \omega})$ solves the linear Boltzmann equation
\begin{equation}
{\boldsymbol \omega} \cdot \nabla  \varphi + \Sigma({\bf r}) \varphi = \int \Sigma_s({\boldsymbol \omega}' \to {\boldsymbol \omega},{\bf r} )\varphi({\bf r}, {\boldsymbol \omega}') d{\boldsymbol \omega}' + S,
\label{boltzmann}
\end{equation}
${\bf r}$ and ${\boldsymbol \omega}$ denoting the position and direction variables, respectively, $\Sigma({\bf r})$ being the total cross section, $\Sigma_s({\boldsymbol \omega}' \to {\boldsymbol \omega},{\bf r} )$ the differential scattering cross section, and $S=S({\bf r}, {\boldsymbol \omega})$ the source term. For isotropic scattering, the differential scattering cross section simplifies to $\Sigma_s({\boldsymbol \omega}' \to {\boldsymbol \omega},{\bf r} )=\Sigma_s({\bf r}) / \Omega_d$, where $\Omega_d$ is the surface of the unit sphere in dimension $d$. For the sake of simplicity, we have here focused our attention to the case of mono-energetic transport in non-fissile media, in stationary (i.e., time-independent) conditions. However, these hypotheses are not restrictive (see the discussion in~\cite{pomraning}). The stochastic nature of particle transport stems from the materials composing the traversed medium being randomly distributed according to some statistical law. Hence, the quantities $\Sigma({\bf r})$, $\Sigma_s({\boldsymbol \omega}' \to {\boldsymbol \omega},{\bf r} )$ and $S({\bf r}, {\boldsymbol \omega})$ are in principle random variables.

A physical realization of the system under analysis will be denoted by a state $q$, associated to some stationary probability ${\cal P}(q)$ of observing the state $q$. To each state $q$ thus correspond the functions $\Sigma^{(q)}({\bf r})$, $\Sigma^{(q)}_s({\boldsymbol \omega}' \to {\boldsymbol \omega},{\bf r} )$ and $S^{(q)}({\bf r}, {\boldsymbol \omega})$ for the material properties~\cite{renewal, pomraning}. The ensemble-averaged angular flux is then formally defined as
\begin{equation}
\langle \varphi({\bf r}, {\boldsymbol \omega}) \rangle = \int {\cal P}(q) \varphi^{(q)}({\bf r}, {\boldsymbol \omega}) dq,
\end{equation}
where $\varphi^{(q)}({\bf r}, {\boldsymbol \omega})$ is the solution of the Boltzmann equation~\eqref{boltzmann} corresponding to a single realization $q$. The ensemble-averaged angular flux can be decomposed as
\begin{equation}
\langle \varphi({\bf r}, {\boldsymbol \omega}) \rangle = \sum_i p_i({\bf r}) \langle \varphi_i({\bf r}, {\boldsymbol \omega}) \rangle,
\end{equation}
where $p_i({\bf r}) = \int {\cal P}(q) \chi_i({\bf r}) dq$ is the probability of finding the material of index $i$ at position ${\bf r}$ (we denote by $\chi_i({\bf r})$ the marker function of material $i$ at position ${\bf r}$), and $\langle \varphi_i({\bf r}, {\boldsymbol \omega}) \rangle$ is restricted to those realizations that have material $i$ at position ${\bf r}$:
\begin{equation}
p_i({\bf r}) \langle \varphi_i({\bf r}, {\boldsymbol \omega}) \rangle =\int {\cal P}(q) \chi_i({\bf r}) \varphi^{(q)}({\bf r}, {\boldsymbol \omega}) dq .
\end{equation}

A widely adopted model of random media is the so-called binary stochastic mixing, where only two immiscible materials (say $\alpha$ and $\beta$) are present~\cite{pomraning}. Then, by averaging Eq.~\eqref{boltzmann} over realizations having material $\alpha$ at ${\bf r}$, we obtain the following equation for $\langle \varphi_\alpha({\bf r}, {\boldsymbol \omega}) \rangle$
\begin{equation}
\left[ {\boldsymbol \omega} \cdot \nabla  + \Sigma_\alpha \right] p_\alpha \langle \varphi_\alpha \rangle = \frac{p_\alpha \Sigma_{s,\alpha} }{\Omega_d} \int \langle \varphi_\alpha({\bf r}, {\boldsymbol \omega}') \rangle d{\boldsymbol \omega}' +p_\alpha S_\alpha + U_{\alpha,\beta},
\label{boltzmann_ave}
\end{equation}
where
\begin{equation}
U_{\alpha,\beta} = p_{\beta,\alpha} \langle \varphi_{\beta, \alpha} \rangle - p_{\alpha,\beta} \langle \varphi_{\alpha, \beta} \rangle,
\label{pomraning_exchange}
\end{equation}
with $p_{i,j}=p_{i,j}({\bf r}, {\boldsymbol \omega})$ denoting the probability per unit length of crossing the interface from material $i$ to material $j$ for a particle located at ${\bf r}$ and travelling in direction ${\boldsymbol \omega}$, and $\langle \varphi_{i, j} \rangle$ denoting the angular flux averaged over those realizations where there is a transition from material $i$ to material $j$ for a particle located at ${\bf r}$ and travelling in direction ${\boldsymbol \omega}$. The cross sections $\Sigma_\alpha$ and $\Sigma_{s,\alpha}$ are those of material $\alpha$. The equation for $\langle \varphi_\beta({\bf r}, {\boldsymbol \omega}) \rangle$ is immediately obtained from Eq.~\eqref{boltzmann_ave} by permuting the indexes $\alpha$ and $\beta$.

The set of equations in Eq.~\eqref{boltzmann_ave} (whose derivation contains no approximations so far) can be shown to form an infinite hierarchy, since the terms $\langle \varphi_\alpha \rangle$ in Eq.~\eqref{pomraning_exchange} would involve equations for the conditional averages $\langle \varphi_{\beta, \alpha} \rangle$ and $\langle \varphi_{\alpha, \beta} \rangle$, which in turn would further involve additional conditional averages~\cite{pomraning, renewal}. Generally speaking, it is necessary to truncate the infinite set of equations with some appropriate model leading to a closure formula, depending on the underlying mixing statistics. The celebrated Levermore-Pomraning model assumes for instance $\langle \varphi_{\alpha, \beta} \rangle = \langle \varphi_{\alpha} \rangle$ for homogeneous Markov mixing statistics~\cite{pomraning, levermore}, which is defined by
\begin{equation}
p_{i,j}({\bf r}, {\boldsymbol \omega}) = \frac{p_{i}}{\Lambda_i({\boldsymbol \omega})},
\end{equation}
depending on the starting position alone, where $\Lambda_i({\boldsymbol \omega})$ is the mean chord length for trajectories crossing material $i$ in direction ${\boldsymbol \omega}$. Several generalisations of this model have been later proposed, including higher-order closure schemes~\cite{pomraning, su}. In parallel, Monte Carlo algorithms such as the Chord Length Sampling have been conceived in order to formally solve the Levermore-Pomraning model, and have been further extended so as to include partial memory effects due to correlations for particles crossing back and forth the same materials~\cite{zimmerman, zimmerman_adams}. Their common feature is that they allow a simpler, albeit approximate, treatment of transport in stochastic mixtures, which might be convenient in practical applications where a trade-off between computational time and precision can be worth considering. Originally formulated for Markov statistics, these models have been largely applied also to random inclusions of disks or spheres into background matrices, with application to pebble-bed and very high temperature gas-cooled reactors~\cite{sutton, donovan}.

In order to assess the accuracy of the various approximate models it is therefore mandatory to compute reference solutions for the exact Eqs.~\eqref{boltzmann_ave}. Such solutions can be obtained in the following way: first, a realization of the medium is sampled from the underlying mixing statistics; then, the linear transport equations corresponding to this realization are solved by either deterministic or Monte Carlo methods, and the physical observables of interest are determined; this procedure is repeated several times so as to create a sufficiently large collection of realizations, and ensemble averages are finally taken for the physical observables. For this purpose, a number of benchmark problems for Markov mixing have been proposed in the literature so far~\cite{benchmark_adams, renewal, brantley_benchmark, brantley_conf, brantley_conf_2, vasques_suite2}, with focus exclusively on $1d$ geometries, either of the rod or slab type.

The aim of this work is two-fold. First, we will revisit the classical benchmark problem proposed by Adams, Larsen and Pomraning for transport in stochastic media~\cite{benchmark_adams}. We will present reference solutions obtained by Monte Carlo particle transport simulation through $1d$ slab, $2d$ extruded and $3d$ tessellations of a finite-size box with Markov mixing. We will compute the particle flux $\langle \varphi \rangle$, the transmission coefficient $\langle T \rangle$ and the reflection coefficient $\langle R \rangle$ by taking ensemble averages over the realizations; the dispersion of the physical observables around their average values will be assessed by evaluating their full distributions. Second, we will discuss the impact of dimension on the obtained results, since benchmark solutions for transport in $2d$ extruded and $3d$ tessellations have never been addressed before~\cite{somaini}.

This paper is organized as follows. In Sec.~\ref{benchmark_definition} we recall the benchmark specifications and set up the required notation. In Sec.~\ref{color_poisson} we discuss the algorithms needed in order to generate the material configurations corresponding to homogeneous Markov mixing, by resorting to the so-called colored Poisson tessellations. Then, in Sec.~\ref{simulation_results} we will present our simulation results for the physical observables of interest, and discuss the obtained findings. Conclusions will be finally drawn in Sec.~\ref{conclusions}.

\section{Benchmark specifications}
\label{benchmark_definition}

The benchmark specifications for our work are essentially taken from those originally proposed in~\cite{benchmark_adams} and~\cite{renewal}, and later extended in~\cite{vasques_suite2, brantley_benchmark, brantley_conf, brantley_conf_2}. We consider single-speed linear particle transport through a stochastic binary medium with homogeneous Markov mixing. The medium is non-multiplying, with isotropic scattering. The geometry consists of a cubic box of side $L=10$, with reflective boundary conditions on all sides of the box except two opposite faces (say those perpendicular to the $x$ axis), where leakage boundary conditions are imposed: particles that leave the domain through these faces can not re-enter. Lengths are expressed in arbitrary units. In~\cite{benchmark_adams} and~\cite{renewal}, system sizes $L=0.1$ and $L=1$ were also considered, but in this work we will focus on the case $L=10$, which leads to more physically relevant configurations. Two kinds of non-stochastic sources will be considered: either an imposed normalized incident angular flux on the leakage surface at $x=0$ (with zero interior sources), or a distributed homogeneous and isotropic normalized interior source (with zero incident angular flux on the leakage surfaces). Following the notation in~\cite{brantley_benchmark}, the benchmark configurations pertaining to the former kind of source will be called {\em suite} I, whereas those pertaining to the latter will be called {\em suite} II. The material properties for the Markov mixing are entirely defined by assigning the average chord length for each material $i = \alpha, \beta$, namely $\Lambda_i$, which in turn allows deriving the homogeneous probability $p_i$ of finding material $i$ at an arbitrary location within the box, namely
\begin{equation}
p_{i}= \frac{\Lambda_i}{\Lambda_i + \Lambda_j}.
\end{equation}
Note that the material probability $p_i$ defines the volume fraction for material $i$. The cross sections for each material will be denoted as customary $\Sigma_i$ for the total cross section and $\Sigma_{s,i}$ for the scattering cross section. The average number of particles surviving a collision in material $i$ will be denoted by $c_i = \Sigma_{s,i} / \Sigma_i \le 1$. The physical parameters for the benchmark configurations are recalled in Tabs.~\ref{tab_param1} and~\ref{tab_param2}: three cases (numbered $1$, $2$ and $3$) are considered, each containing three sub-cases (noted $a$, $b$ and $c$). The case numbers correspond to permutation of materials, whereas the sub-cases represents varying ratios of $c_i $ for each material.

\begin{table}[!ht]
\begin{center}
\begin{tabular}{lcccc}
\toprule
Case & $\Sigma_\alpha$ & $\Lambda_\alpha$ & $\Sigma_\beta$ & $\Lambda_\beta$ \\
\midrule
1 & 10/99 & 99/100 & 100/11 & 11/100 \\
2 & 10/99 & 99/10 & 100/11 & 11/10 \\
3 & 2/101 & 101/20 & 200/101 & 101/20 \\
\bottomrule
\end{tabular}
\end{center}
\caption{Material parameters for the three cases of the benchmark configurations.}
\label{tab_param1}
\end{table}

\begin{table}[!ht]
\begin{center}
\begin{tabular}{lccc}
\toprule
Sub-case & a & b & c \\
\midrule
$c_\alpha$ & 0 & 1 & 0.9 \\
$c_\beta$ & 1 & 0 & 0.9 \\
\bottomrule
\end{tabular}
\end{center}
\caption{Material parameters for the three sub-cases of the benchmark configurations.}
\label{tab_param2}
\end{table}

The physical observables of interest for the proposed benchmark will be the ensemble-averaged outgoing particle currents $\langle J \rangle $ on the two surfaces with leakage boundary conditions, and the ensemble-averaged scalar particle flux $\langle \varphi(x) \rangle= \langle \int \int \int\varphi({\bf r}, {\boldsymbol \omega}) d {\boldsymbol \omega} dy dz \rangle$ along $0 \le x \le L$. For the {\em suite} I configurations, the outgoing particle current on the side opposite to the imposed current source will represent the ensemble-averaged transmission coefficient, namely, $\langle T \rangle = \langle J_{x=L} \rangle $, whereas the outgoing particle current on the side of the current source will represent the ensemble-averaged reflection coefficient, namely, $\langle R \rangle = \langle J_{x=0} \rangle $. For the {\em suite} II configurations, the outgoing currents on opposite faces are expected to be equal (within statistical fluctuations), for symmetry reasons. In this case, we also introduce the average leakage current $\langle J_\text{ave} \rangle = \langle (T+R)/2 \rangle$.

For the sake of completeness, we have also considered the so-called atomic mixing model~\cite{pomraning}, where one assumes that the statistical disorder can be approximated by simply taking a full homogenization of the physical properties based on the ensemble-averaged cross sections. The atomic mixing approximation is known to fail whenever the linear size of the material chunks composing the stochastic mixture is not small compared to the mean free path of the particles~\cite{pomraning}. The ensemble-averaged scattering cross section $\langle \Sigma_s \rangle$ needed for the atomic mixing are resumed in Tab.~\ref{tab_param3}. The ensemble-averaged total cross section $\langle \Sigma \rangle = p_\alpha \Sigma_\alpha + p_\beta \Sigma_\beta$ for all cases and sub-cases of the benchmark is $\langle \Sigma \rangle = 1$.

\begin{table}[!ht]
\begin{center}
\begin{tabular}{lc}
\toprule
Configuration & $\langle \Sigma_s \rangle$ \\
\midrule
1a, 2a & 10/11 \\
1b, 2b & 1/11 \\
3a & 100/101\\
3b & 1/101\\
1c, 2c, 3c & 9/10\\
\bottomrule
\end{tabular}
\end{center}
\caption{Ensemble-averaged scattering cross section $\langle \Sigma_s \rangle$ for the benchmark configurations.}
\label{tab_param3}
\end{table}

\section{Poisson geometries}
\label{color_poisson}

Poisson geometries form a prototype process of isotropic stochastic tessellations: a portion of a $d$-dimensional space is partitioned by randomly generated $(d-1)$-dimensional hyper-planes drawn from an underlying Poisson process~\cite{santalo}.

In order for this paper to be self-contained, in this Section we will recall the strategy for the Poisson tessellation of a $d$-dimensional box~\cite{larmier}. An explicit construction amenable to Monte Carlo realizations for two-dimensional homogeneous and isotropic Poisson geometries of finite size has been established in~\cite{switzer}. The case $d=3$, which is key for real-world applications, has comparatively received less attention. A generalization of the construction algorithm to three-dimensional (and in principle $d$-dimensional) domains has nonetheless recently been proposed~\cite{serra, mikhailov}.

The method proposed by~\cite{mikhailov} is based on a spatial tessellation of the $d$-hypersphere of radius $R$ centered at the origin by generating a random number $m$ of $(d-1)$-hyperplanes with random orientation and position. Any given $d$-dimensional region within the $d$-hypersphere will therefore undergo the same tessellation procedure. The number $m$ of $(d-1)$-hyperplanes is sampled from a Poisson distribution with parameter $\rho R {\cal A}_d(1)/{\cal V}_{d-1}(1) $. Here ${\cal A}_{d}(1)=2\pi^{d/2}/\Gamma(d/2)$ denotes the surface of the $d$-dimensional unit sphere~\footnote{$\Gamma(a)$ being the Gamma function~\cite{special_functions}}, ${\cal V}_{d}(1)=\pi^{d/2}/\Gamma(1+d/2)$ denotes the volume of the $d$-dimensional unit sphere, and $\rho$ is the arbitrary density of the tessellation, carrying the units of an inverse length. This normalization of the density $\rho$ corresponds to the convention used in~\cite{santalo}, and is such that $\rho s$ yields the mean number of $(d-1)$-hyperplanes intersected by an arbitrary segment of length $s$.

\begin{figure}[t]
\begin{center}
\,\,\,\, Case 1 \,\,\,\,\\
\includegraphics[width=0.42\columnwidth]{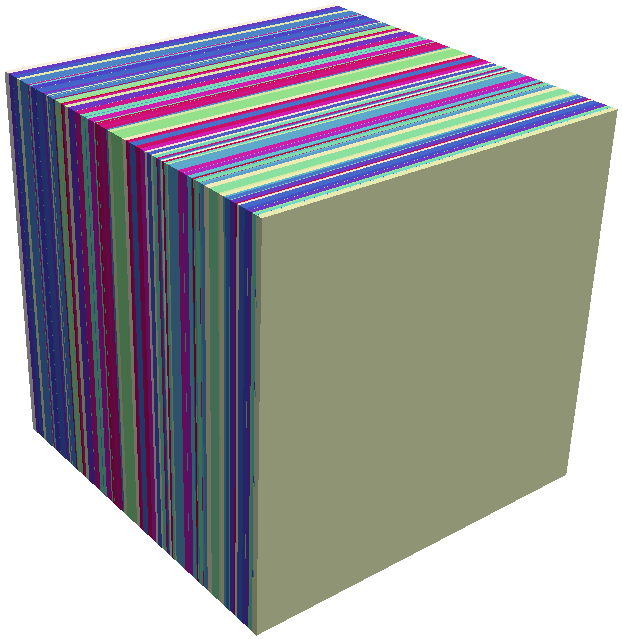}\,\,\,\,
\includegraphics[width=0.42\columnwidth]{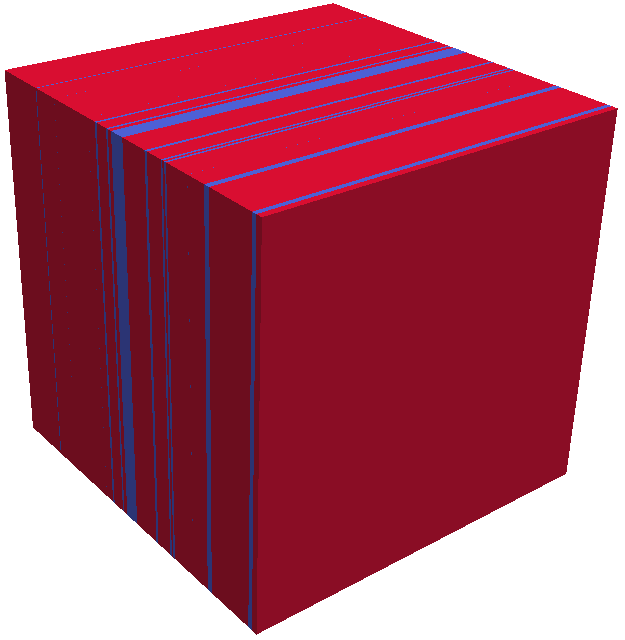}\\
\,\,\,\, Case 2 \,\,\,\,\\
\includegraphics[width=0.42\columnwidth]{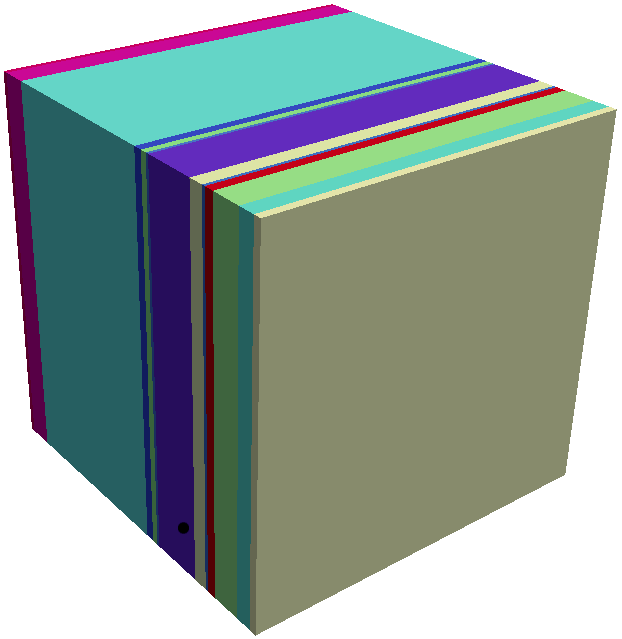}\,\,\,\,
\includegraphics[width=0.42\columnwidth]{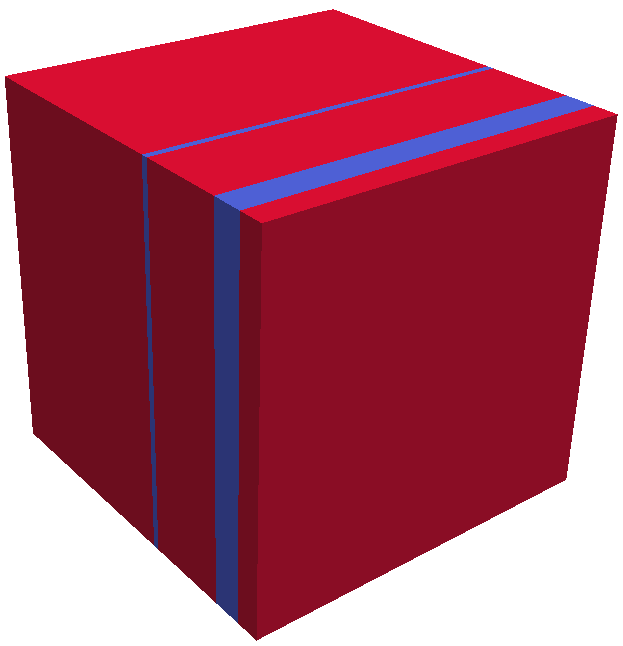}\\
\,\,\,\, Case 3 \,\,\,\,\\
\includegraphics[width=0.42\columnwidth]{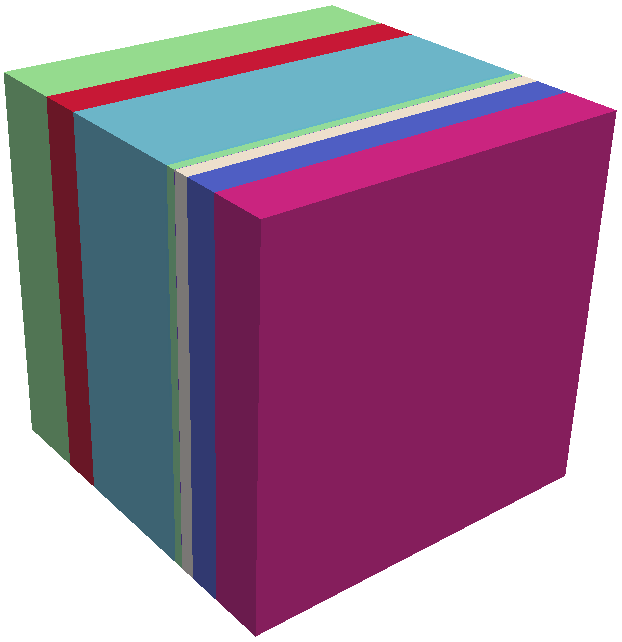}\,\,\,\,
\includegraphics[width=0.42\columnwidth]{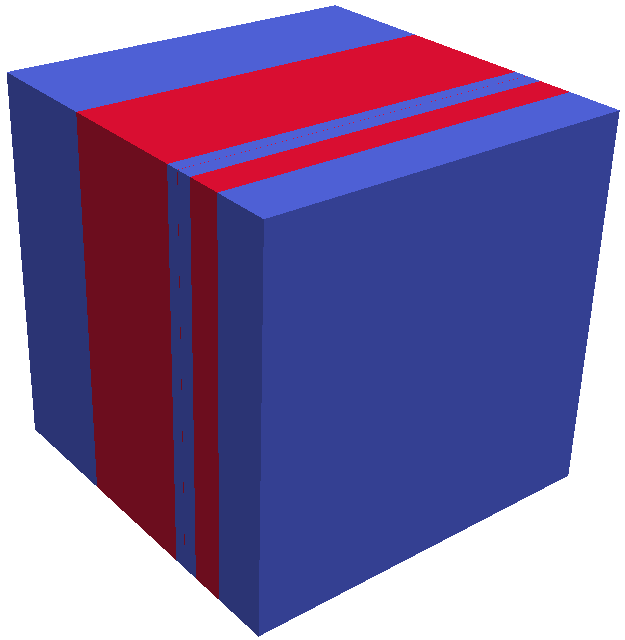}\\
\end{center}
\caption{Examples of realizations of Poisson geometries corresponding to the benchmark specifications for the $1d$ slab tessellations, before (left) and after (right) attributing the material label. Red denotes label $\alpha$ and blue denotes label $\beta$. For cases $1$ and $2$, $p_\alpha=0.9$, whereas for case $3$ $p_\alpha = 0.5$.}
\label{fig1}
\end{figure}

For the $1d$ slab tessellations, the algorithm proceeds then as follows. The first step consists in sampling a random number of points $m$ from a Poisson distribution of parameter $2 \rho R$, where the factor $2$ stems from ${\cal A}_1(1)/{\cal V}_0(1) = 2$. The points are uniformly distributed on a segment aligned over the $x$ axis, in the interval $0 \le  x \le L = R$ (the $1$-sphere circumscribed to the segment actually coincides with the segment). Then, the cube of side $L$ will be cut by planes passing through the generated points and orthogonal to the $x$ axis (see Fig.~\ref{fig1}).

\begin{figure}[t]
\begin{center}
\,\,\,\, Case 1 \,\,\,\,\\
\includegraphics[width=0.42\columnwidth]{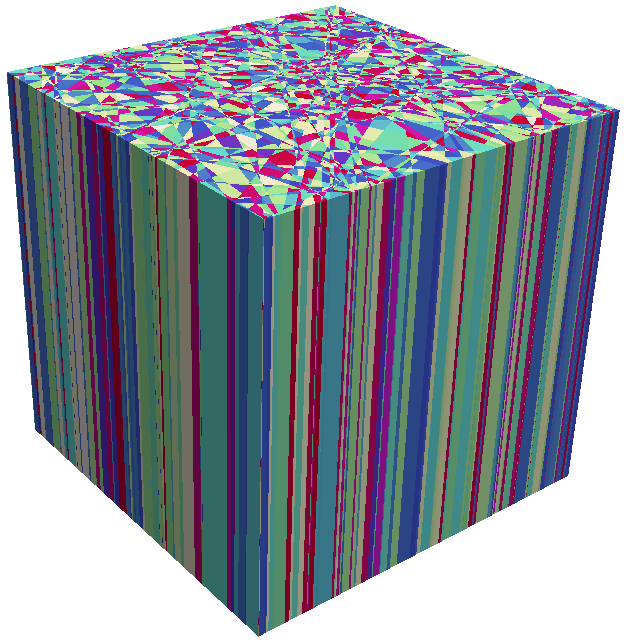}\,\,\,\,
\includegraphics[width=0.42\columnwidth]{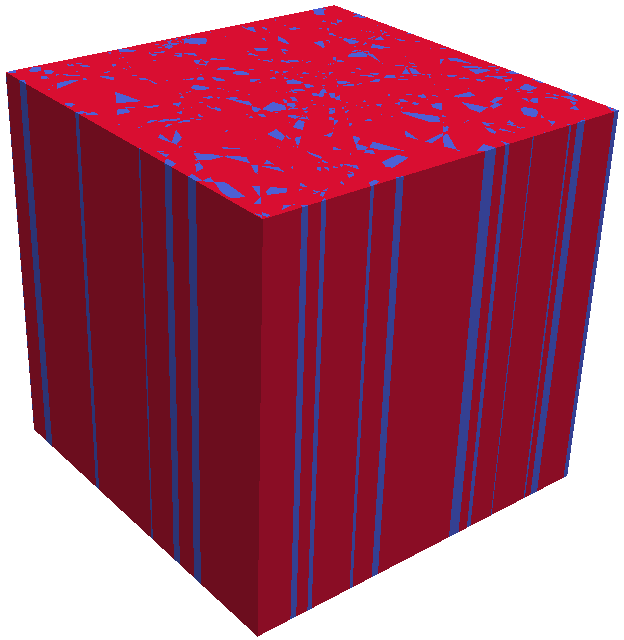}\\
\,\,\,\, Case 2 \,\,\,\,\\
\includegraphics[width=0.42\columnwidth]{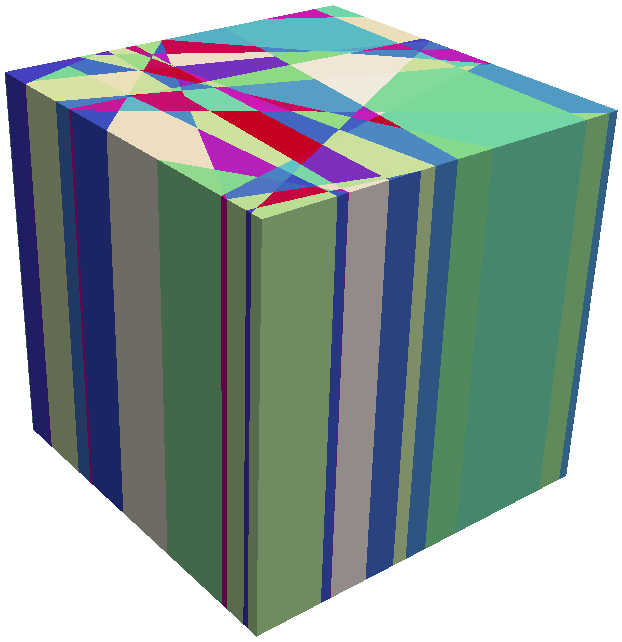}\,\,\,\,
\includegraphics[width=0.42\columnwidth]{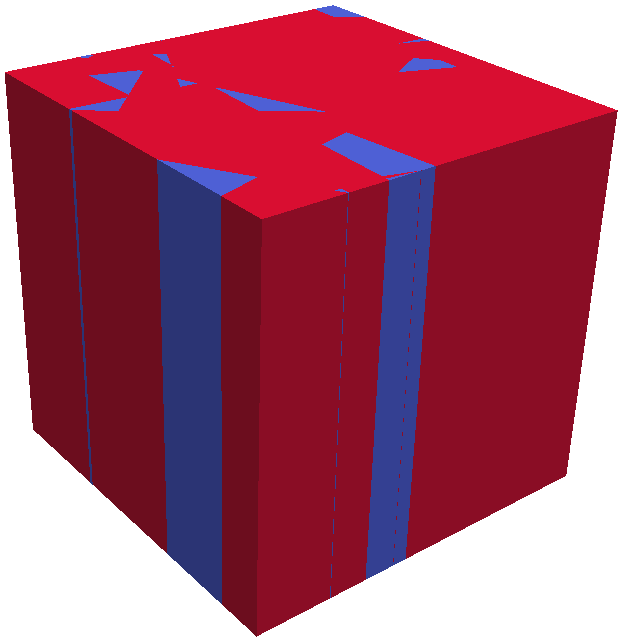}\\
\,\,\,\, Case 3 \,\,\,\,\\
\includegraphics[width=0.42\columnwidth]{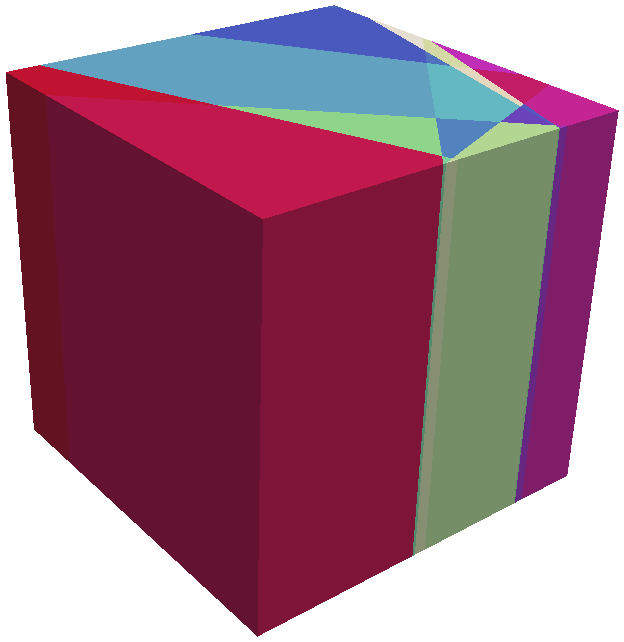}\,\,\,\,
\includegraphics[width=0.42\columnwidth]{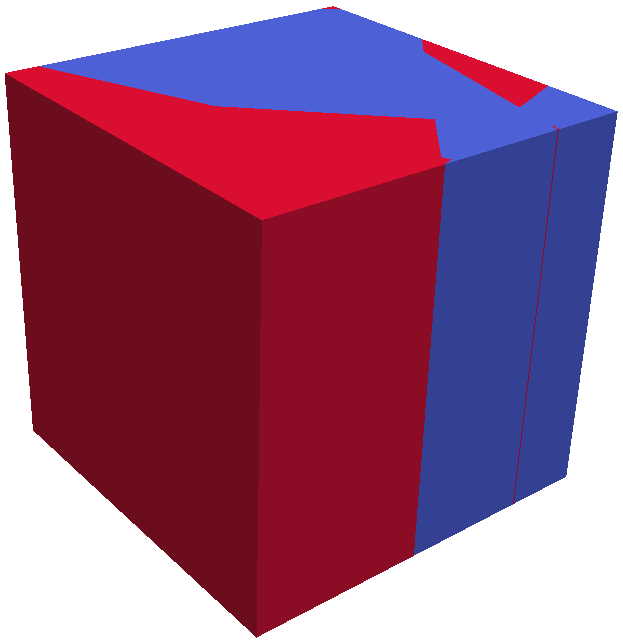}\\
\end{center}
\caption{Examples of realizations of Poisson geometries corresponding to the benchmark specifications for the $2d$ extruded tessellations, before (left) and after (right) attributing the material label. Red denotes label $\alpha$ and blue denotes label $\beta$. For cases $1$ and $2$, $p_\alpha=0.9$, whereas for case $3$ $p_\alpha = 0.5$.}
\label{fig2}
\end{figure}

For the $2d$ extruded tessellations, we begin by creating an isotropic tessellation of a square of side $L$. Suppose for the sake of simplicity that the square is centered in the origin $O$. We denote by $R$ the radius of the circle circumscribed to the square. We sample a random number of points $m$ from a Poisson distribution of parameter $\pi \rho R $, where we have used ${\cal A}_2(1)/{\cal V}_1(1) = \pi$. We then sample the random lines that will cut the square: we choose a radius $r$ uniformly in the interval $[0,R]$ and then draw $\theta$ uniformly in $[0, 2\pi]$. Based on these two random parameters, a unit vector ${\mathbf n}=(n_1,n_2)^T$ is generated, with components
\begin{align}
n_1&=\cos{\theta}\nonumber \\
n_2&=\sin{\theta}.\nonumber
\end{align}
Let now $\mathbf M$ be the point such that ${\mathbf{ OM}}=r {\mathbf n}$. The random line will be finally defined by the equation $n_1 x + n_2 y =r$, passing trough $\mathbf M$ and having normal vector ${\mathbf n}$. By construction, this line does intersect the circumscribed disk of radius $R$ but not necessarily the square. It can be shown that the probability for a random line to fall within the square is $2\sqrt{2}/\pi \simeq 0.900$ ~\cite{santalo}. This procedure is iterated until $m$ random lines have been generated. By construction, the polygons defined by the intersection of such random lines are convex. Once the square has been tessellated, the full geometrical description for the cube is simply achieved by extruding the random polyhedra (which lie on the $x-y$ plane) along the orthogonal $z$ axis (see Fig.~\ref{fig2}).

\begin{figure}[t]
\begin{center}
\,\,\,\, Case 1 \,\,\,\,\\
\includegraphics[width=0.42\columnwidth]{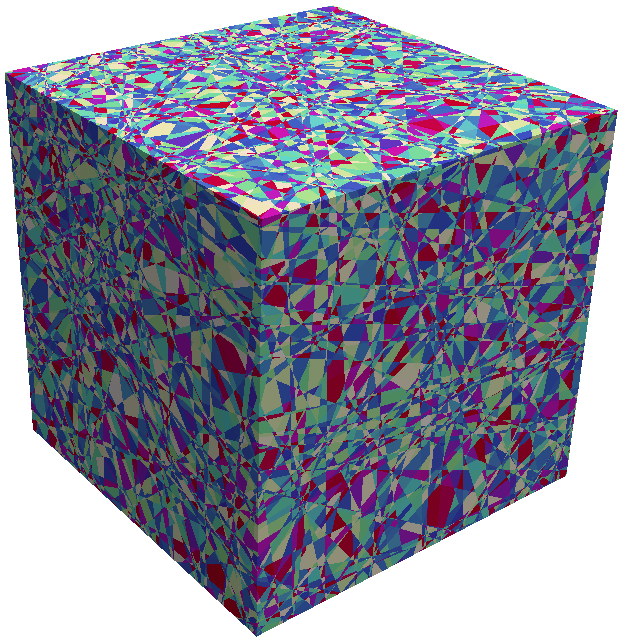}\,\,\,\,
\includegraphics[width=0.42\columnwidth]{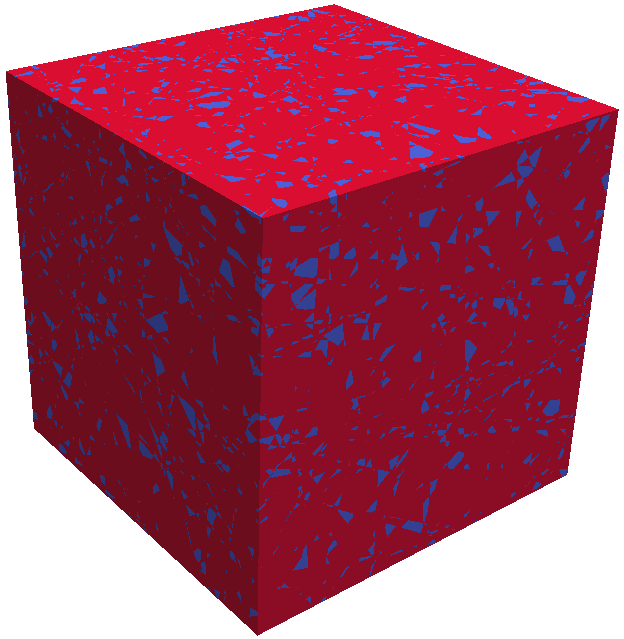}\\
\,\,\,\, Case 2 \,\,\,\,\\
\includegraphics[width=0.42\columnwidth]{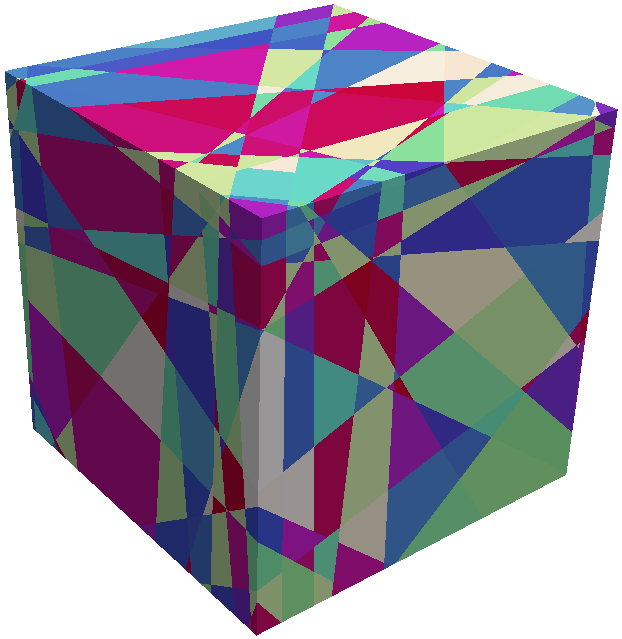}\,\,\,\,
\includegraphics[width=0.42\columnwidth]{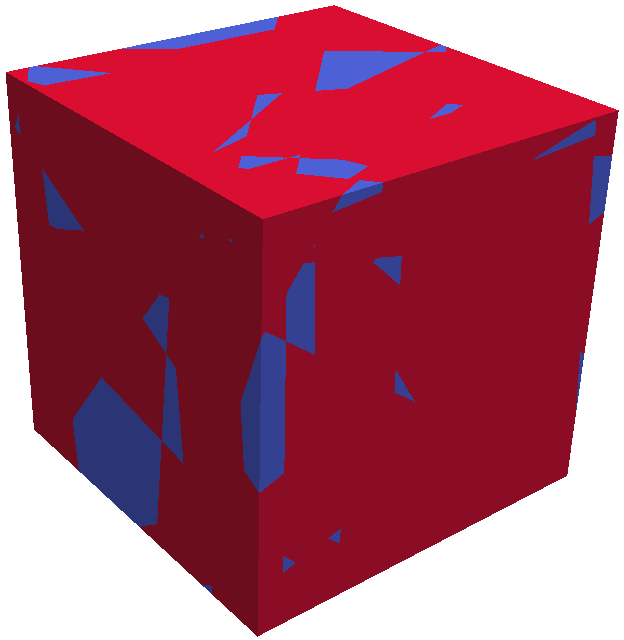}\\
\,\,\,\, Case 3 \,\,\,\,\\
\includegraphics[width=0.42\columnwidth]{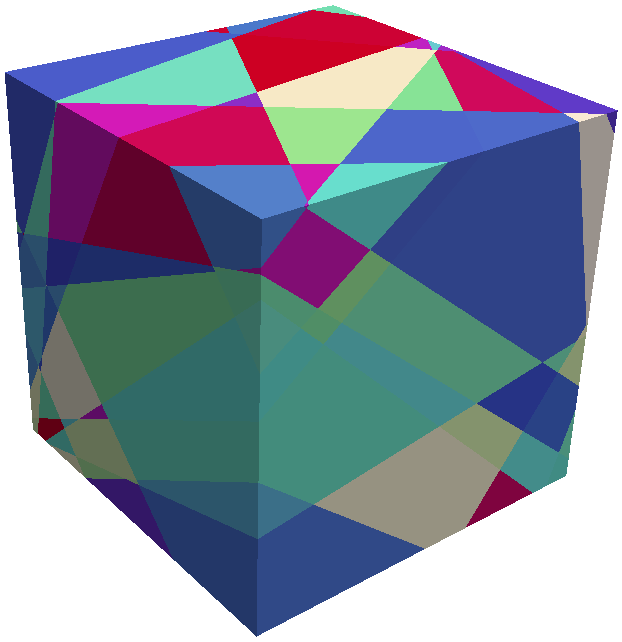}\,\,\,\,
\includegraphics[width=0.42\columnwidth]{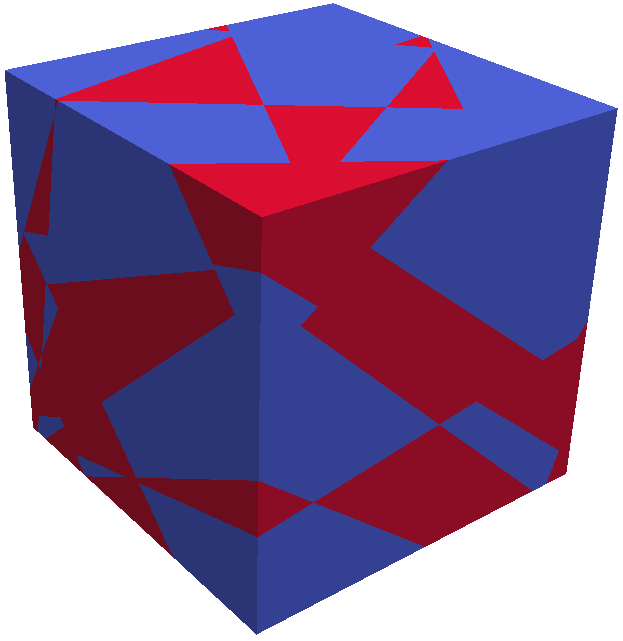}\\
\end{center}
\caption{Examples of realizations of Poisson geometries corresponding to the benchmark specifications for the $3d$ tessellations, before (left) and after (right) attributing the material label. Red denotes label $\alpha$ and blue denotes label $\beta$. For cases $1$ and $2$, $p_\alpha=0.9$, whereas for case $3$ $p_\alpha = 0.5$.}
\label{fig3}
\end{figure}

Let us now focus on $3d$ tessellations. We denote by $R$ the radius of the sphere circumscribed to the cube, and suppose that the cube is centered in the origin $O$. We start again by sampling a random number of points $m$ from a Poisson distribution of parameter $4 \rho R$, where we have used ${\cal A}_3(1)/{\cal V}_2(1) = 4$. Then we generate the planes that will cut the cube. We choose a radius $r$ uniformly in the interval $[0,R]$ and then sample two additional parameters, namely, $\xi_1$ and $\xi_2$, from two independent uniform distributions in the interval $[0,1]$. A unit vector ${\mathbf n}=(n_1,n_2,n_3)^T$ with components
\begin{align}
n_1&=1-2\xi_1 \nonumber \\
n_2&=\sqrt{1-n_1^2}\cos{(2 \pi \xi_2)}\nonumber \\
n_3&=\sqrt{1-n_1^2}\sin{(2 \pi \xi_2)}\nonumber
\end{align}
is generated. Denoting again $\mathbf M$ the point such that ${\mathbf{ OM}}=r {\mathbf n}$, the random plane will finally obey $n_1 x + n_2 y +n_3 z =r$, passing trough $\mathbf M$ and having normal vector ${\mathbf n}$. By construction, this plane does intersect the circumscribed sphere of radius $R$ but not necessarily the cube: the probability that the plane intersects both the sphere and the cube can be shown to be $\sqrt{3}/2 \simeq 0.866$~\cite{santalo}. The procedure is iterated until $m$ random planes have been generated. The polyhedra defined by the intersection of such random planes are convex (see Fig.~\ref{fig3}).

\subsection{Statistical properties of Poisson geometries}
\label{uncolored_geo}

The physical observables of interest associated to the Poisson geometries, such as for instance the volume of a polyhedron, its surface, the number of edges, and so on, are clearly random variables, whose exact distributions are unfortunately unknown, with a few remarkable exceptions~\cite{santalo}. The number $N_p$ of polyhedra in $d$-dimensional Poisson geometries at finite size $L$ is known to obey $\langle N_p|L\rangle \sim L^d$ for large $L$~\cite{larmier}. The quantity $N_p$ provides a measure of the complexity of the resulting geometries: this means that the computational cost to generate a realization of a Poisson geometry is an increasing function of the system size and of the dimension. The dispersion factor scales as $\sigma[N_p|L]/\langle N_p|L\rangle \sim 1/\sqrt{L}$~\cite{larmier}: for large systems, the distribution of $N_p$ will be then peaked around the average value $\langle N_p|L\rangle$.

Poisson geometries satisfy a Markov property: for domains of infinite size, arbitrary drawn lines will be cut by the $(d-1)$-surfaces of the $d$-polyhedra into segments whose lengths $\ell$ are exponentially distributed, i.e.,
\begin{equation}
P(\ell)  = \rho e^{-\rho \ell},
\label{asy_length}
\end{equation}
with average $\langle \ell \rangle = \int \ell P(\ell) d\ell = 1/\rho$~\cite{santalo}. The quantity $\Lambda = 1/\rho$ intuitively defines the correlation length of the Poisson geometry, i.e, the typical linear size of a volume composing the random tessellation. For small $L$, finite-size effects will come into play, due to the fact that the longest line that can be drawn across a box of linear size $L$ is $\sqrt{d} L$, which implies a cut-off on the distribution. However, for $\Lambda \ll L$, the finite-size effects fade away and the probability density $P(\ell)$ eventually converge to the expected exponential behaviour~\cite{larmier}.

\subsection{Assigning material properties: colored geometries}

Binary Markov mixtures required for the benchmark specifications are obtained as follows: first, a $d$-dimensional Poisson tessellation is constructed as described above. Then, each polyhedron of the geometry is assigned a material composition by formally attributing a distinct `label' (also called `color'), say `$\alpha$' or `$\beta$', with associated complementary probabilities $p_\alpha$ and $p_\beta = 1-p_\alpha$. Adjacent polyhedra sharing the same label are finally merged. This gives rise to (generally) non-convex $\alpha$ and $\beta$ clusters, each composed of a random number of convex polyhedra. The statistical features of Poisson binary mixtures, including percolation probabilities and exponents, have been previously addressed in~\cite{lepage} for $2d$ geometries and in~\cite{larmier} for $3d$ geometries.

Due to the Markov property, it can be shown that the average chord length $\Lambda_\alpha$ through clusters with composition $\alpha$ is related to the correlation length $\Lambda$ of the geometry via
\begin{equation}
\Lambda = (1-p_\alpha) \Lambda_\alpha,
\end{equation}
and for $\Lambda_\beta$ we similarly have
\begin{equation}
\Lambda =  p_\alpha \Lambda_\beta.
\end{equation}
This yields $1 / \Lambda_\alpha + 1 / \Lambda_\beta =  1 / \Lambda$, and we recover
\begin{equation}
p_\alpha = \frac{\Lambda}{\Lambda_\beta} = \frac{\Lambda_\alpha}{\Lambda_\alpha + \Lambda_\beta}.
\end{equation}
Thus, based on the formulas above, and using $\rho = 1/\Lambda$, the parameters of the colored Poisson geometries corresponding to the benchmark specifications provided in Tab.~\ref{tab_param1} are easily derived. Their values are resumed in Tab.~\ref{tab_param4}.

\begin{table}[!ht]
\begin{center}
\begin{tabular}{lcc}
\toprule
Case & $\rho$ & $p_\alpha$ \\
\midrule
1 & 1000/99 & 0.9 \\
2 & 100/99 & 0.9 \\
3 & 40/101 & 0.5 \\
\bottomrule
\end{tabular}
\end{center}
\caption{Parameters for the colored Poisson geometries corresponding to the three cases of the benchmark configurations.}
\label{tab_param4}
\end{table}

For the purpose of illustration, examples of realizations of Poisson geometries for the benchmark configurations are displayed in Fig.~\ref{fig1} for the $1d$ slab tessellations, in Fig.~\ref{fig2} for the $2d$ extruded tessellations, and in Fig.~\ref{fig3} for the $3d$ tessellations. The impact of varying the correlation lengths $\Lambda$ and the volumetric fractions $p_\alpha$ is apparent.

\subsection{Description of the benchmark configurations}

Among the benchmark configurations, case $1$ displays the largest fragmentation, since the material chord lengths $\Lambda_\alpha$ and $\Lambda_\beta$ are both much smaller than the linear size $L$ of the box (see Tab.~\ref{tab_param1}). This is mirrored in a small correlation length $\Lambda \simeq 0.1$, and a very large number of $d$-polyhedra $\langle N_p |L \rangle$ composing the tessellation (see Tab.~\ref{tab_geo}). The volume fractions are $p_\alpha = 0.9$ and $p_\beta = 0.1$, respectively, so that it is much more probable to cross material $\alpha$ than material $\beta$: chunks of material $\beta$ are small and well mixed within material $\alpha$. For case $1a$, material $\alpha$ is purely absorbing, but with a small cross section, whereas material $\beta$ is purely scattering, and is opaque due to its large cross section: in this case, a typical realization will consist in small chunks of opaque material dispersed in larger chunks of an almost transparent material. In the atomic mixing approximation, case $1a$ is mainly scattering.

Case $1b$ is the opposite of case $1a$, with material $\alpha$ being now purely scattering, and material $\beta$ being purely absorbing. A typical realization will then consist in small absorbing chunks, with a large cross section, dispersed in larger chunks of an almost transparent material. In the atomic mixing approximation, case $1b$ is mainly absorbing.

For case $1c$, the materials have $c_i \neq 1$, so that the chunks have intermediate properties between absorption and scattering.

The features of case $2$ are such that the total cross sections and the volume fractions are the same as in case $1$, but with larger mean chord lengths $\Lambda_\alpha$ and $\Lambda_\beta$ (see Tab.~\ref{tab_param1}). Correspondingly, the correlation length $\Lambda \simeq 1$ is ten times larger than that of case $1$, which leads to a considerably lower number of $d$-polyhedra $\langle N_p |L \rangle$ with respect to case $1$ (see Tab.~\ref{tab_geo}). The chunks of material $\alpha$ are rather large, whereas those of material $\beta$ are smaller. The volume fractions are the same as those of case $1$, and so are the total cross sections. The scattering and absorbing cross sections for all the sub-cases are then equal to those of case $1$.

For case $3$, materials $\alpha$ and $\beta$ share the same mean chord length, and their respective volume fractions are also equal, i.e., $p_\alpha = p_\beta = 0.5$ (see Tab.~\ref{tab_param1}). Chunks are larger than for case $2$, with a correlation length $\Lambda \simeq 2.5$, and the number of $d$-polyhedra $\langle N_p |L \rangle$ is thus lower than for case $2$ (see Tab.~\ref{tab_geo}). A typical realization in $d=1$ will be therefore dominated by a single material, which means that the entire box will be typically either colored in red or in blue, but this effect fades away with increasing dimension, as apparent from Fig.~\ref{fig2} for the $2d$ extruded tessellations and from Fig.~\ref{fig3} for the $3d$ tessellations. The total cross sections of $\alpha$ and $\beta$ are smaller than those of case $1$ and $2$, with different scattering and absorbing cross sections for each sub-case.

In case $3a$, material $\alpha$ is purely absorbing, with a small cross section, whereas material $\beta$ is purely scattering, with a large cross section. In case $3b$, the role of the absorbing and scattering material are inverted, as before. For case $3c$, both materials are partly absorbing and reflecting, and the global cross sections are the same of cases $1c$ and $2c$.

\subsection{Finite size effects}

The chord lengths $\Lambda_\alpha$ and $\Lambda_\beta$ given in the benchmark specifications apply strictly speaking to infinite-size Poisson tessellations, i.e., in the limit $L \gg \Lambda$. For any given system size $L$, finite-size effects will in principle come into play. In order to assess the impact of such effects on the benchmark configurations, we have computed the chord lengths within each material by Monte Carlo ray tracing. A large number of random straight lines traversing the tessellations have been sampled by imposing an isotropic and homogeneous incident flux on the surface of the box (the so-called $\mu$-randomness~\cite{santalo}), and we have thus estimated the corresponding average chord lengths $\Lambda_\alpha(L)$ and $\Lambda_\beta(L)$ by measuring the portion of the lines spent in material $\alpha$ and $\beta$, respectively. Simulation results are reported in Tab.~\ref{tab_geo}, where the measured $\Lambda_\alpha(L)$ and $\Lambda_\beta(L)$ are compared to the expected values $\Lambda_\alpha$ and $\Lambda_\beta$ for infinite tessellations. For case $1$, the discrepancy between the measured average chords $\Lambda_\alpha(L)$ and $\Lambda_\beta(L)$ and the asymptotic limits $\Lambda_\alpha$ and $\Lambda_\beta$ is quite small, which is due to the high fragmentation of the tessellation. The discrepancy increases for cases $2$ and $3$, as expected on physical grounds, since the material chunks are larger, and become comparable to the box size. We have separately computed the correlation length $\Lambda(L)$ for the tessellation by Monte Carlo ray tracing (based again on the $\mu$-randomness): not surprisingly the discrepancies with respect to the asymptotic value $\Lambda$ for infinite-size geometries follow the same pattern as for the material chord lengths (see Tab.~\ref{tab_geo}).

\section{Monte Carlo simulation results}
\label{simulation_results}

The reference solutions for the ensemble-averaged scalar particle flux $\langle \varphi(x) \rangle$ and the currents $\langle R \rangle$ and $\langle T \rangle$ have been computed as follows. For each benchmark case and sub-case, a large number $M$ of geometries has been generated, and the material properties have been attributed to each volume as described above. Then, for each realization $k$ of the ensemble, linear particle transport has been simulated by resorting to the production Monte Carlo code \tripoli{}, developed at CEA~\cite{T4}. \tripoli{} is a general-purpose stochastic transport code capable of simulating the propagation of neutral and charged particles with continuous-energy cross sections in arbitrary geometries. In order to comply with the benchmark specifications, constant cross sections adapted to mono-energetic transport and isotropic angular distributions have been prepared. The number of simulated particle histories per configuration is $10^6$. For a given physical observable ${\cal O}$, the benchmark solution is obtained as the ensemble average
\begin{equation}
\langle {\cal O} \rangle = \frac{1}{M} \sum_{k=1}^M {\cal O}_k,
\end{equation}
where ${\cal O}_k$ is the Monte Carlo estimate for the observable ${\cal O}$ obtained for the $k$-th realization. Specifically, currents $R_k$ and $T_k$ at a given surface are estimated by summing the statistical weights of the particles crossing that surface. Scalar fluxes $\varphi_k(x)$ have been recorded by resorting to the standard track length estimator over a pre-defined spatial grid containing $10^2$ uniformly spaced meshes along the $x$ axis.

The error affecting the average observable $\langle {\cal O} \rangle$ results from two separate contributions, the dispersion
\begin{equation}
\sigma^2_G = \frac{1}{M} \sum_{k=1}^M {{\cal O}_k}^2 - {\langle {\cal O} \rangle}^2
\end{equation}
of the observables exclusively due to the stochastic nature of the geometries and of the material compositions, and 
\begin{equation}
\sigma^2_{{\cal O}}=\frac{1}{M} \sum_{k=1}^M \sigma_{{\cal O}_k}^2,
\end{equation}
which is an estimate of the variance due to the stochastic nature of the Monte Carlo method for the particle transport, $\sigma_{{\cal O}_k}^2$ being the dispersion of a single calculation~\cite{donovan, sutton}. The statistical error on $\langle {\cal O} \rangle$ is then estimated as
\begin{equation}
\sigma[ \langle {\cal O}\rangle ] = \sqrt{\frac{\sigma^2_G}{M}+\sigma^2_{{\cal O}}}.
\end{equation}

Depending on the correlation lengths and on the volumetric fractions, the physical observables might display a larger or smaller dispersion around their average values. In order to assess the impact of such dispersion, we have also computed the full distributions of $T$ and $R$ based on the available realizations.

The number $M$ of realizations that have been used for the Monte Carlo simulations has been chosen as follows: for $1d$ slab tessellations, we have taken $M=10^4$ (except for the case $2a$ for the {\em suite} II, where the number of geometries has been increased to $M=5\times 10^4$ in order to reduce the statistical fluctuations); for the $2d$ extruded tessellations, we have taken $M=4 \times 10^3$; finally, for the $3d$ tessellations we have taken $M=10^3$. As a general remark, increasing the dimension implies an increasing computational burden (each realization takes longer both for generation and for Monte Carlo transport), but also a better statistical mixing (a single realization is more representative of the average behaviour).

Concerning particle transport, it is important to stress that for the simulations discussed here we have largely benefited from a feature that has been recently implemented in the code \tripoli{}, namely the possibility of reading pre-computed connectivity maps for the volumes composing the geometry. During the generation of the Poisson tessellations, care has been taken so as to store the indexes of the neighbouring volumes for each realization, which means that during the geometrical tracking a particle will have to find the following crossed volume in a list that might be considerably smaller than the total number of random volumes composing the box (depending on the features of the random geometry). To provide and example, a typical realization of a $3d$ geometry for case $1$ will be composed of $\sim 10^5$ volumes, whereas the typical number of neighbours for each volume will of the order of $\sim 10$. When fed to the transport code, such connectivity maps allow thus for considerable speed-ups for the most fragmented geometries, up to one hundred.

Transport calculations have been run on a cluster based at CEA, with Xeon E5-2680 V2 2.8 GHz processors. An overview of the average computer time $\langle t \rangle$ for each benchmark configuration is provided in Tabs.~\ref{tab_times1} and \ref{tab_times2}. Dispersions $\sigma[ t]$ are also given. While an increasing trend for $\langle t \rangle$ as a function of dimension is clearly apparent, subtle effects due to correlation lengths and volume fractions for the material compositions come also into play, and strongly influence the average computer time. For some configurations, the dispersion $\sigma[ t]$ may become very large, and even be comparable to the average $\langle t \rangle$. Atomic mixing simulations are based on a single homogenized realization, and the dispersion is thus trivially zero.

\subsection{Transmission, reflection and integral flux}

The simulation results for the ensemble-averaged transmission coefficient $\langle T \rangle$, the reflection coefficient $\langle R \rangle$ and the integral flux $\langle \varphi  \rangle= \langle \int \int\varphi({\bf r}, {\boldsymbol \omega}) d {\boldsymbol \omega} d {\bf r}\rangle$ are provided in Tabs.~\ref{tab_suite1_case1} to \ref{tab_suite2_case3} for all the benchmark configurations, for both {\em suite} I and {\em suite} II conditions. For the sake of conciseness, for the {\em suite} II we denote by $\langle T \rangle$ the leakage current at $x=L$ and by $\langle R \rangle$ the leakage current at $x=0$. For the case of {\em suite} II, $R$ and $T$ are equivalent within statistical fluctuations, as expected. Atomic mixing results have been also given for reference. For each Monte Carlo transport simulation, the error on the estimated observable was significantly lower than $1\%$.

The computed values for the $1d$ slab configurations and the atomic mixing approximation are in excellent agreement (typically to two or three digits) with those previously reported in~\citep{benchmark_adams, brantley_benchmark, brantley_conf, brantley_conf_2, renewal}, and allow concluding that our choice for the benchmark specifications are coherent. For all examined cases, the atomic mixing approximation generally yields poor results as compared with the benchmark solutions, and in some cases the discrepancy can add up to several orders of magnitude. In addition, the atomic mixing solutions for several cases are strictly identical, since the ensemble-averaged total and scattering cross sections are identical by design. Concerning the benchmark solutions in dimension $d=1,2$ and $3$, the impact of dimension on the transmission and reflection coefficient is stronger between $d=1$ and $d=2$ than between $d=2$ and $d=3$, as expected on physical grounds, and has a large variability between cases (from less than $1\%$ for case $1b$ in {\em suite} I, to almost $100\%$ for case $2b$ in {\em suite} II). For {\em suite} I configurations, the reflection coefficient $\langle R \rangle$ in $d=1$ is always larger than those in $d=2,3$. The transmission coefficient $\langle T \rangle$ is also generally larger, apart from cases $1a$, $1c$, and $3a$, where it is smaller. For {\em suite} II configurations, the leakage coefficient in $d=1$ is generally larger than in $d=2,3$, apart from case $1a$, where it is smaller, and case $3a$, where it is almost constant with respect to dimension.

\subsection{Distributions of transmission and reflection coefficients}

\begin{figure*}
\begin{center}
\,\,\,\, Case 1a \,\,\,\,\\
\includegraphics[width=0.9\columnwidth]{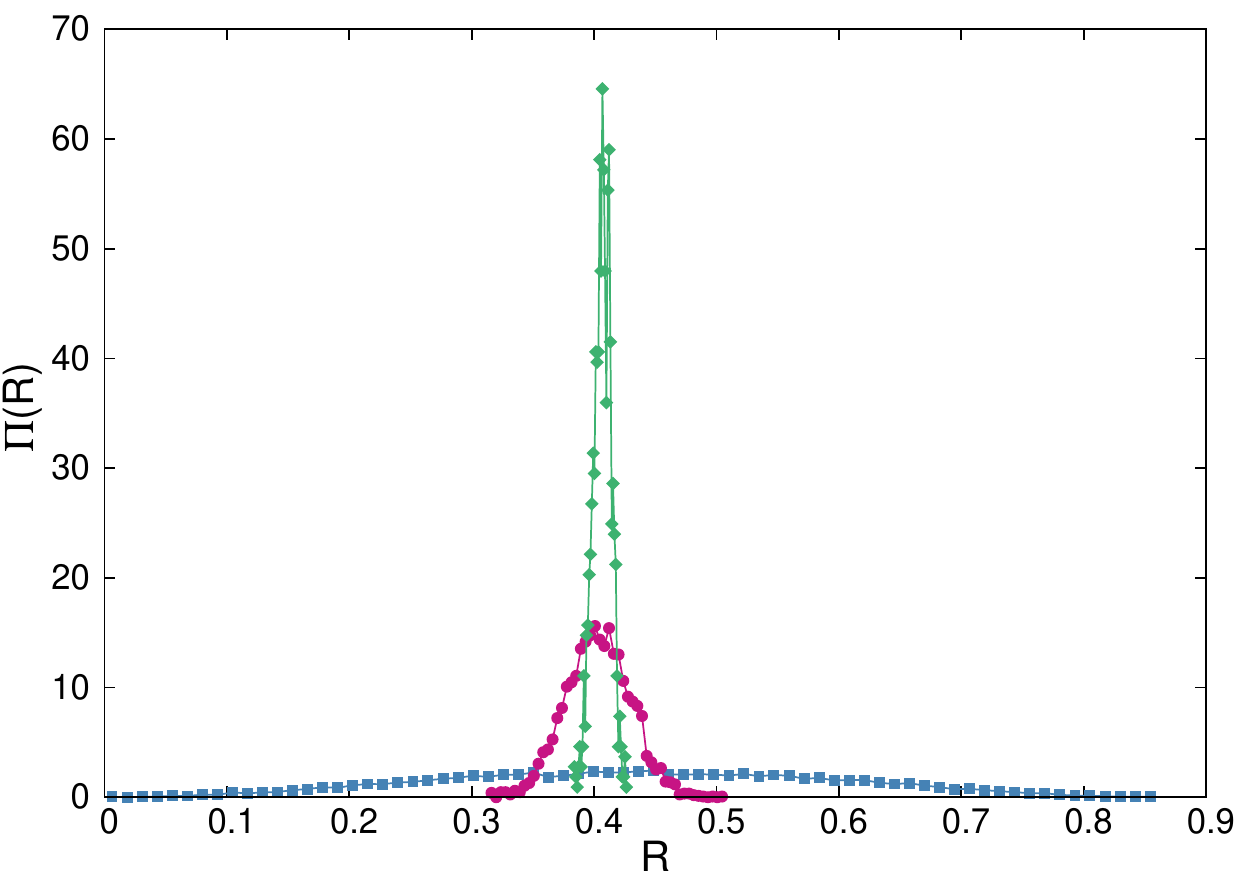}\,\,\,\,
\includegraphics[width=0.9\columnwidth]{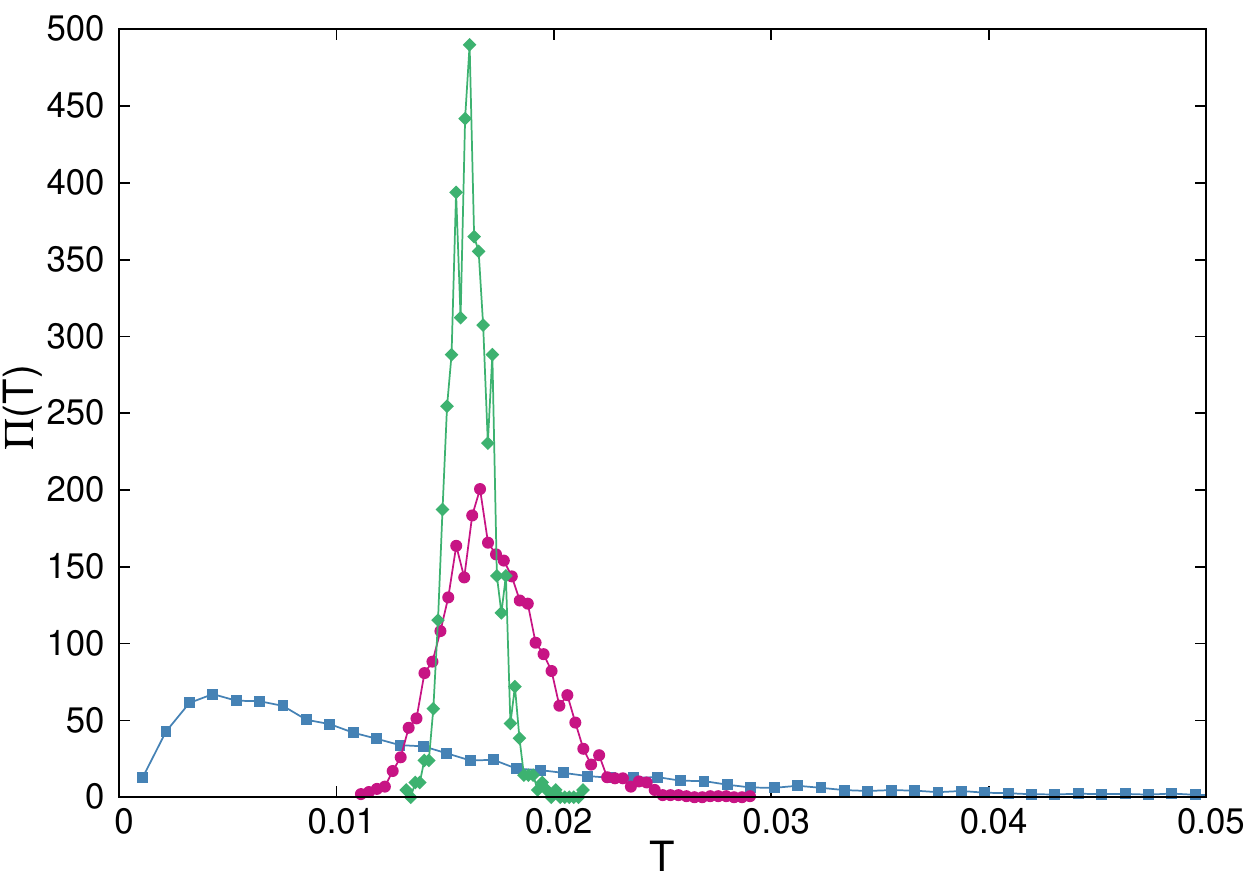}\\
\,\,\,\, Case 1b \,\,\,\,\\
\includegraphics[width=0.9\columnwidth]{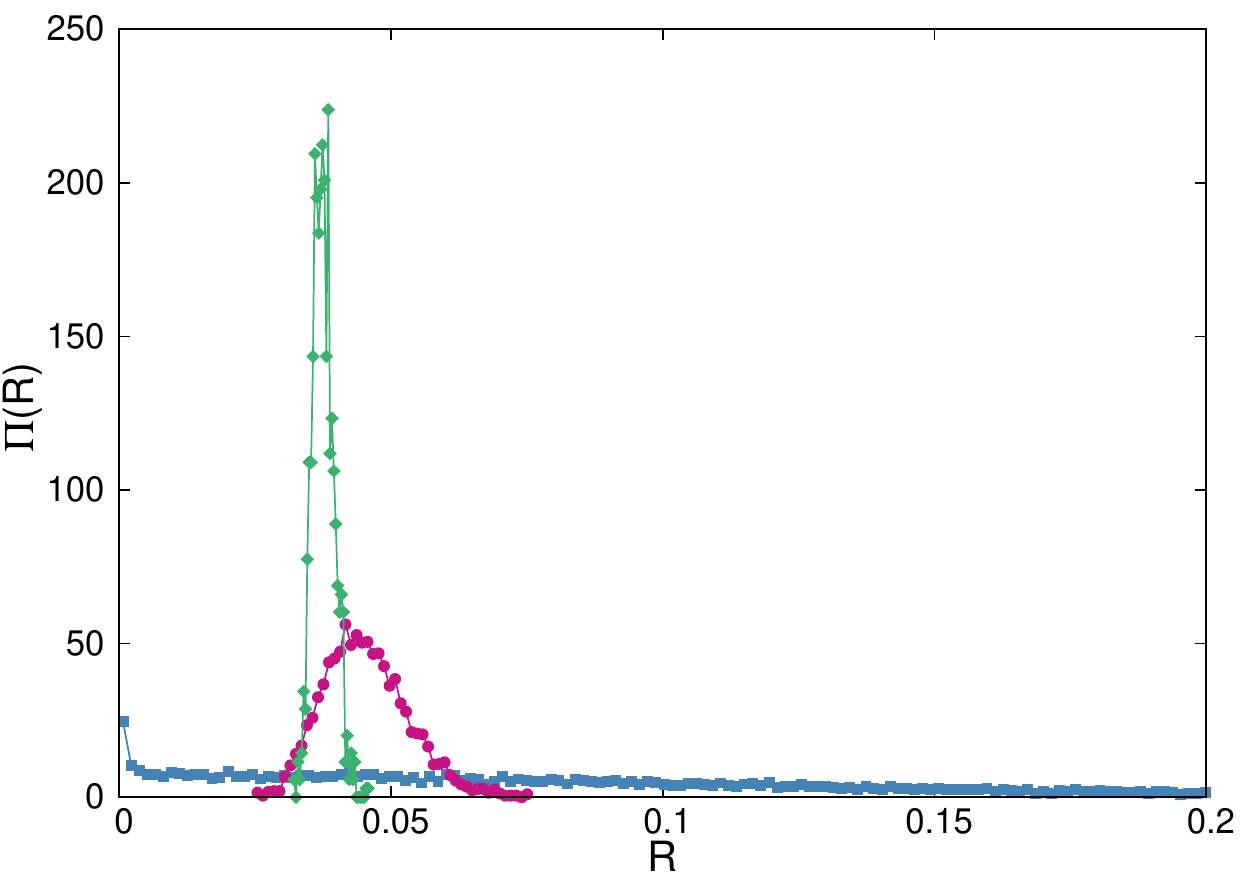}\,\,\,\,
\includegraphics[width=0.9\columnwidth]{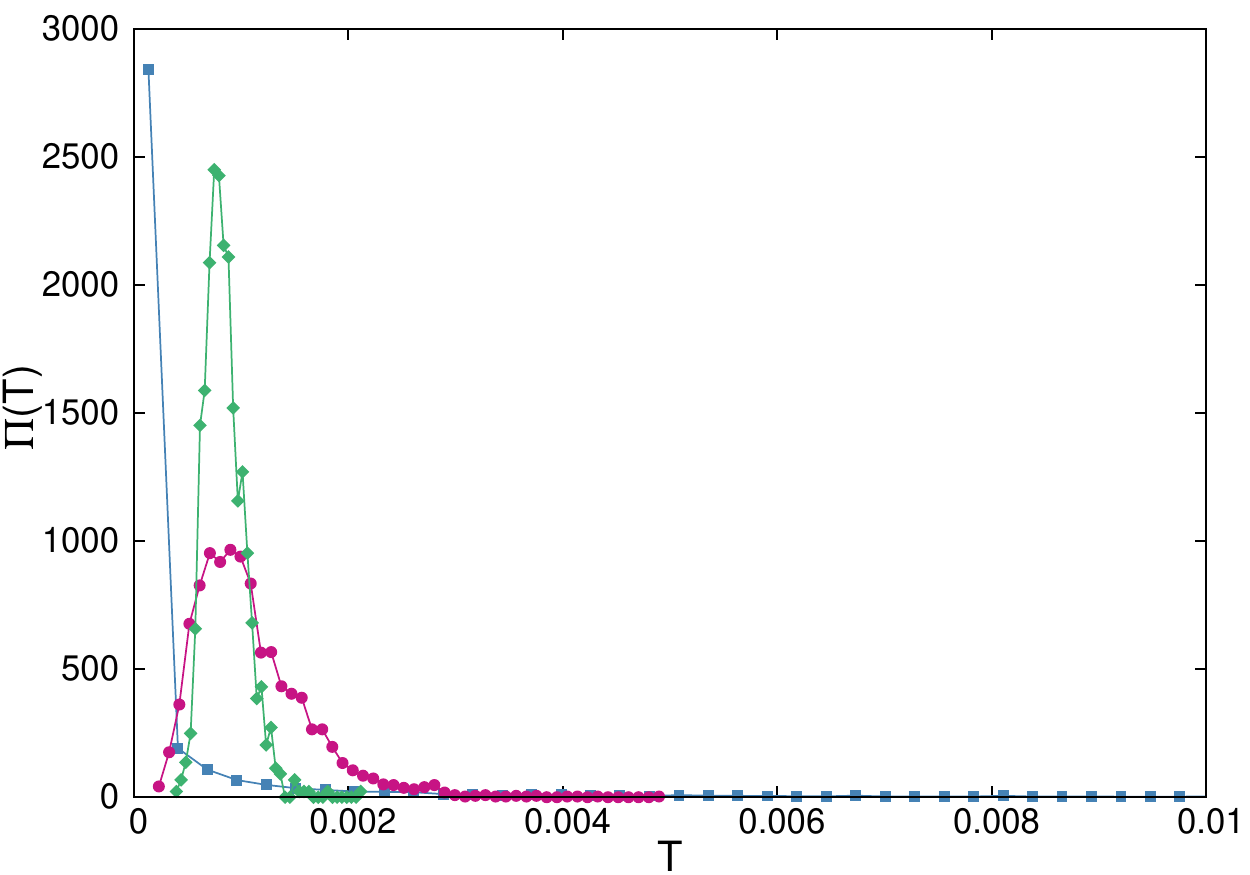}\\
\,\,\,\, Case 1c \,\,\,\,\\
\includegraphics[width=0.9\columnwidth]{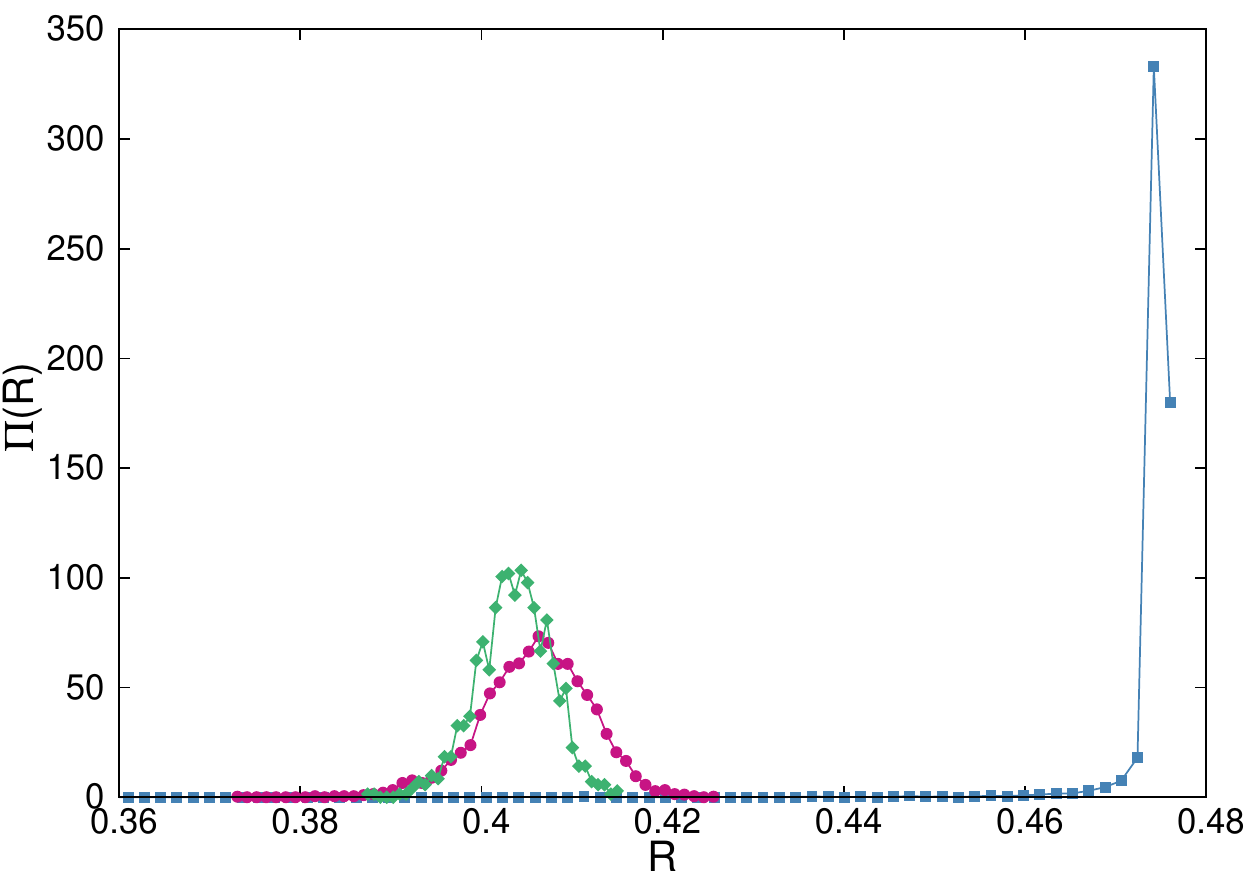}\,\,\,\,
\includegraphics[width=0.9\columnwidth]{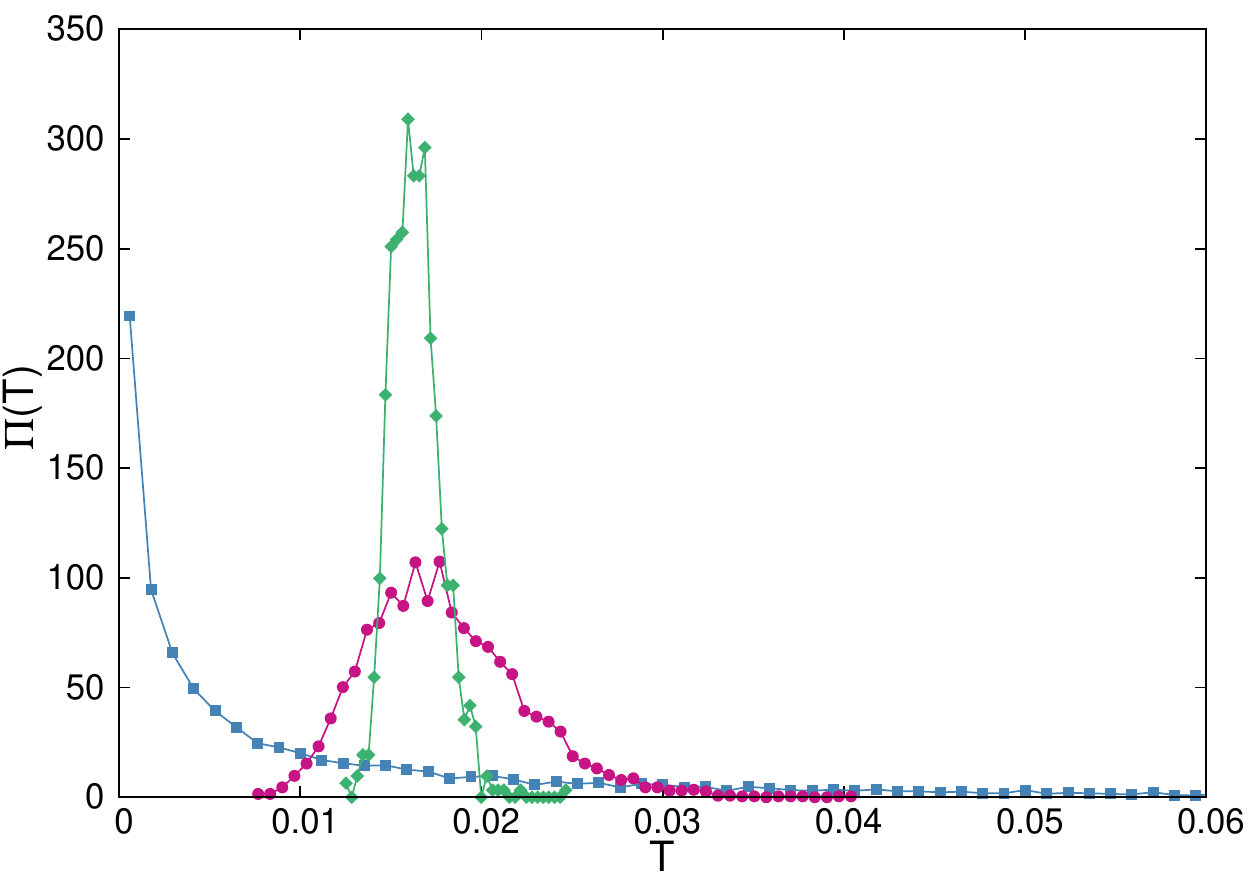}
\end{center}
\caption{Left column: normalized distributions $\Pi(R)$ of the reflection coefficients $R$; right column: normalized distributions $\Pi(T)$ of the transmission coefficients $T$. {\em Suite} I configurations, case $1$. Blue squares represent the $1d$ slab tessellations, red circles the $2d$ extruded tessellations, and green diamonds the $3d$ tessellations.}
\label{fig_histo_1_I}
\end{figure*}

\begin{figure*}
\begin{center}
\,\,\,\, Case 2a \,\,\,\,\\
\includegraphics[width=0.9\columnwidth]{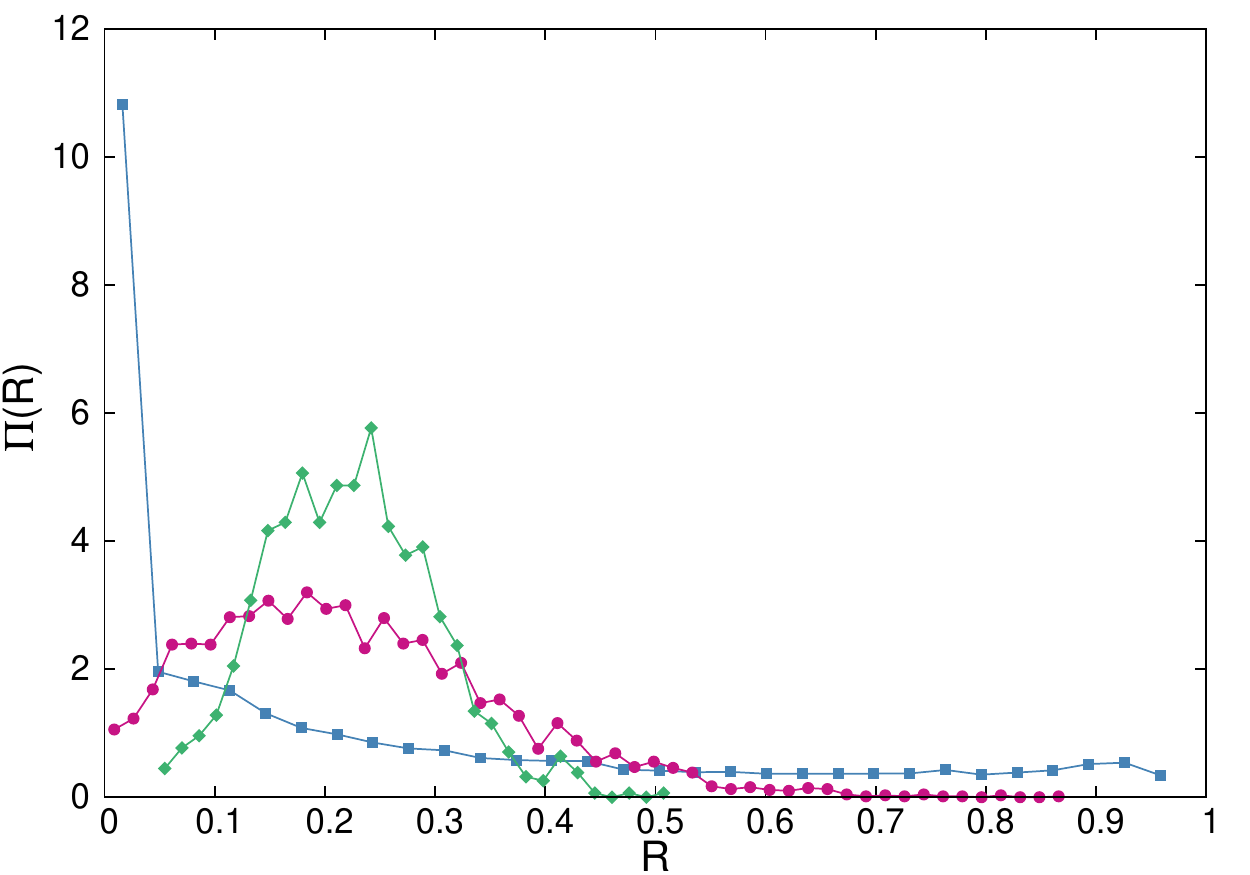}\,\,\,\,
\includegraphics[width=0.9\columnwidth]{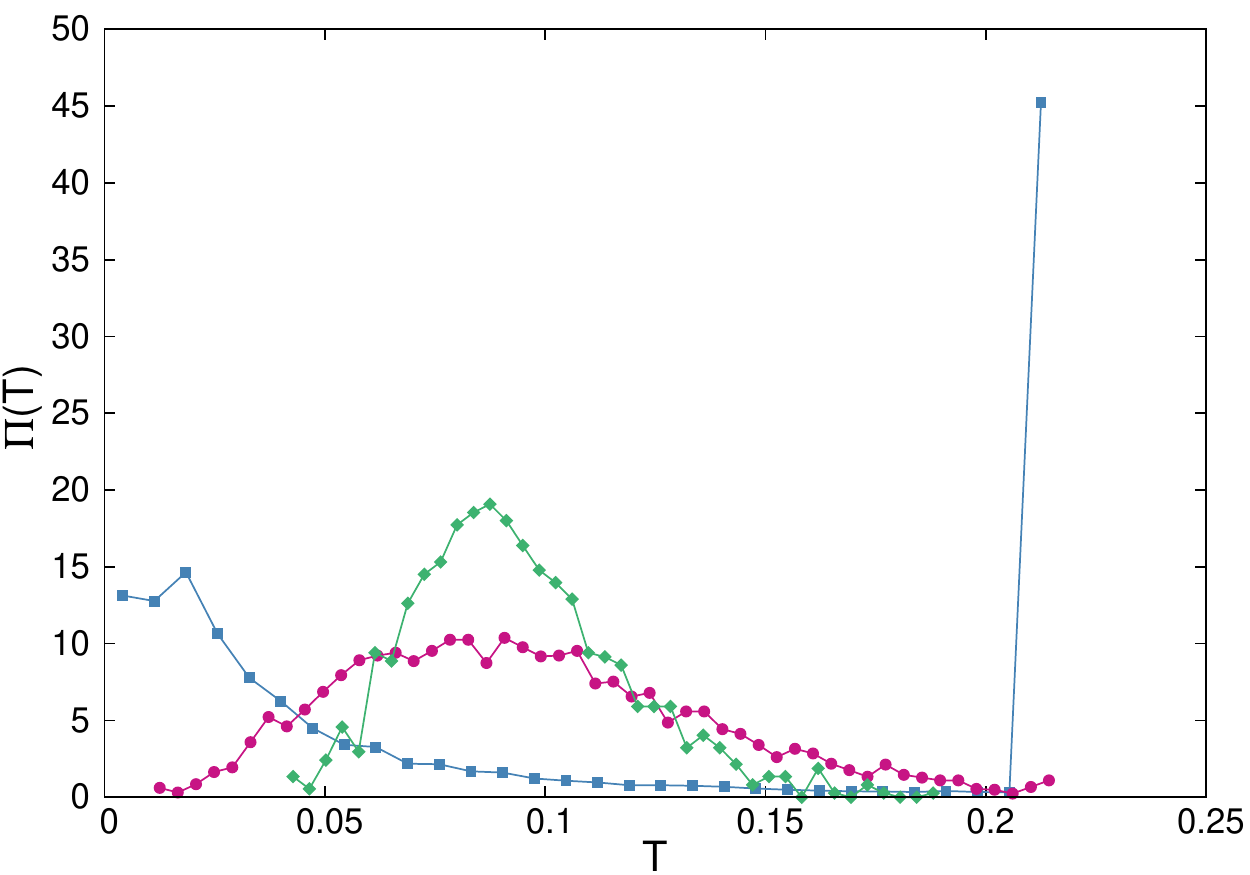}\\
\,\,\,\, Case 2b \,\,\,\,\\
\includegraphics[width=0.9\columnwidth]{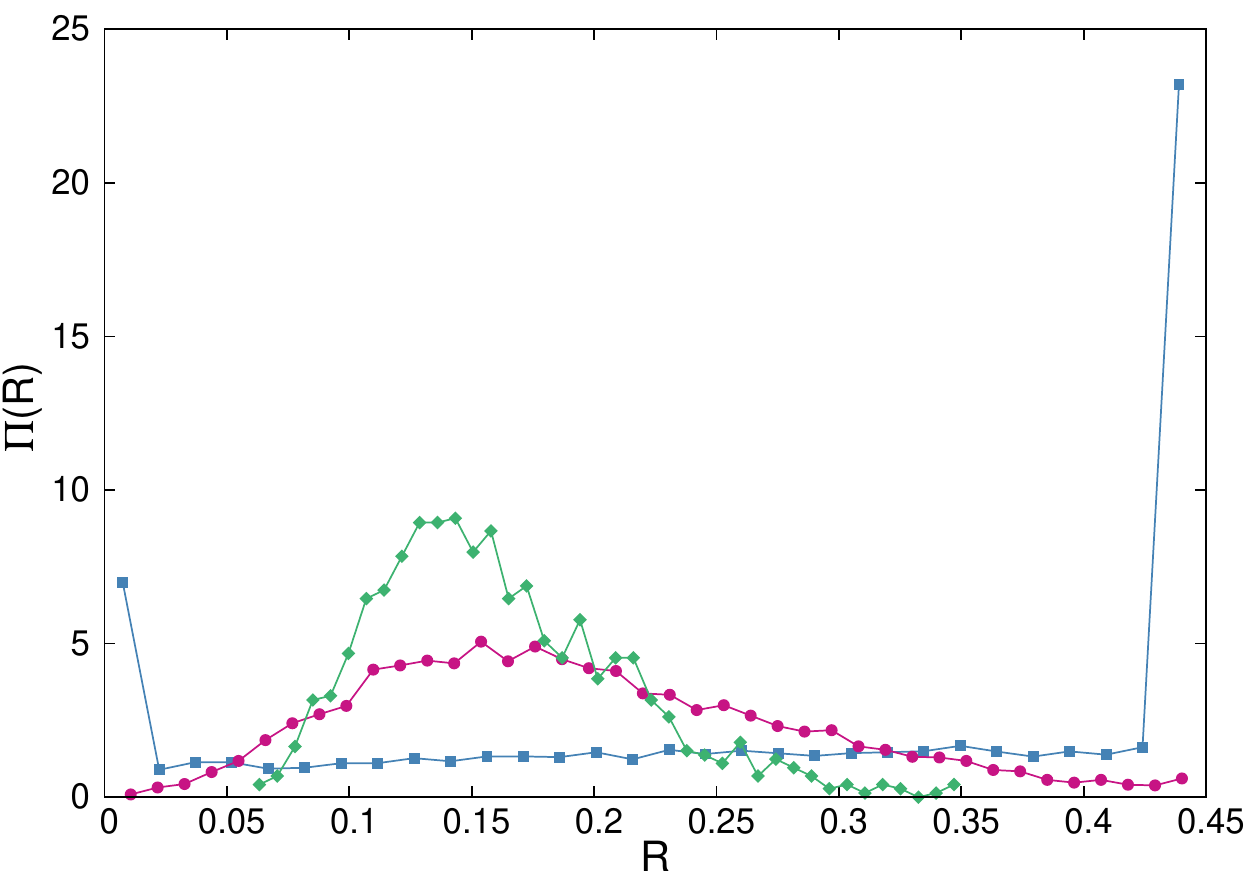}\,\,\,\,
\includegraphics[width=0.9\columnwidth]{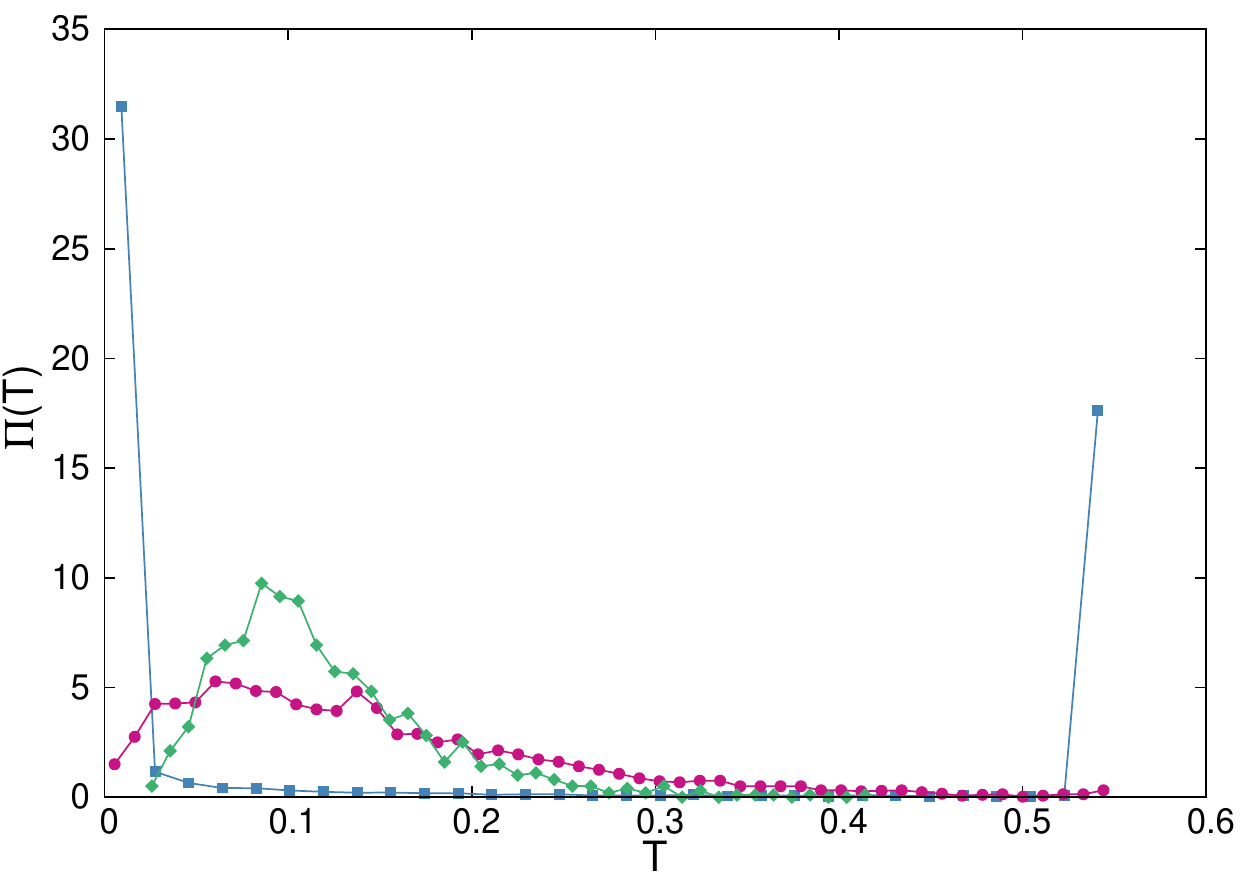}\\
\,\,\,\, Case 2c \,\,\,\,\\
\includegraphics[width=0.9\columnwidth]{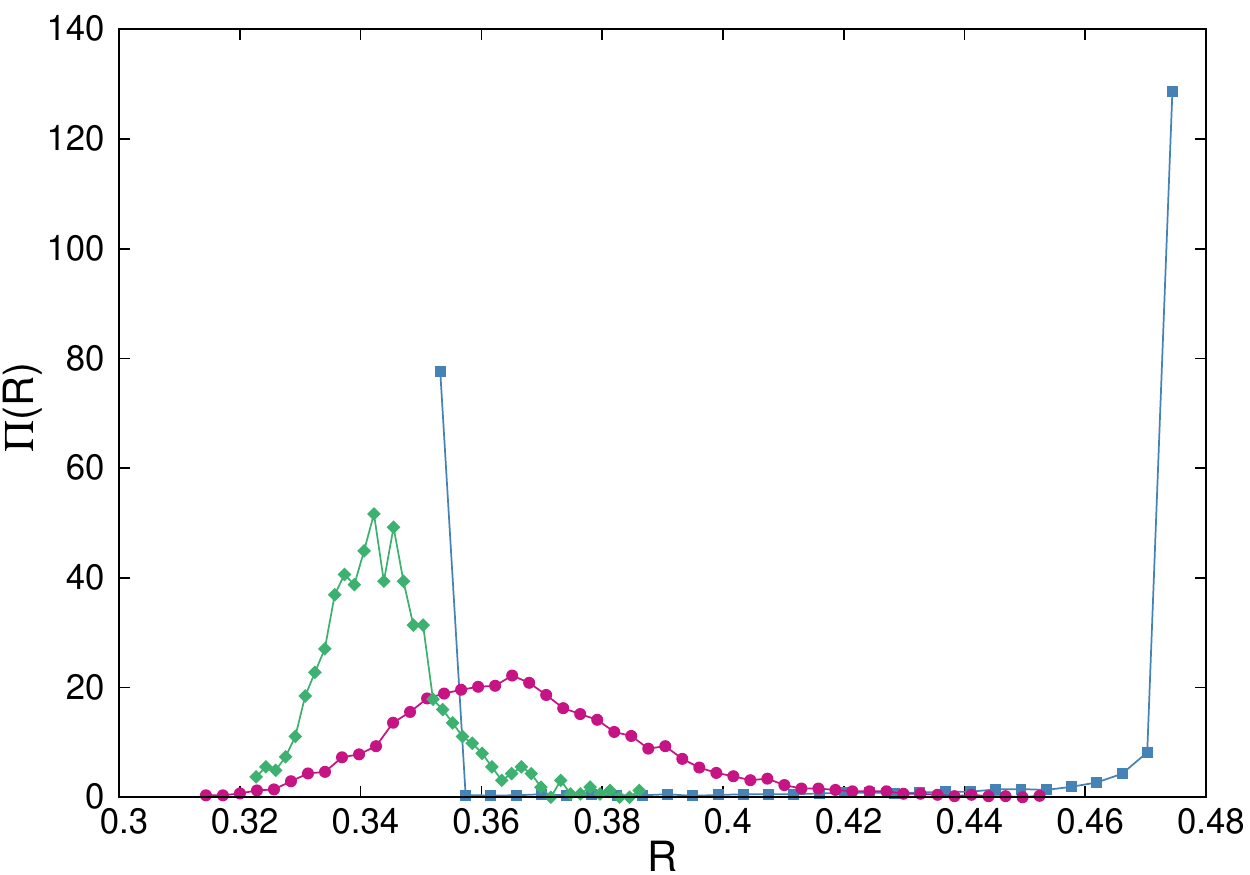}\,\,\,\,
\includegraphics[width=0.9\columnwidth]{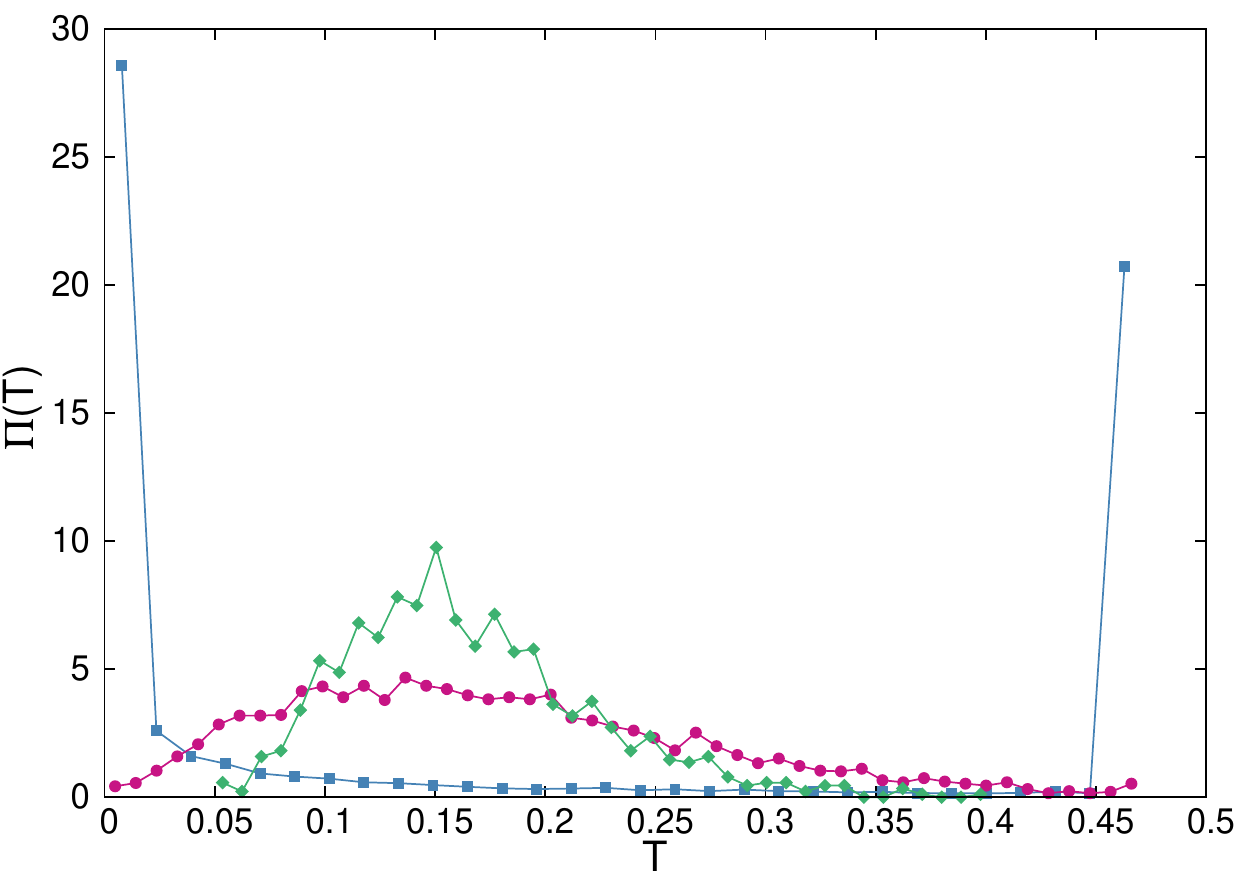}
\end{center}
\caption{Left column: normalized distributions $\Pi(R)$ of the reflection coefficients $R$; right column: normalized distributions $\Pi(T)$ of the transmission coefficients $T$. {\em Suite} I configurations, case $2$. Blue squares represent the $1d$ slab tessellations, red circles the $2d$ extruded tessellations, and green diamonds the $3d$ tessellations.}
\label{fig_histo_2_I}
\end{figure*}

\begin{figure*}
\begin{center}
\,\,\,\, Case 3a \,\,\,\,\\
\includegraphics[width=0.9\columnwidth]{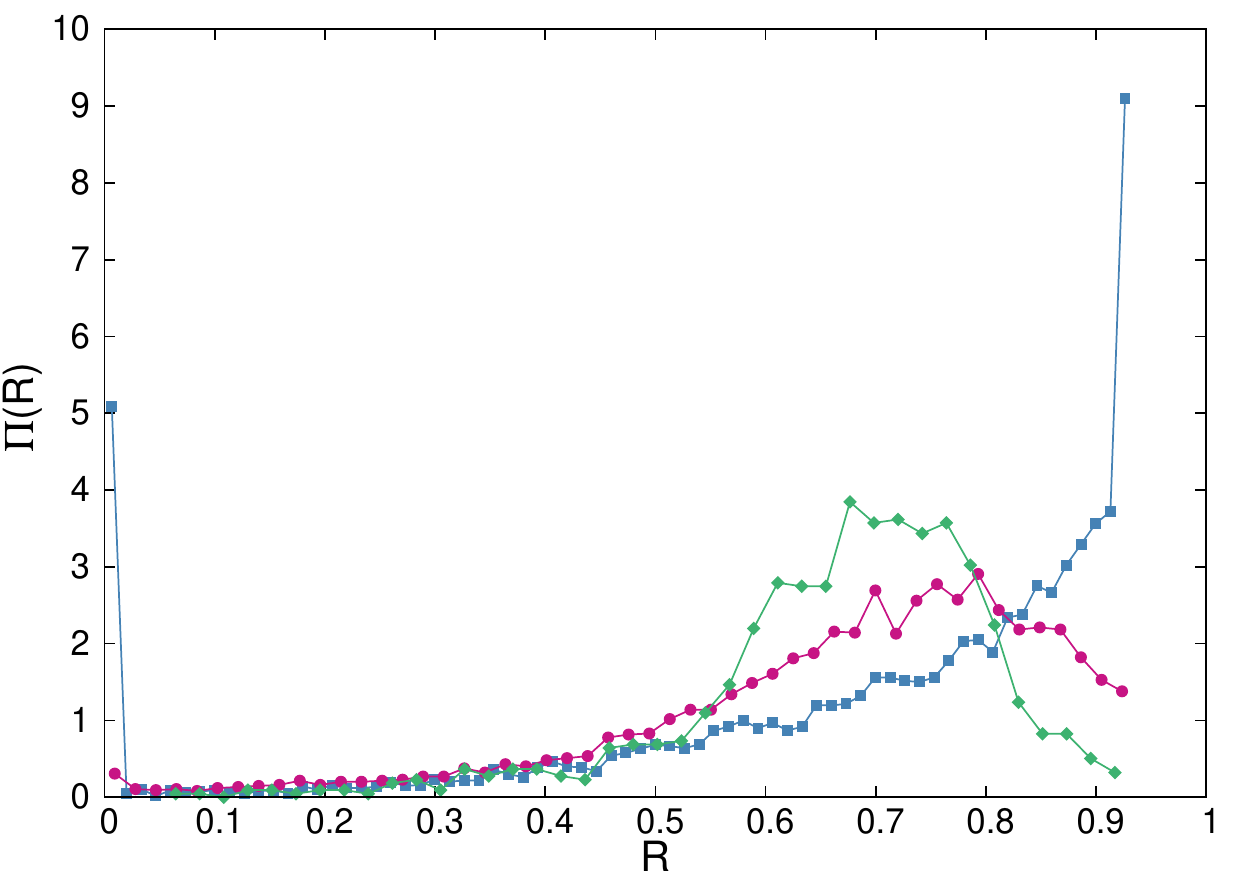}\,\,\,\,
\includegraphics[width=0.9\columnwidth]{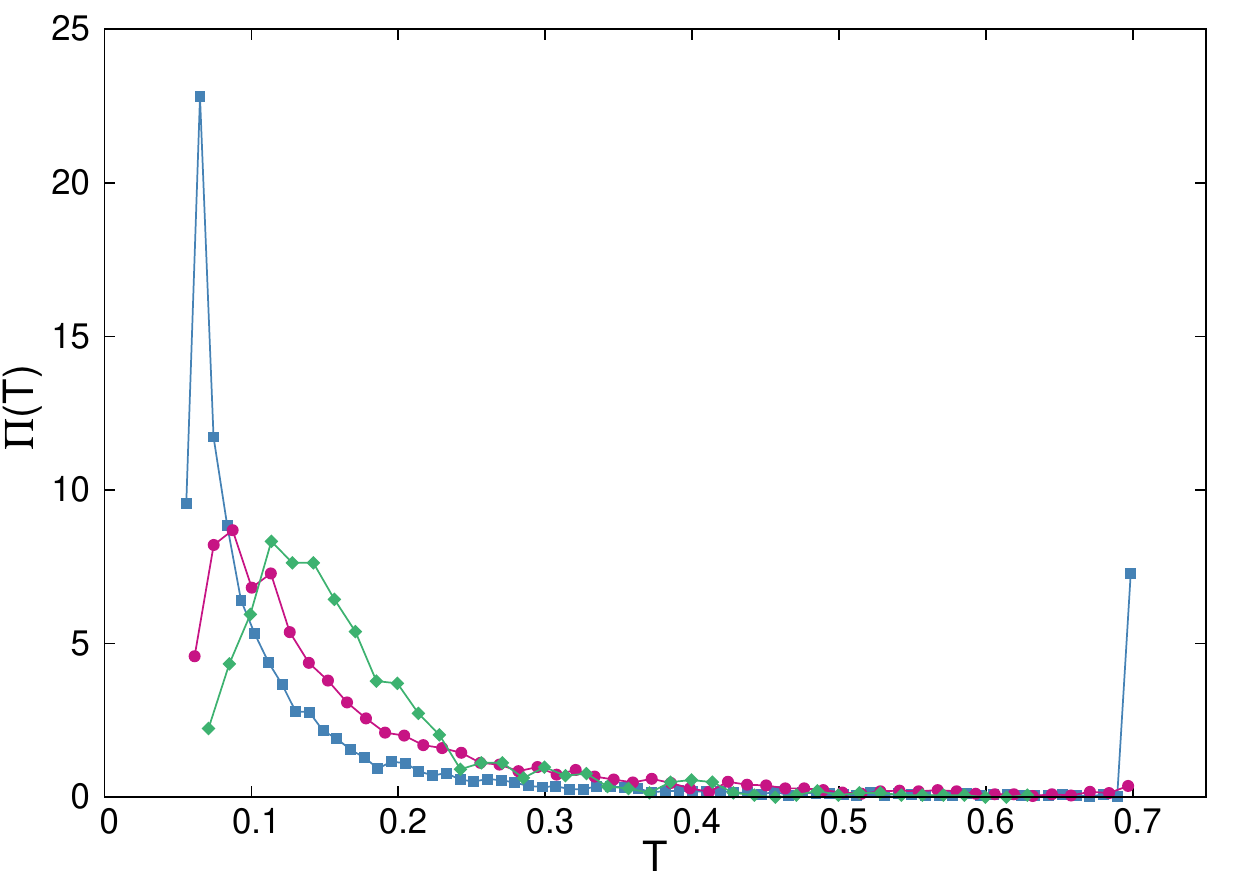}\\
\,\,\,\, Case 3b \,\,\,\,\\
\includegraphics[width=0.9\columnwidth]{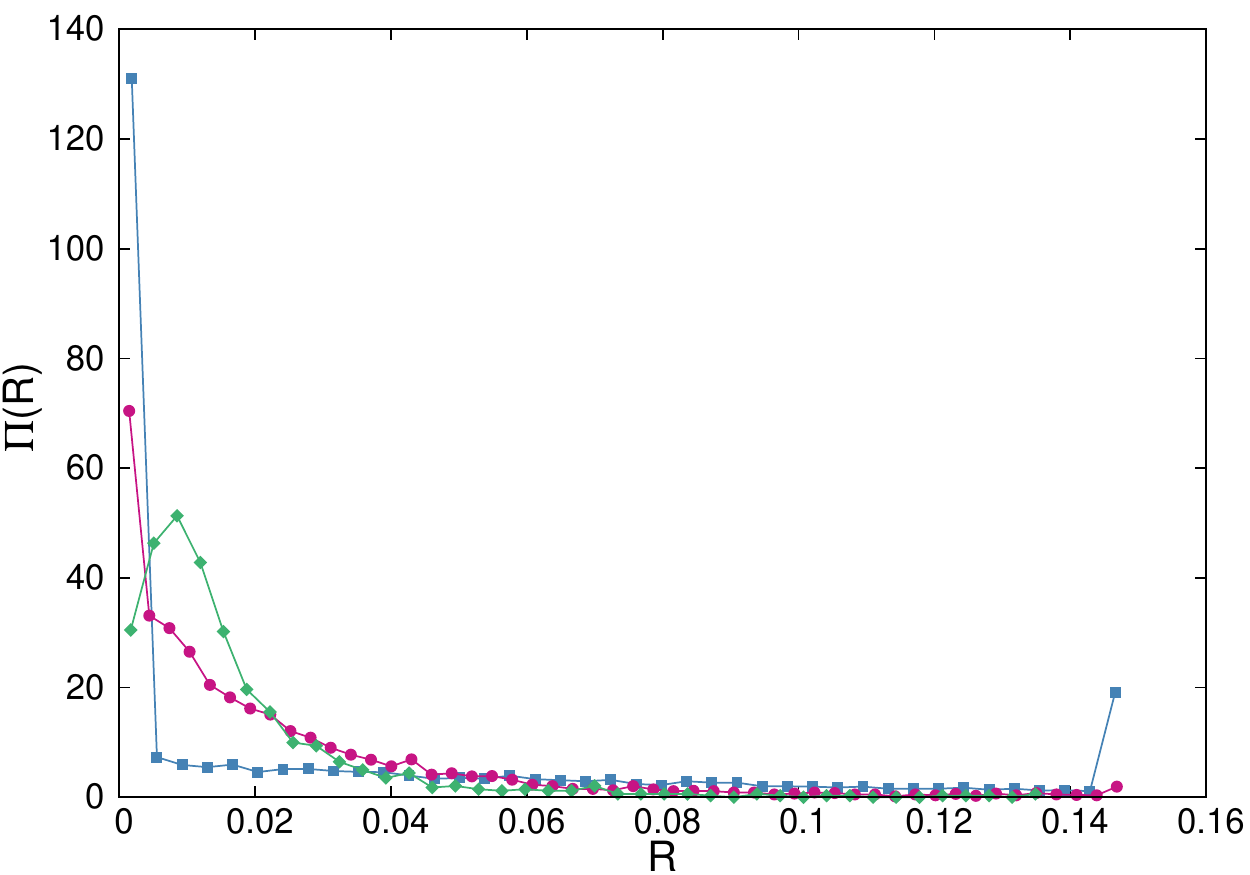}\,\,\,\,
\includegraphics[width=0.9\columnwidth]{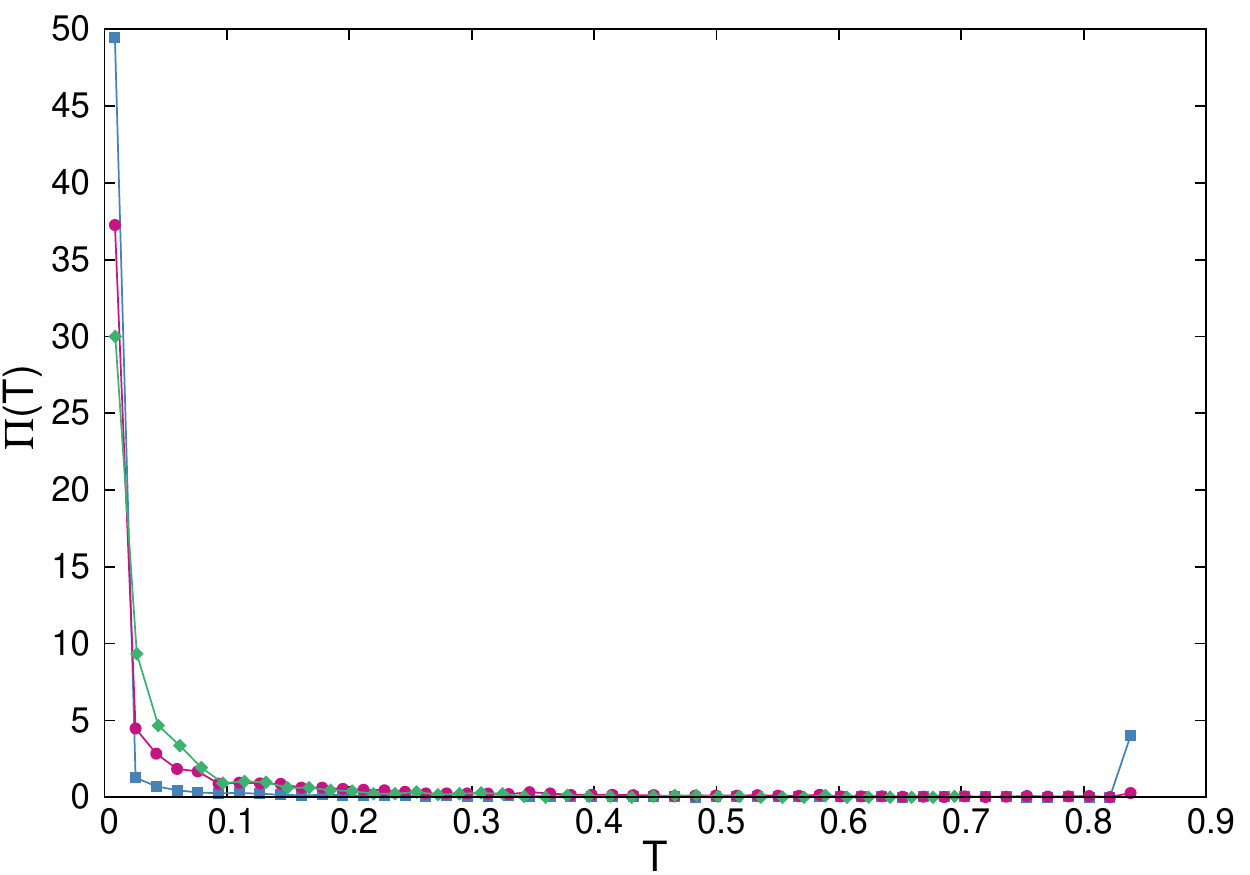}\\
\,\,\,\, Case 3c \,\,\,\,\\
\includegraphics[width=0.9\columnwidth]{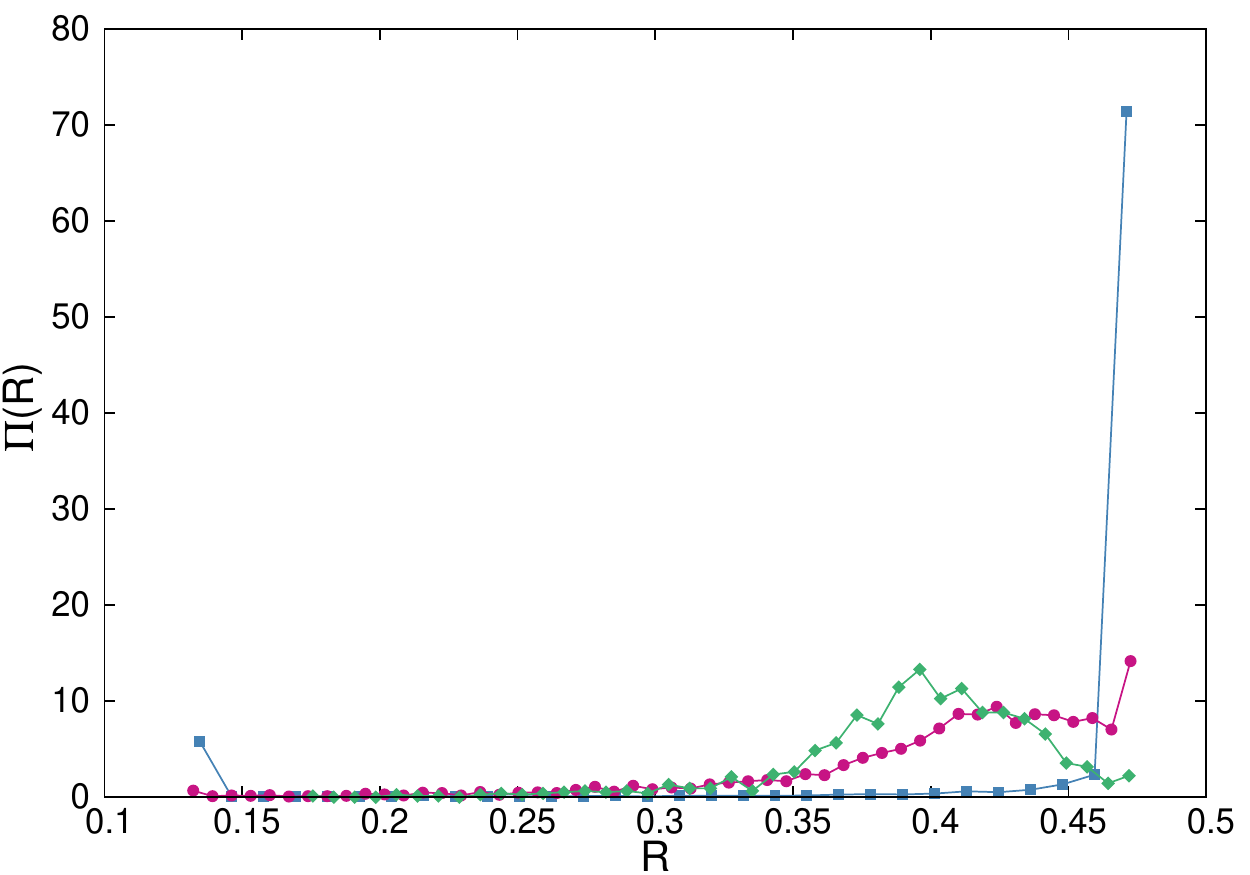}\,\,\,\,
\includegraphics[width=0.9\columnwidth]{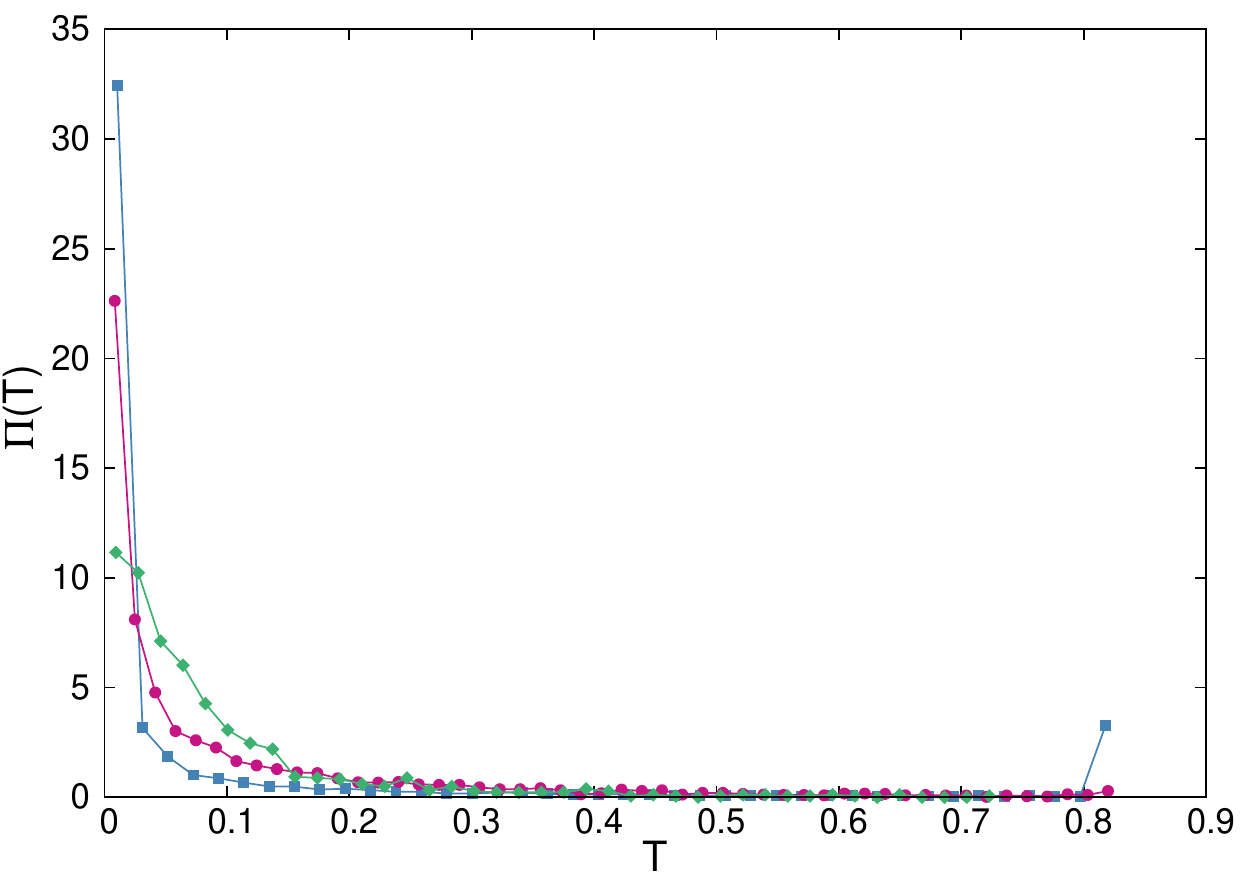}
\end{center}
\caption{Left column: normalized distributions $\Pi(R)$ of the reflection coefficients $R$; right column: normalized distributions $\Pi(T)$ of the transmission coefficients $T$. {\em Suite} I configurations, case $3$. Blue squares represent the $1d$ slab tessellations, red circles the $2d$ extruded tessellations, and green diamonds the $3d$ tessellations.}
\label{fig_histo_3_I}
\end{figure*}

\begin{figure*}
\begin{center}
Case 1a\\
\includegraphics[width=0.9\columnwidth]{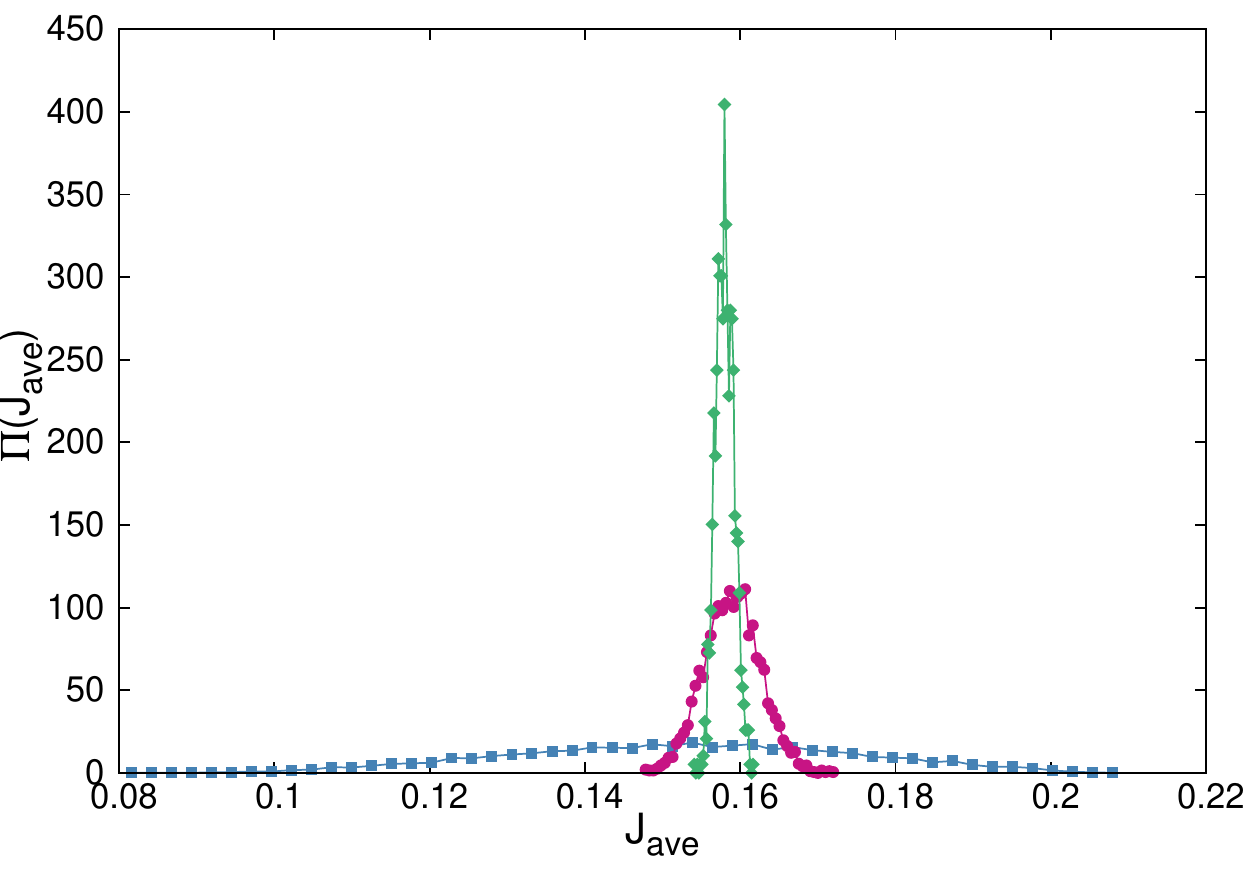}\\
Case 1b\\
\includegraphics[width=0.9\columnwidth]{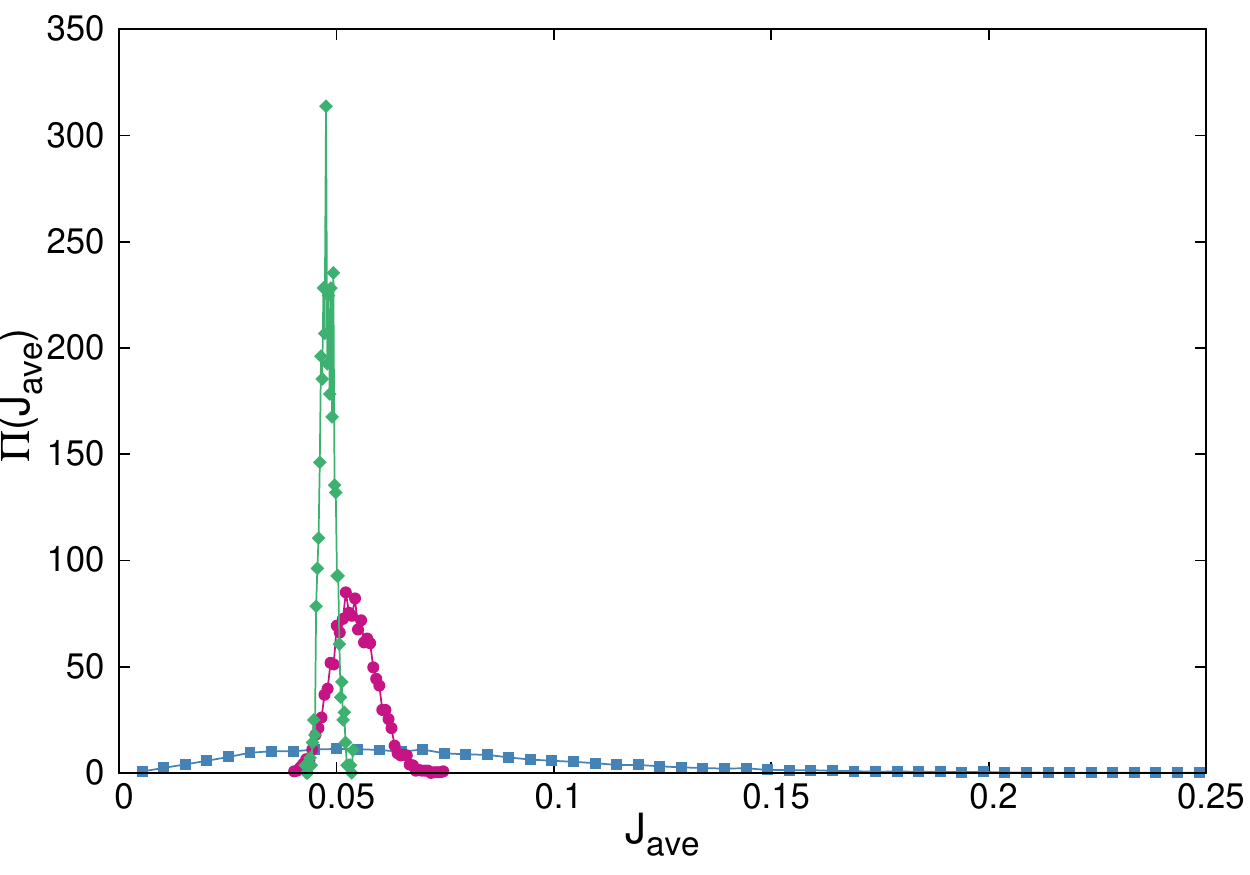}\\
Case 1c\\
\includegraphics[width=0.9\columnwidth]{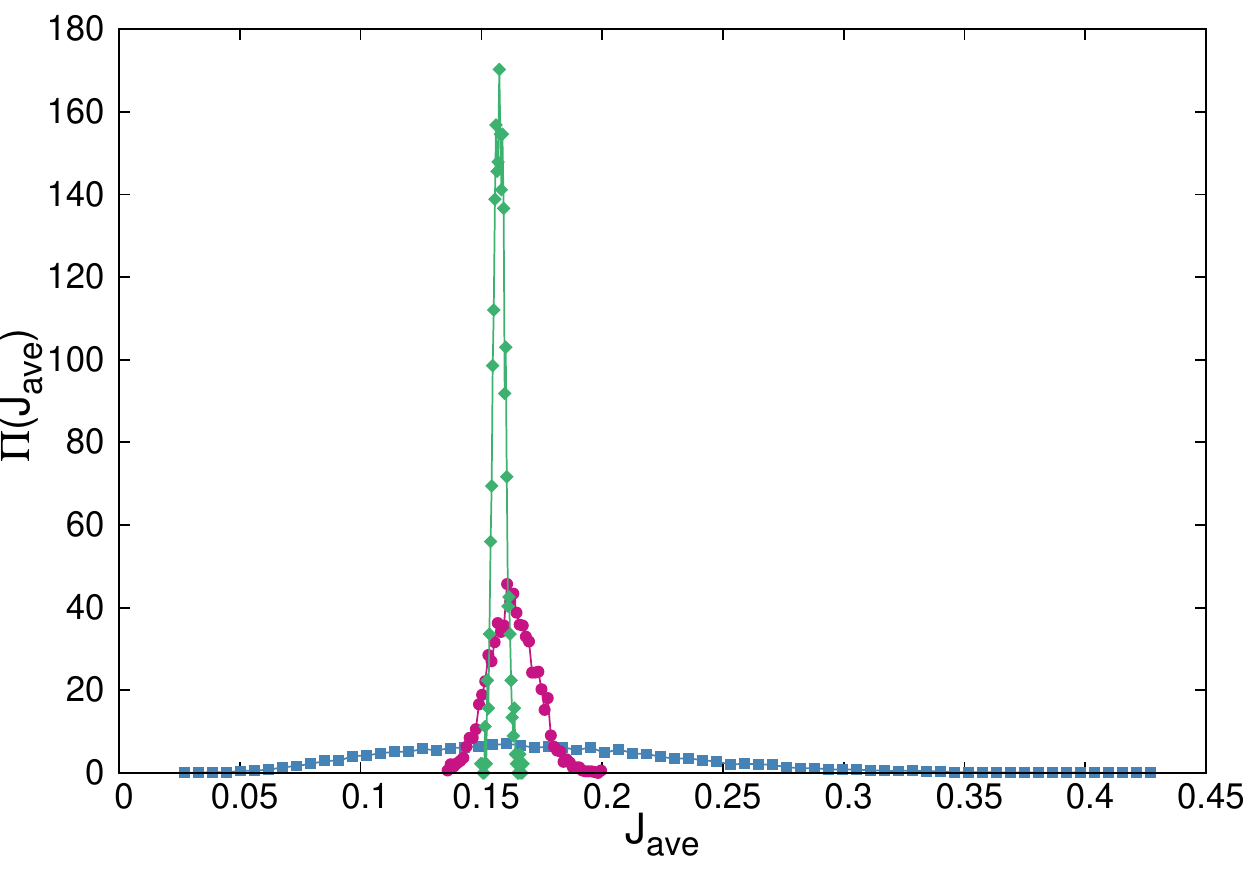}
\end{center}
\caption{Normalized distributions $\Pi(J_\text{ave})$ of the average leakage current $J_\text{ave} = (T+R)/2$ for {\em suite} II configurations, case $1$. Blue squares represent the $1d$ slab tessellations, red circles the $2d$ extruded tessellations, and green diamonds the $3d$ tessellations.}
\label{fig_histo_1_II}
\end{figure*}

\begin{figure*}
\begin{center}
Case 2a\\
\includegraphics[width=0.9\columnwidth]{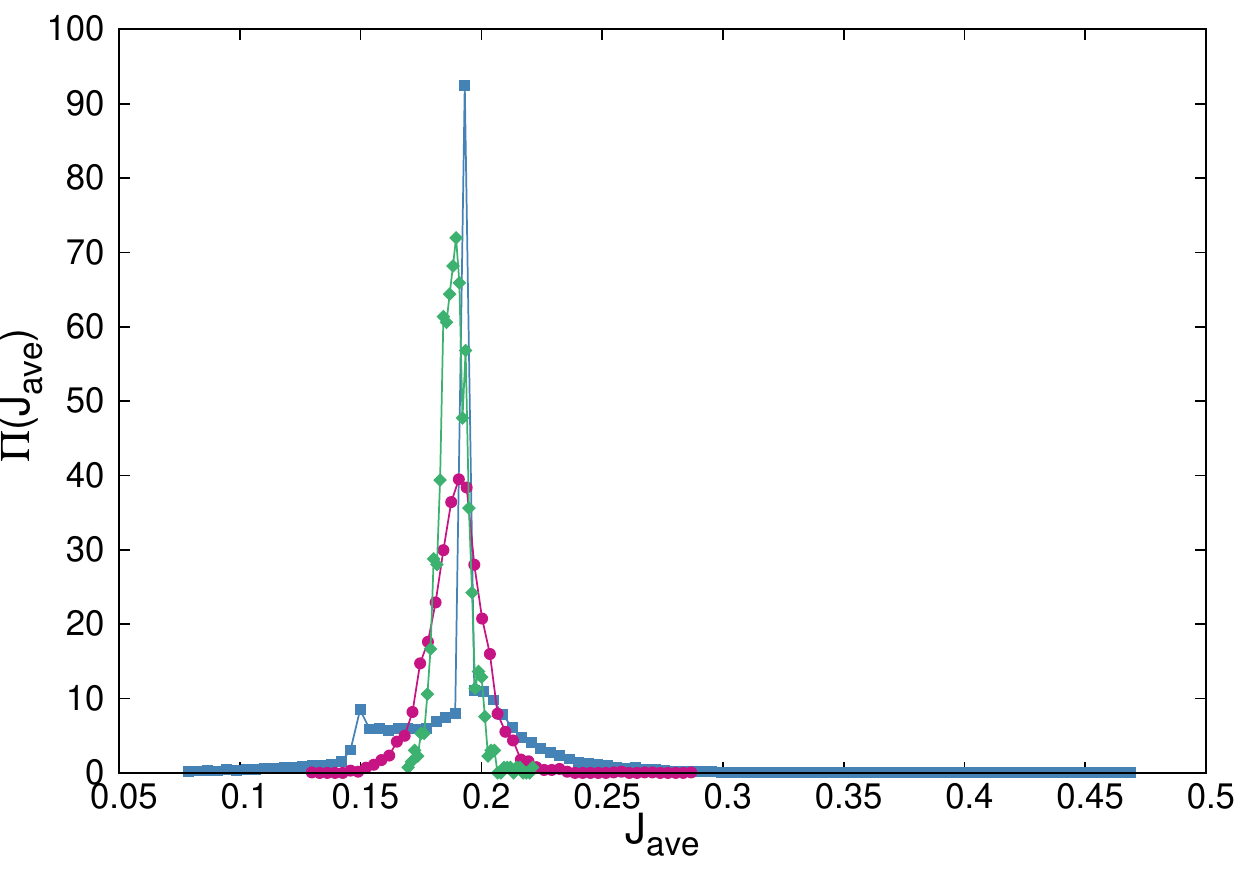}\\
Case 2b\\
\includegraphics[width=0.9\columnwidth]{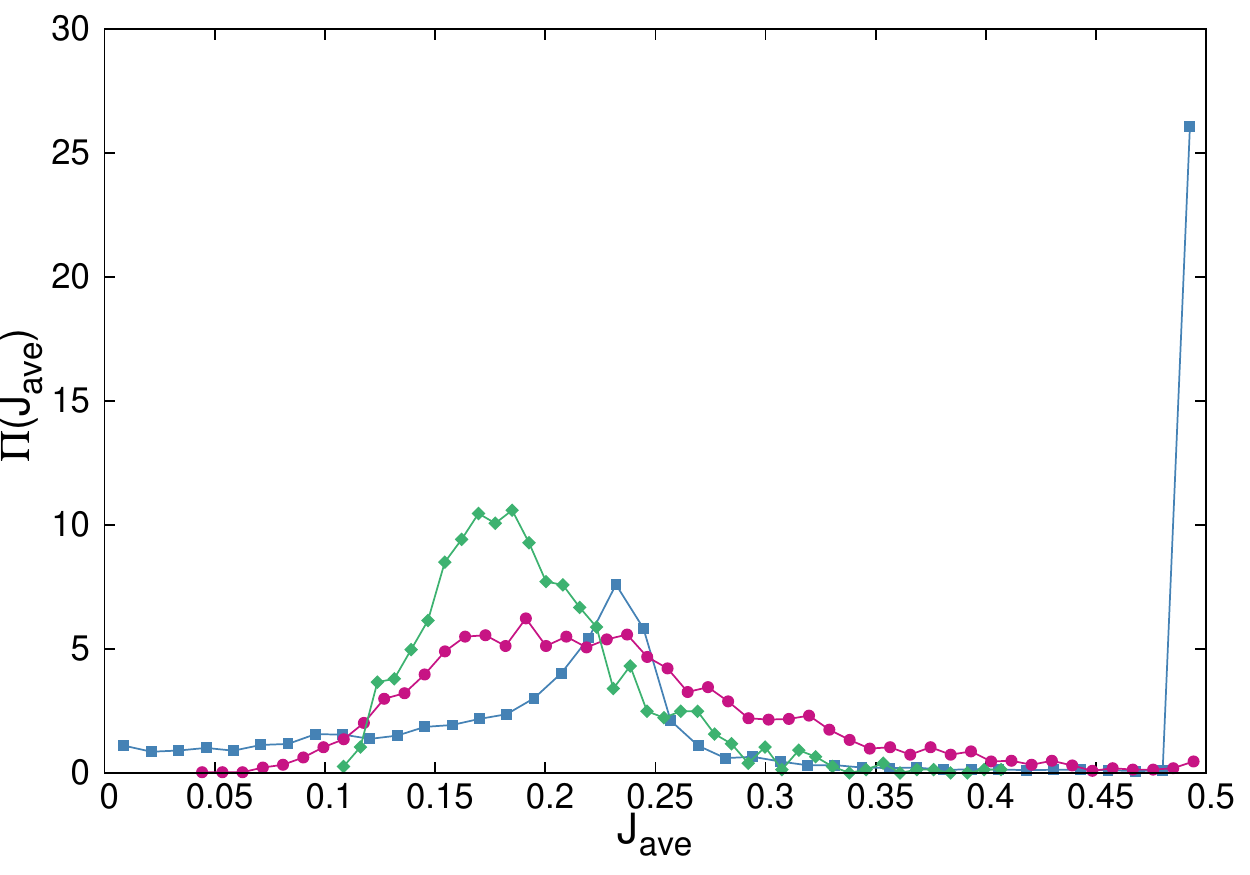}\\
Case 2c\\
\includegraphics[width=0.9\columnwidth]{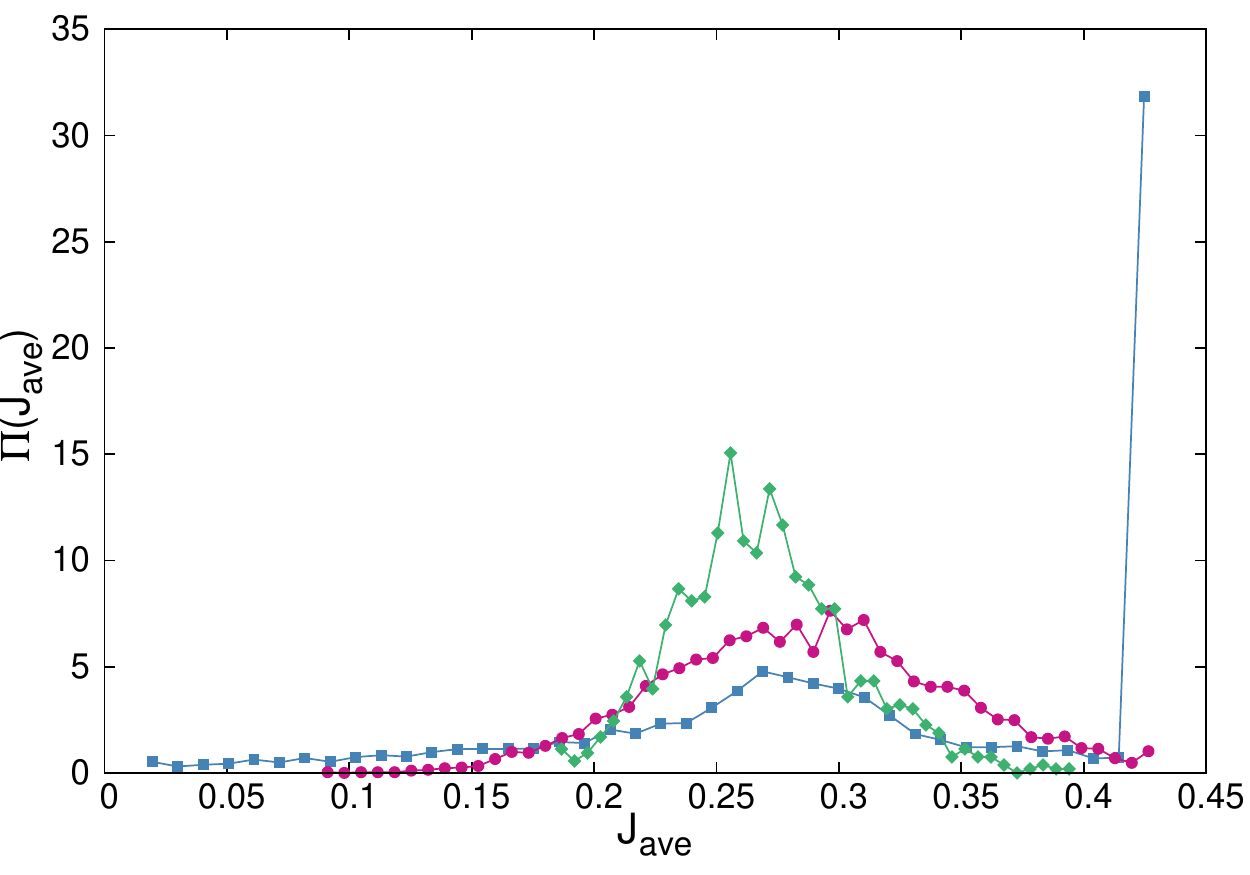}
\end{center}
\caption{Normalized distributions $\Pi(J_\text{ave})$ of the average leakage current $J_\text{ave} = (T+R)/2$ for {\em suite} II configurations, case $2$. Blue squares represent the $1d$ slab tessellations, red circles the $2d$ extruded tessellations, and green diamonds the $3d$ tessellations.}
\label{fig_histo_2_II}
\end{figure*}

\begin{figure*}
\begin{center}
Case 3a\\
\includegraphics[width=0.9\columnwidth]{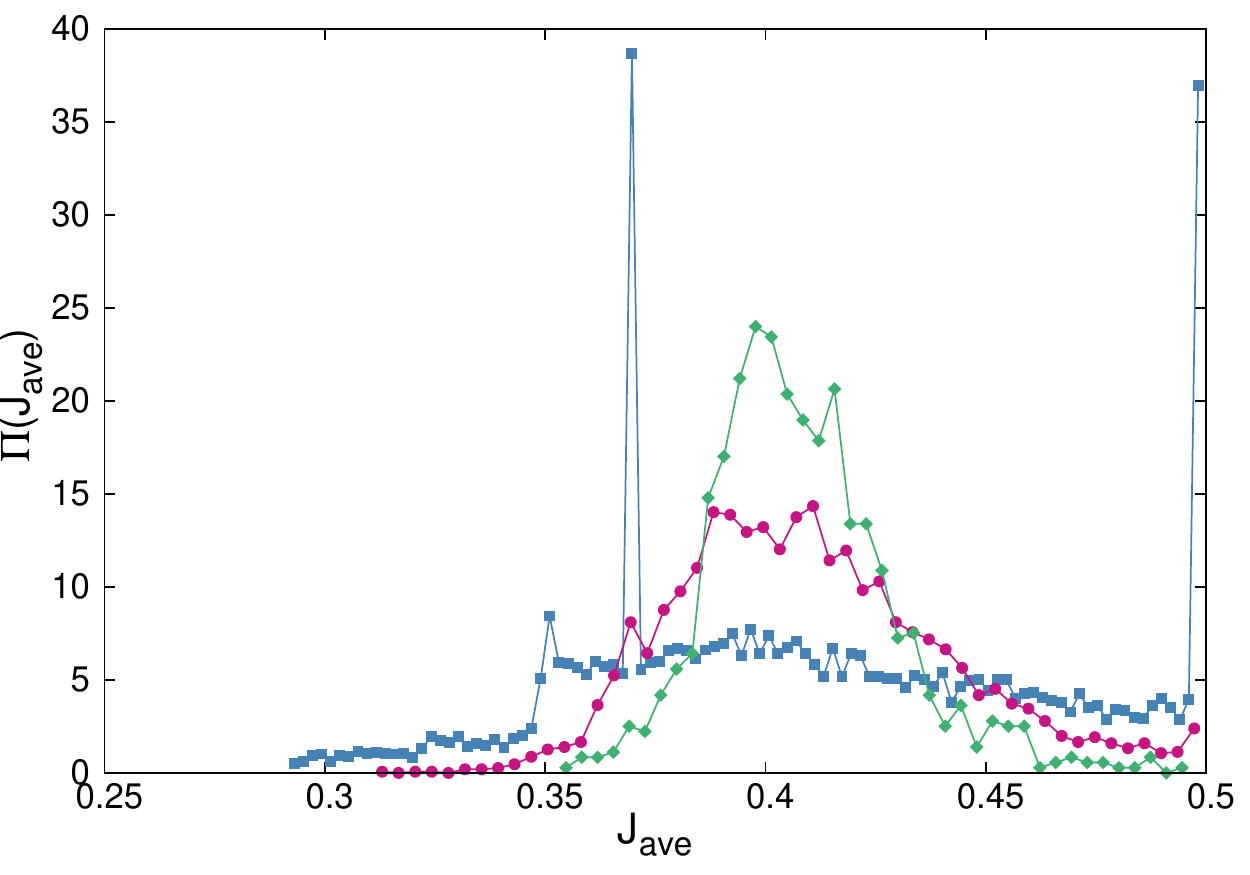}\\
Case 3b\\
\includegraphics[width=0.9\columnwidth]{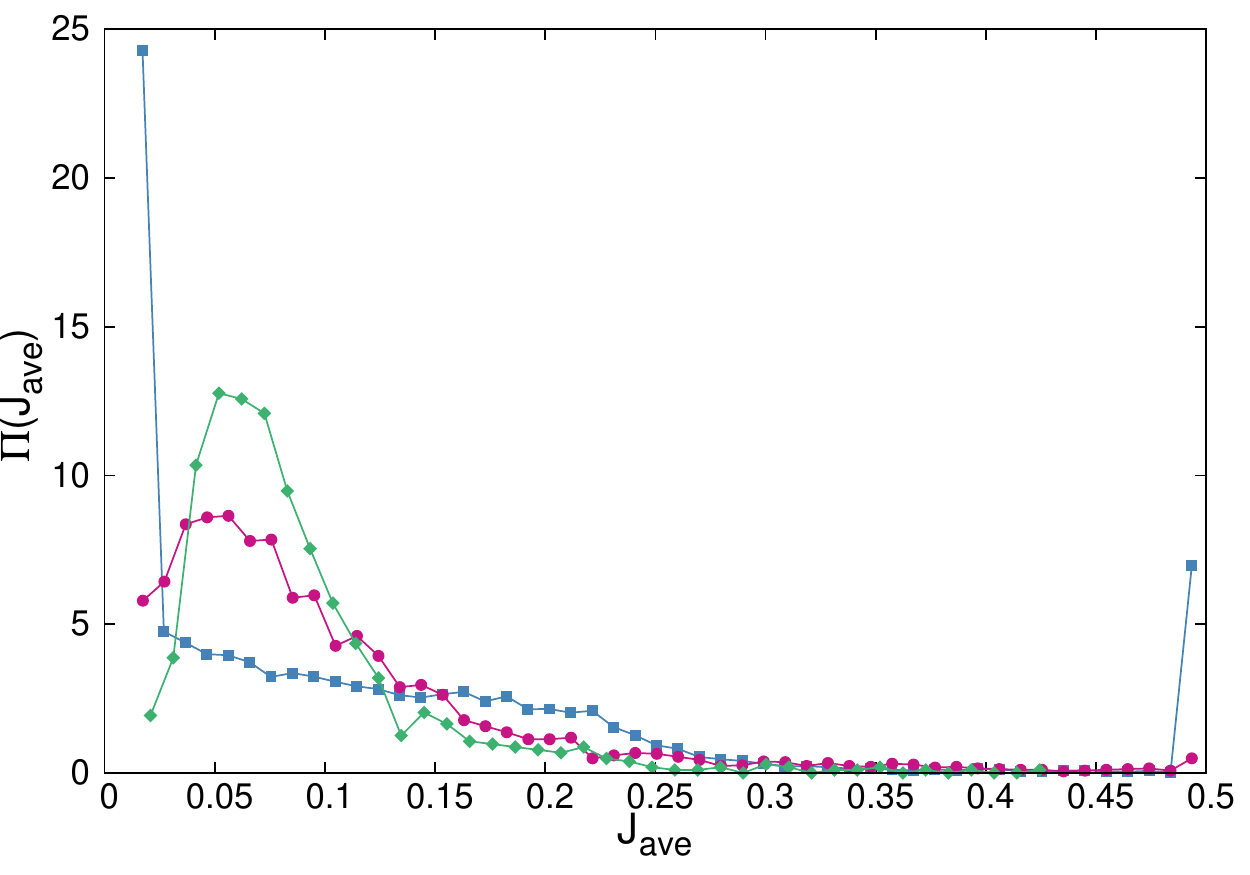}\\
Case 3c\\
\includegraphics[width=0.9\columnwidth]{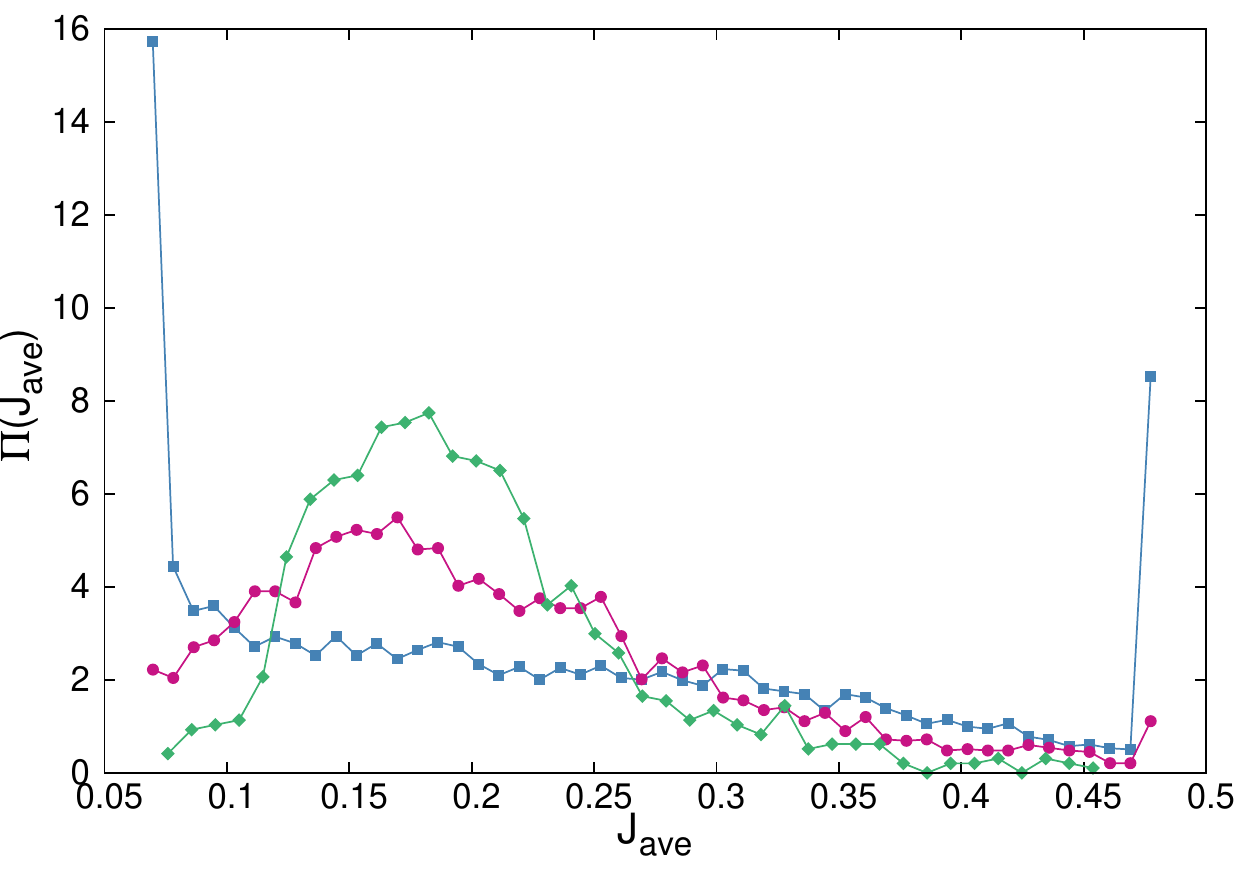}
\end{center}
\caption{Normalized distributions $\Pi(J_\text{ave})$ of the average leakage current $J_\text{ave} = (T+R)/2$ for {\em suite} II configurations, case $3$. Blue squares represent the $1d$ slab tessellations, red circles the $2d$ extruded tessellations, and green diamonds the $3d$ tessellations.}
\label{fig_histo_3_II}
\end{figure*}

In order to better apprehend the variability of the transmission and reflection coefficients (or the average leakage current $J_\text{ave} = (T+R)/2$ in the {\em suite} II) around their average values, we have also computed their full distributions based on the available realizations in the generated ensembles. The resulting normalized histograms are illustrated in Figs.~\ref{fig_histo_1_I} to \ref{fig_histo_3_II}. As a general consideration, the dispersion of the observables decreases with increasing dimension: the mixing is increasingly efficient and the distribution is more peaked around the average, which is expected on physical grounds. However, even for $d=3$ it is apparent that several configurations display highly non-symmetrical shapes, and possible cut-offs due to finite-size effects. Especially in $d=1$, bi-modality may also arise for cases $2$ and $3$, which is due to the aforementioned effect of random geometries being entirely filled with either material $\alpha$ or $\beta$: the peaks observed in the distributions correspond to the values of the transmission or reflection coefficient associated to a fully red or fully blue realization. The data sets of the distributions are available from the authors upon request. For the $1d$ slab tessellations, the variances of the transmission and reflection coefficient have been numerical computed in~\citep{benchmark_adams}: the values obtained in our simulations are in excellent agreement with those previously reported.

\subsection{Spatial flux}

\begin{figure*}
\begin{center}
\,\,\,\, Case 1a \,\,\,\,\\
\includegraphics[width=0.9\columnwidth]{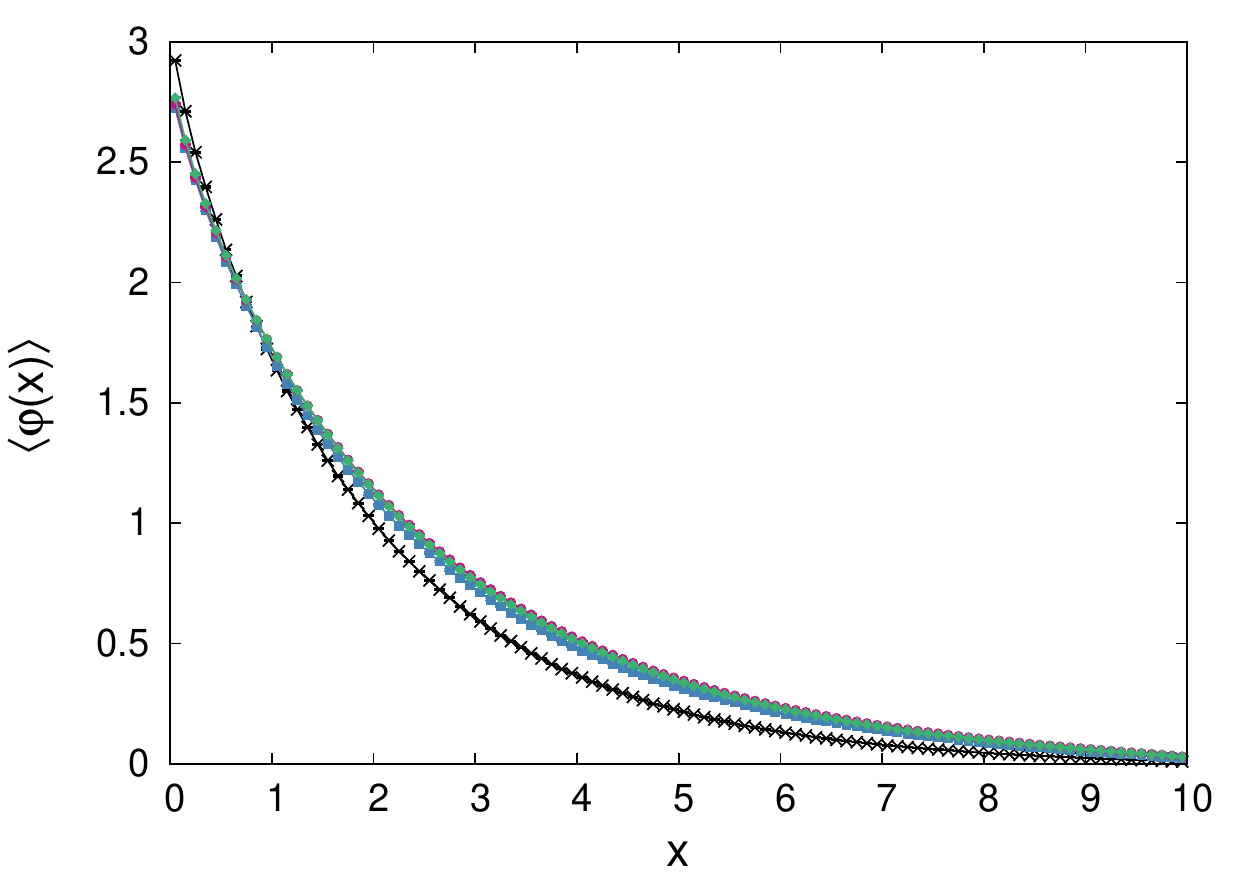}\,\,\,\,
\includegraphics[width=0.9\columnwidth]{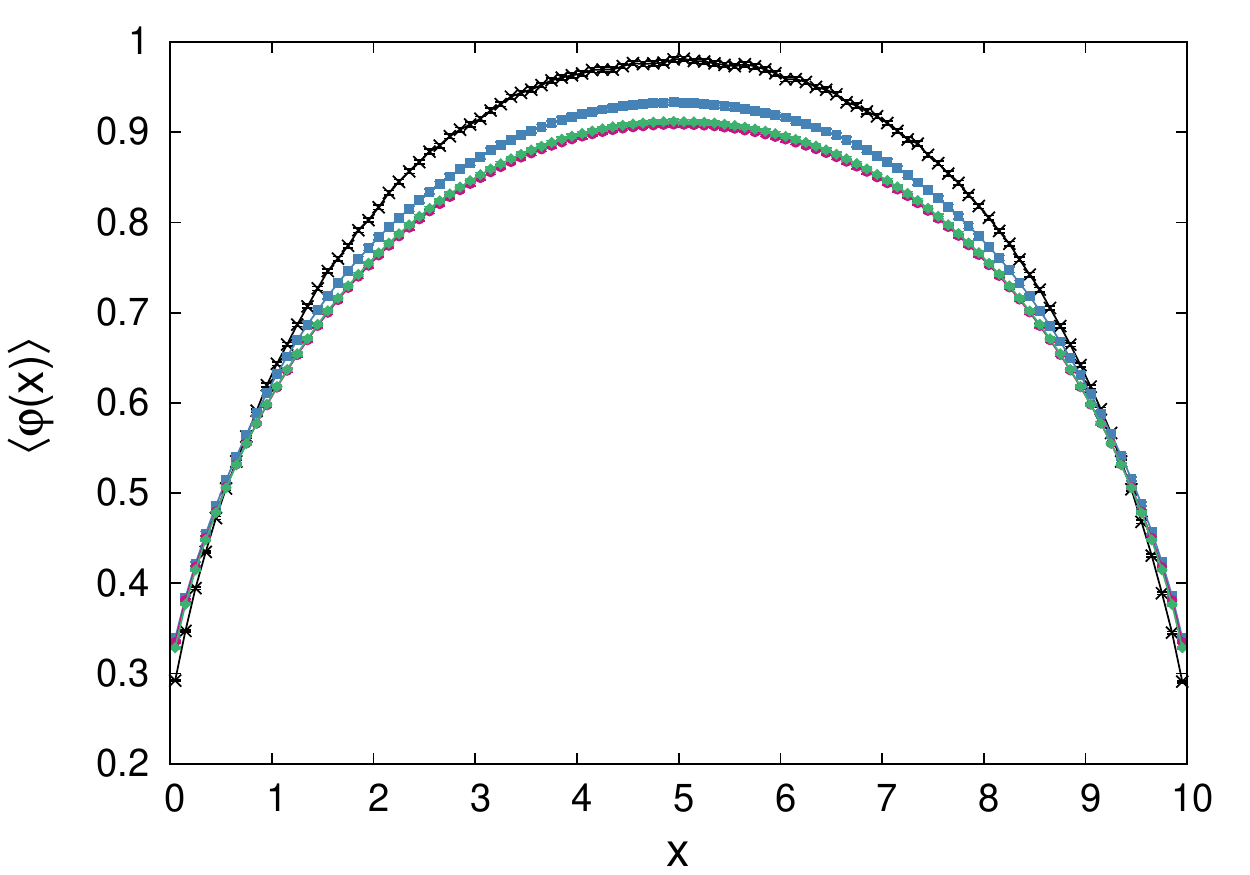}\\
\,\,\,\, Case 1b \,\,\,\,\\
\includegraphics[width=0.9\columnwidth]{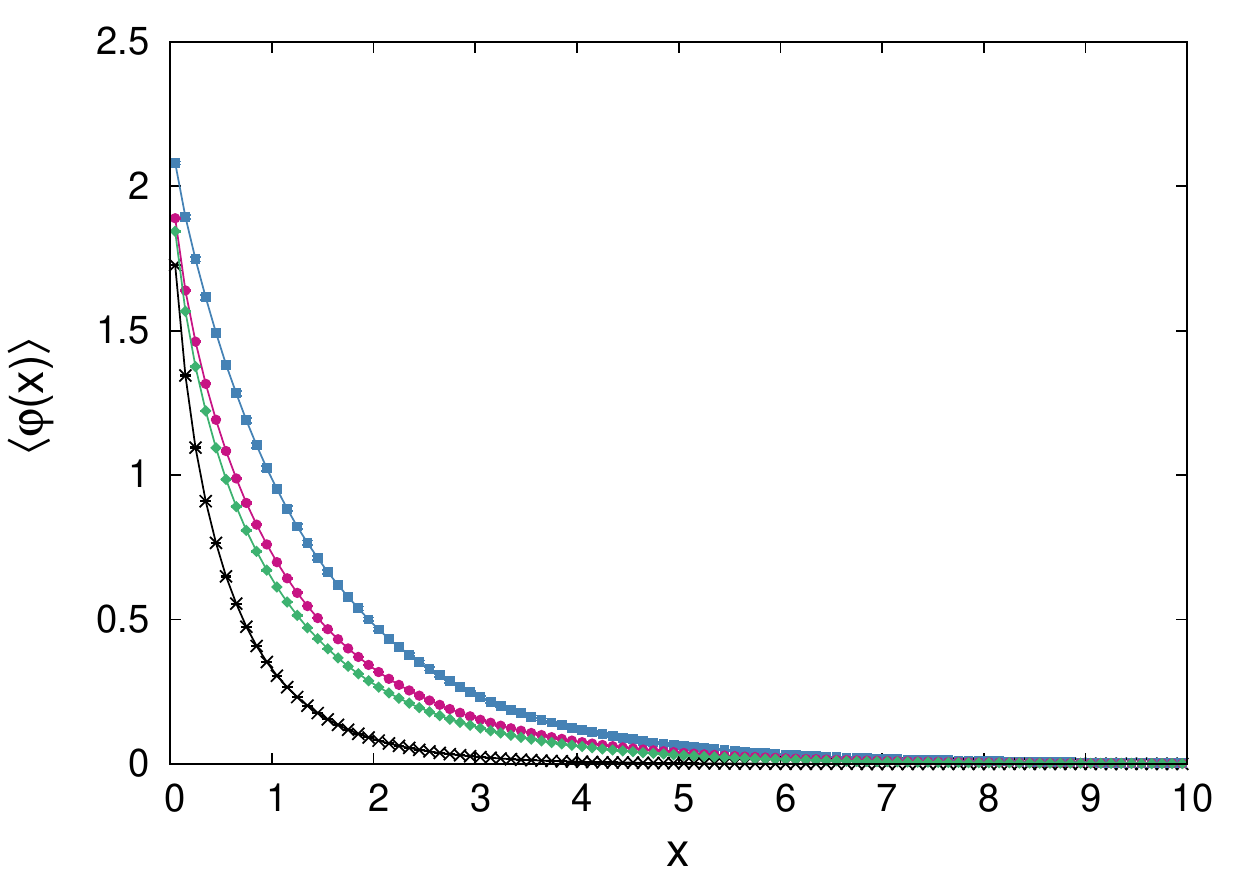}\,\,\,\,
\includegraphics[width=0.9\columnwidth]{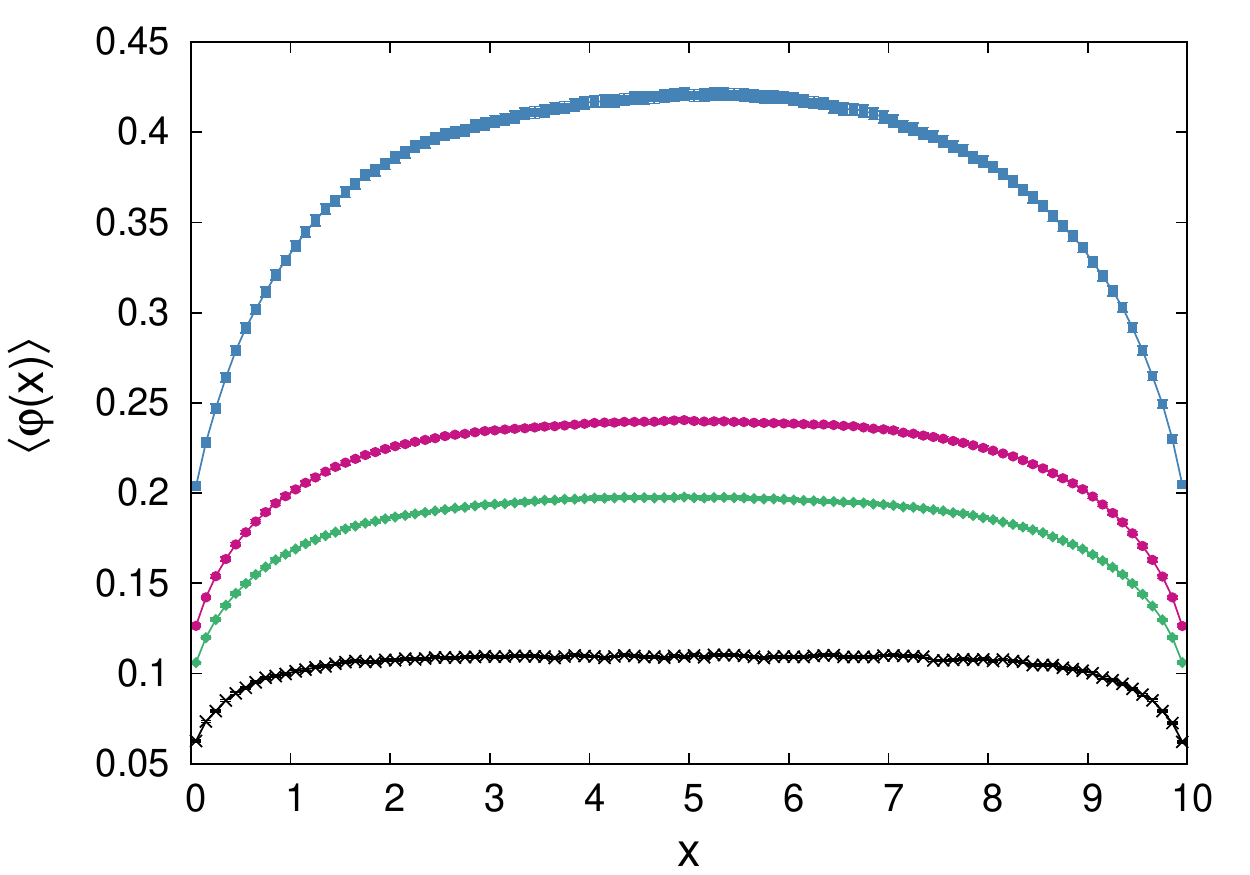}\\
\,\,\,\, Case 1c \,\,\,\,\\
\includegraphics[width=0.9\columnwidth]{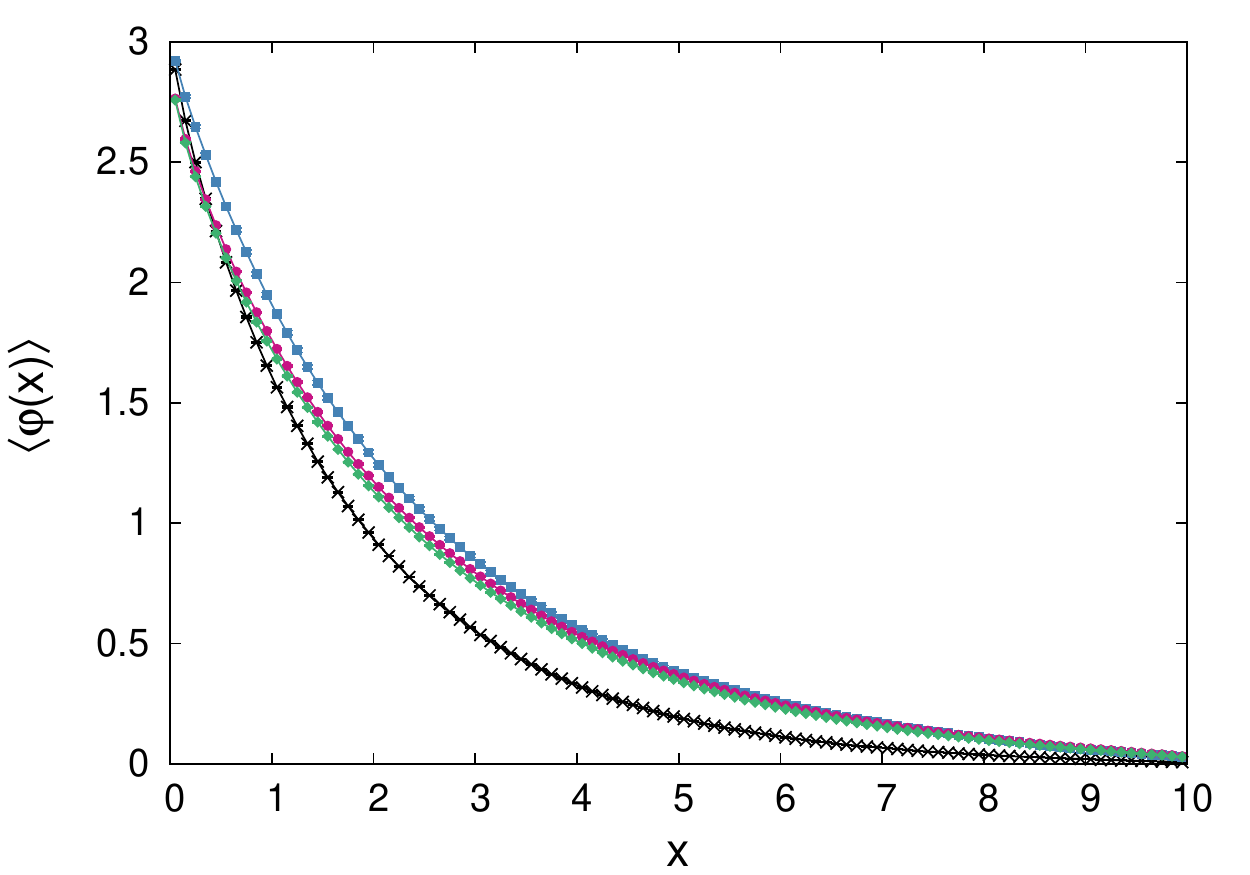}\,\,\,\,
\includegraphics[width=0.9\columnwidth]{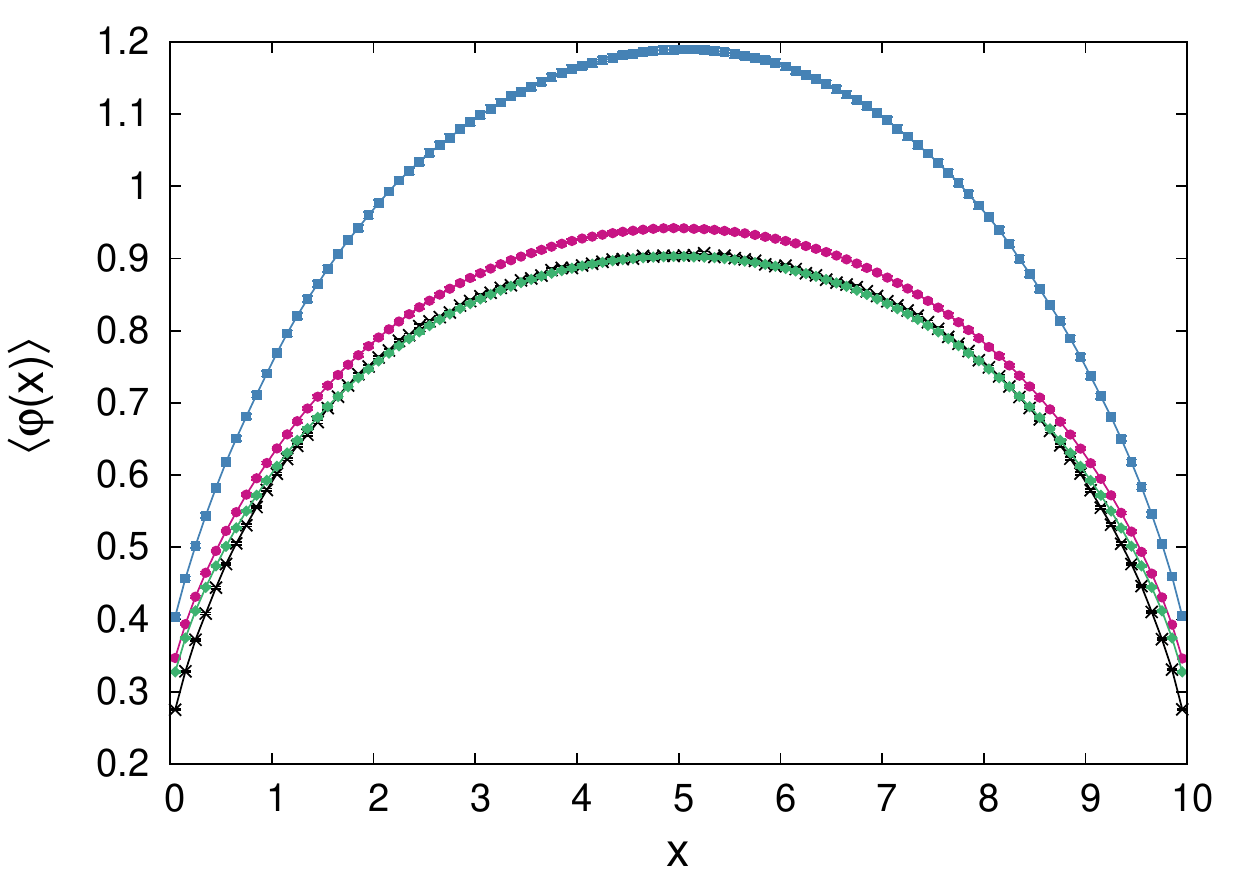}
\end{center}
\caption{Ensemble-averaged spatial scalar flux for the benchmark configurations: Case 1. Left column: {\em suite} I configurations; right column: {\em suite} II configurations. Black crosses represent the atomic mixing approximation, blue squares the $1d$ slab tessellations, red circles the $2d$ extruded tessellations, and green diamonds the $3d$ tessellations.}
\label{fig_space_1}
\end{figure*}

\begin{figure*}
\begin{center}
\,\,\,\, Case 2a \,\,\,\,\\
\includegraphics[width=0.9\columnwidth]{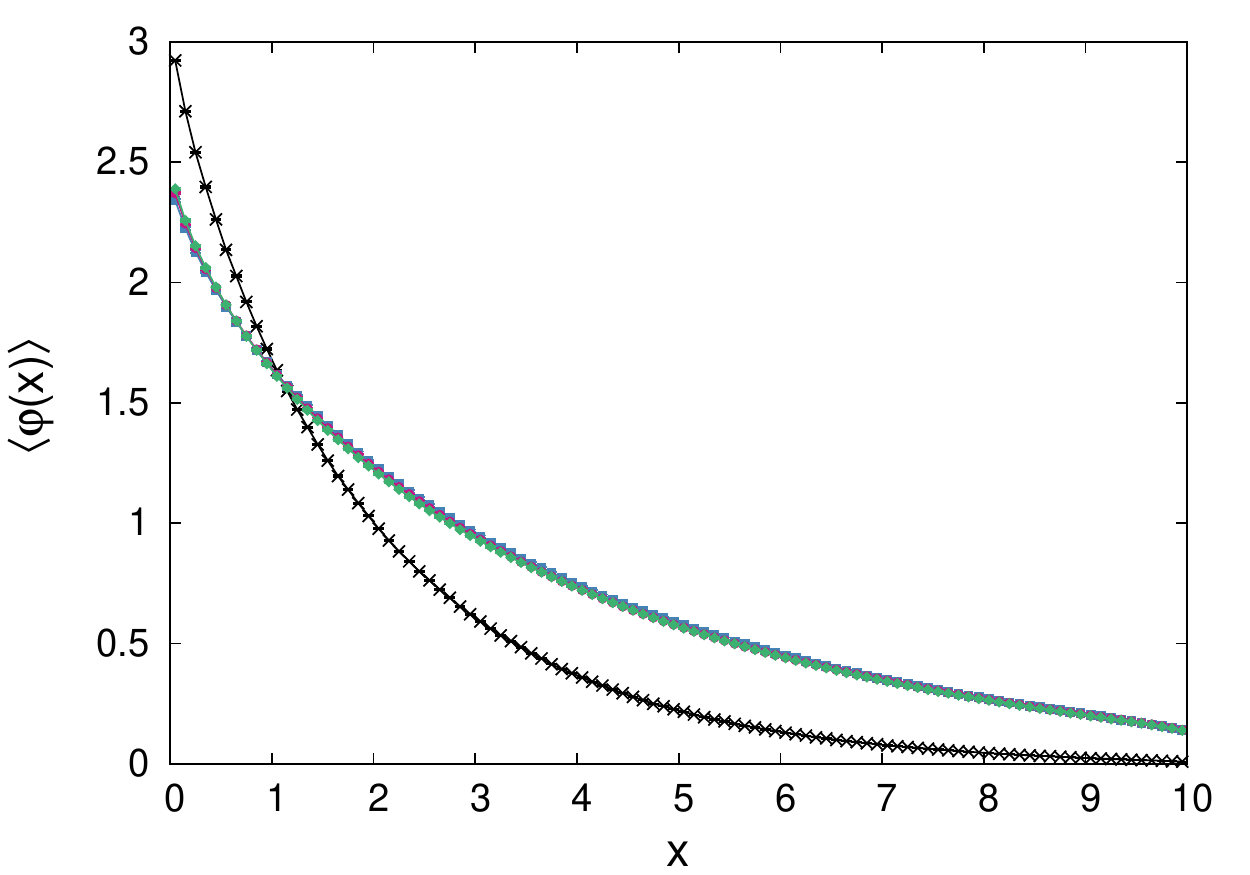}\,\,\,\,
\includegraphics[width=0.9\columnwidth]{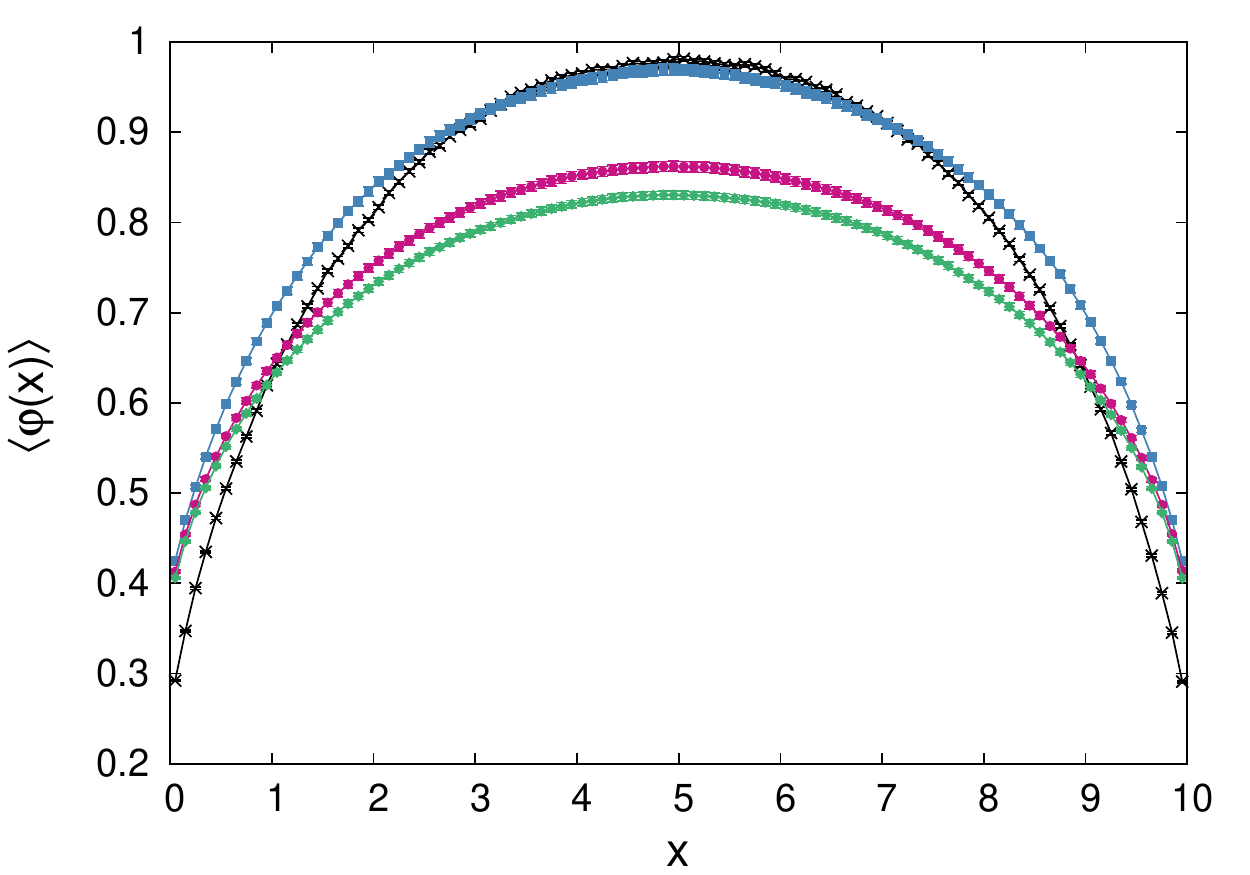}\\
\,\,\,\, Case 2b \,\,\,\,\\
\includegraphics[width=0.9\columnwidth]{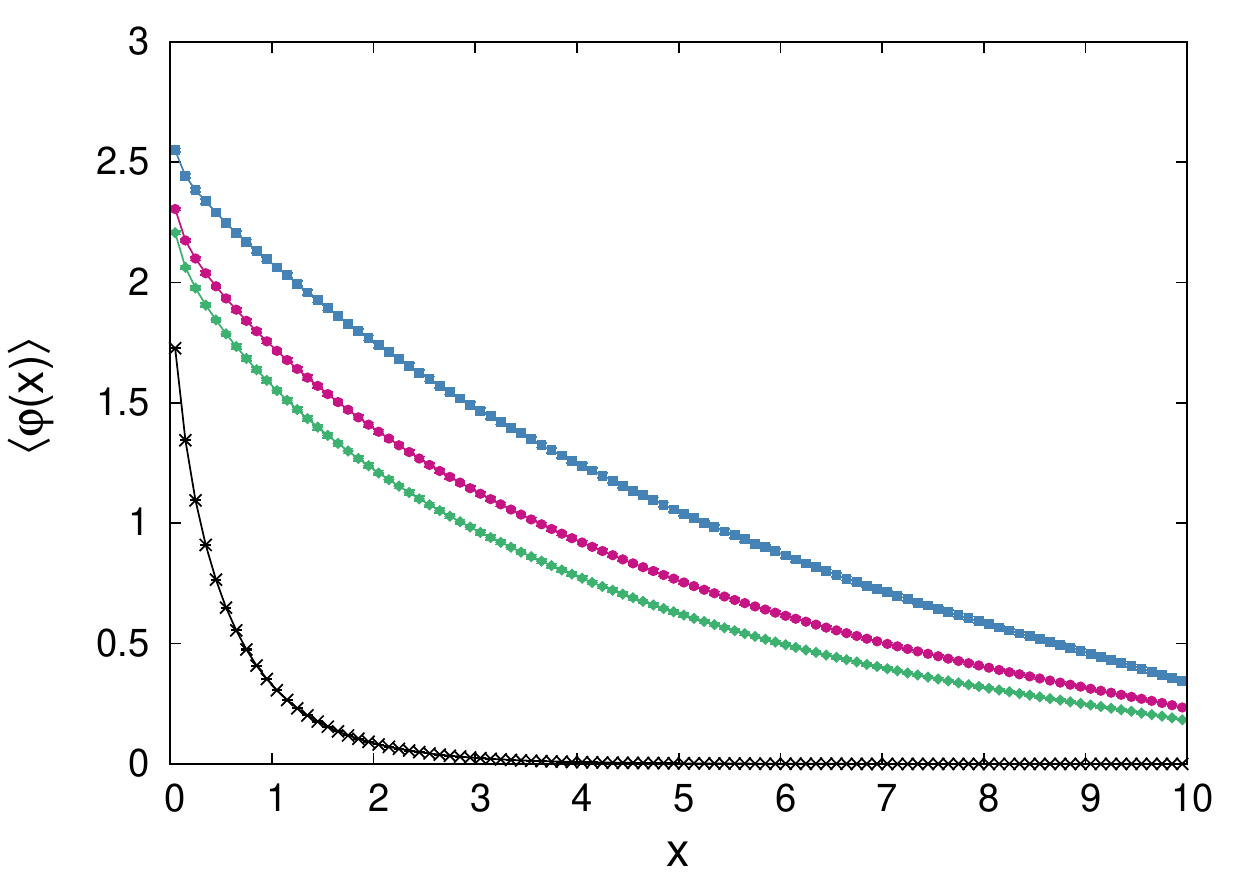}\,\,\,\,
\includegraphics[width=0.9\columnwidth]{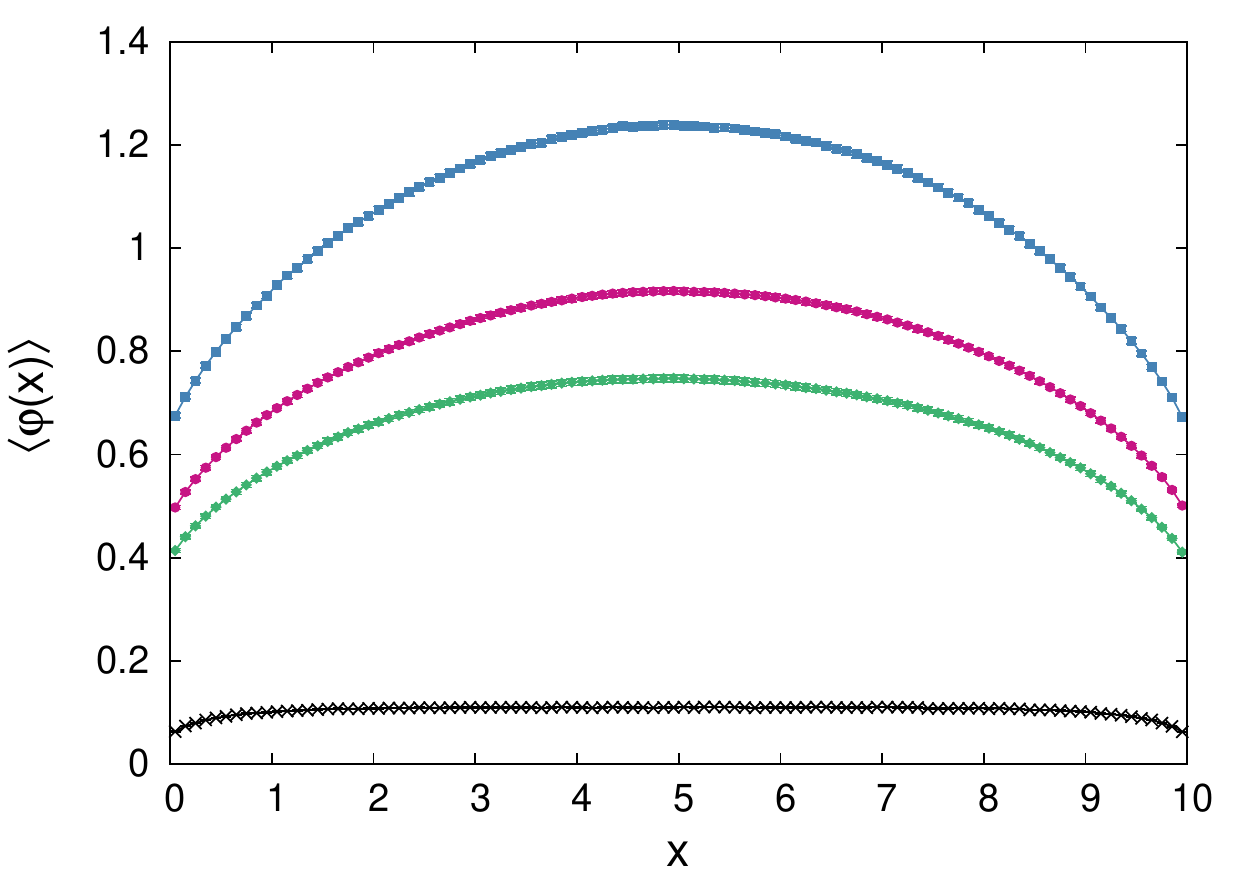}\\
\,\,\,\, Case 2c \,\,\,\,\\
\includegraphics[width=0.9\columnwidth]{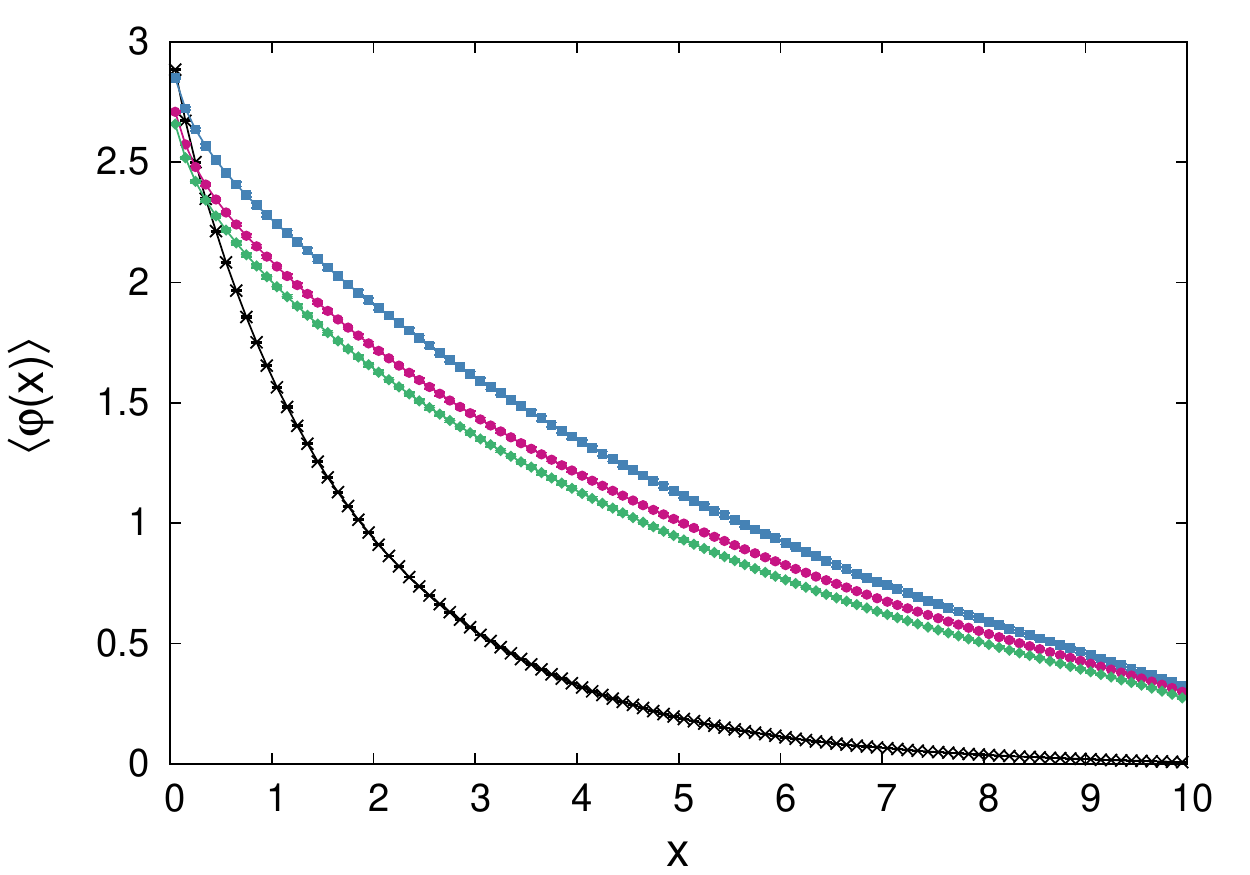}\,\,\,\,
\includegraphics[width=0.9\columnwidth]{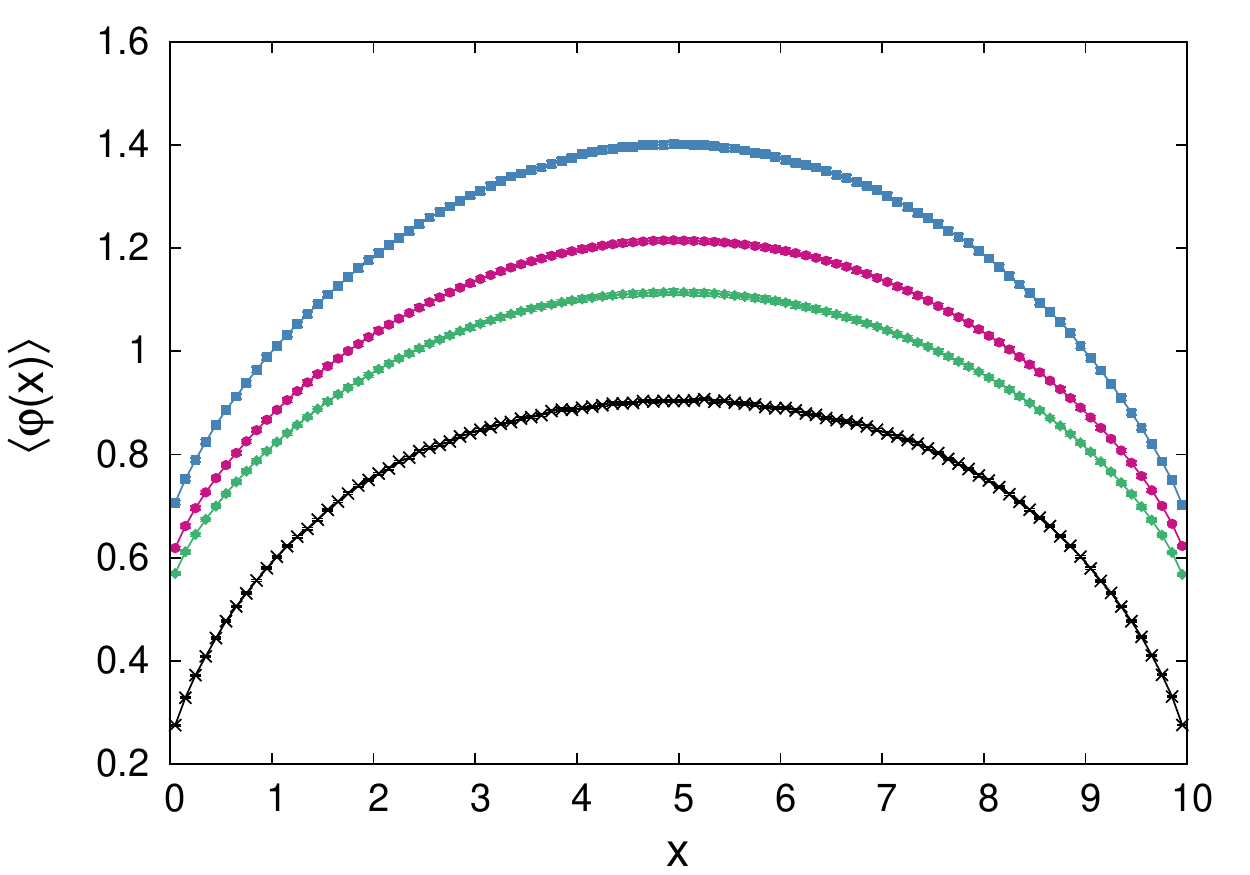}
\end{center}
\caption{Ensemble-averaged spatial scalar flux for the benchmark configurations: Case 2. Left column: {\em suite} I configurations; right column: {\em suite} II configurations. Black crosses represent the atomic mixing approximation, blue squares the $1d$ slab tessellations, red circles the $2d$ extruded tessellations, and green diamonds the $3d$ tessellations.}
\label{fig_space_2}
\end{figure*}

\begin{figure*}
\begin{center}
\,\,\,\, Case 3a \,\,\,\,\\
\includegraphics[width=0.9\columnwidth]{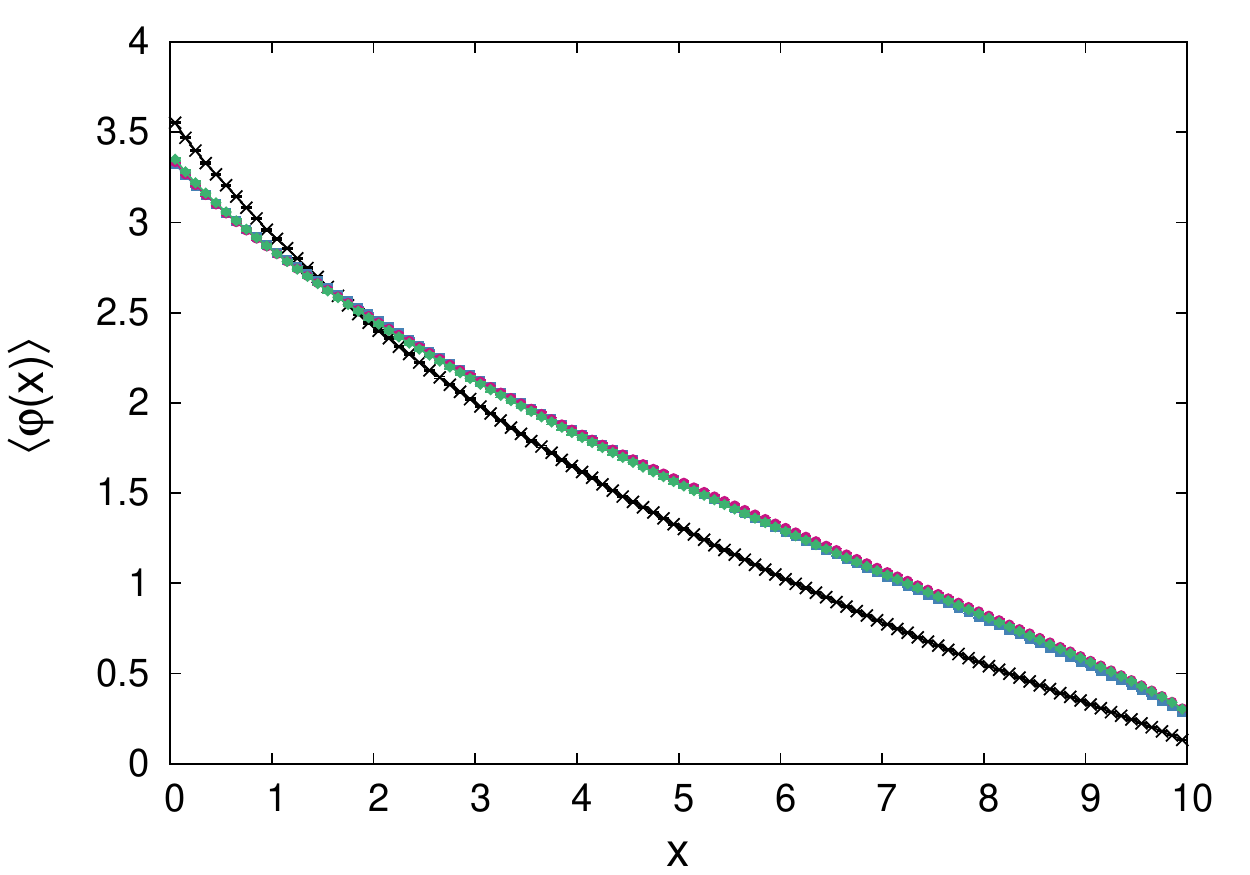}\,\,\,\,
\includegraphics[width=0.9\columnwidth]{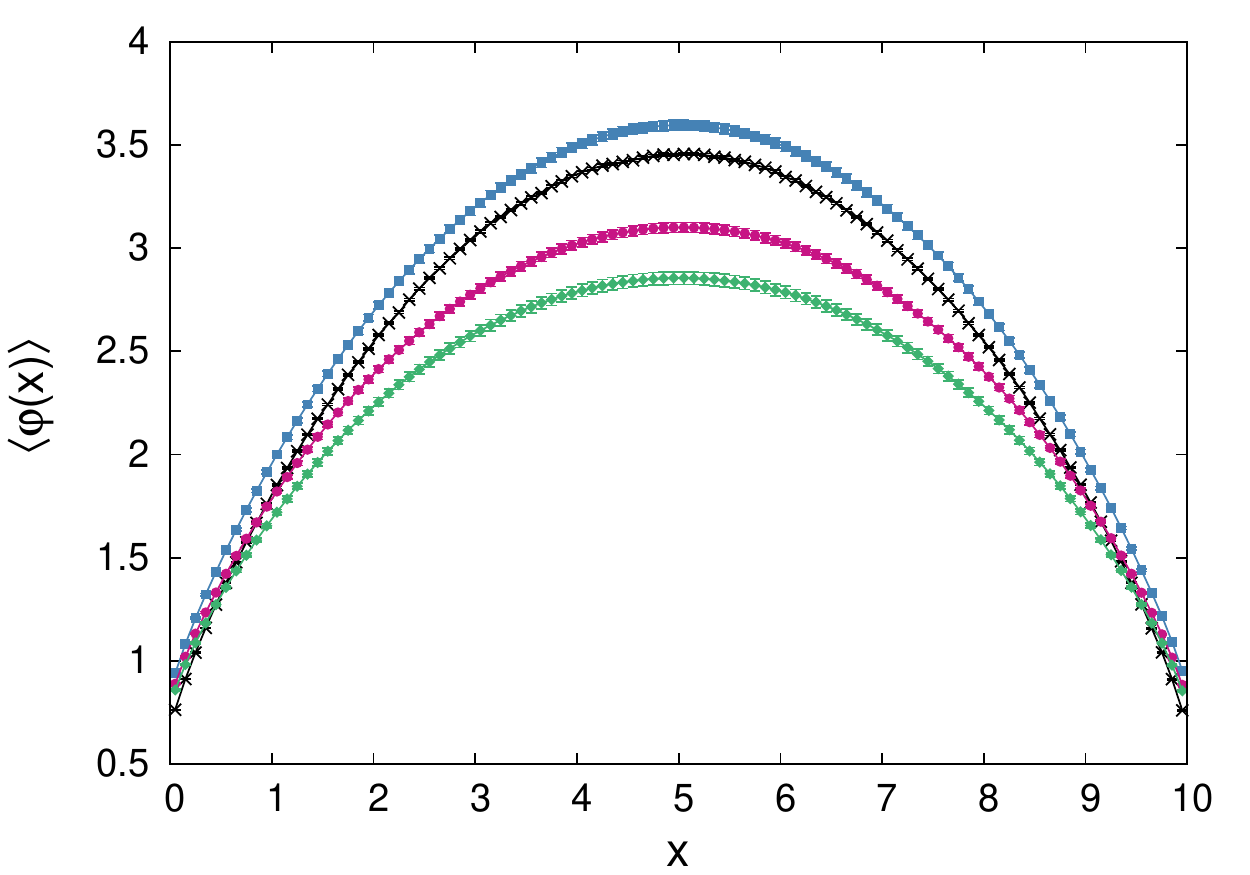}\\
\,\,\,\, Case 3b \,\,\,\,\\
\includegraphics[width=0.9\columnwidth]{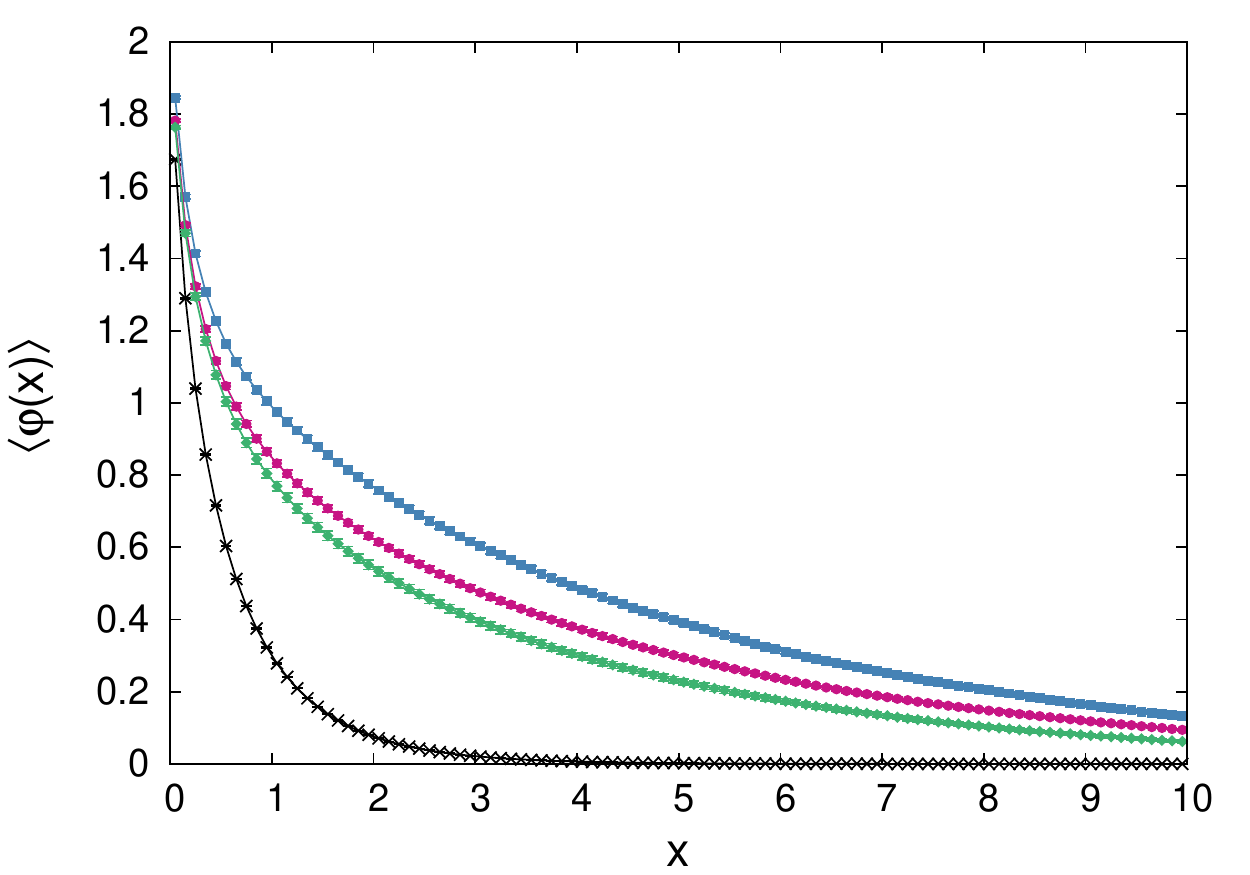}\,\,\,\,
\includegraphics[width=0.9\columnwidth]{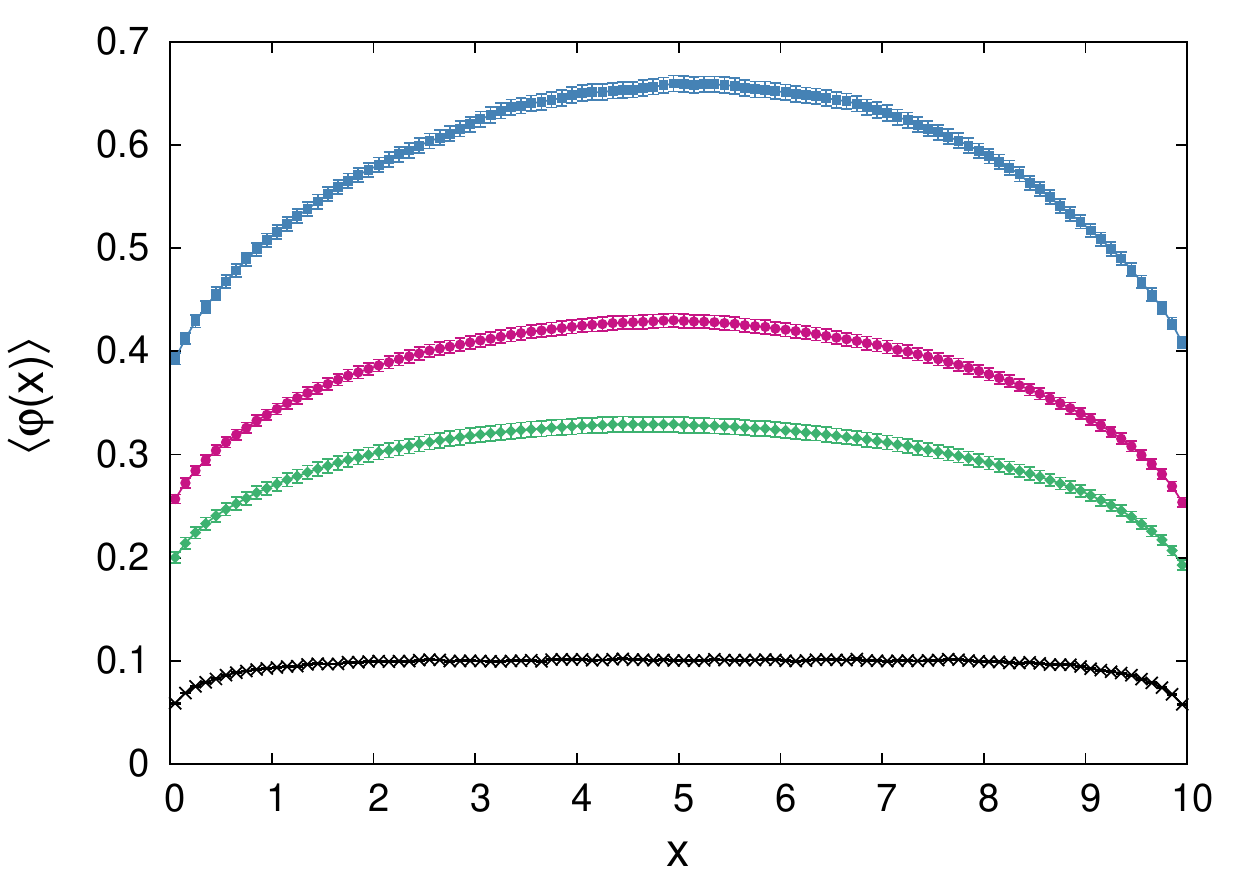}\\
\,\,\,\, Case 3c \,\,\,\,\\
\includegraphics[width=0.9\columnwidth]{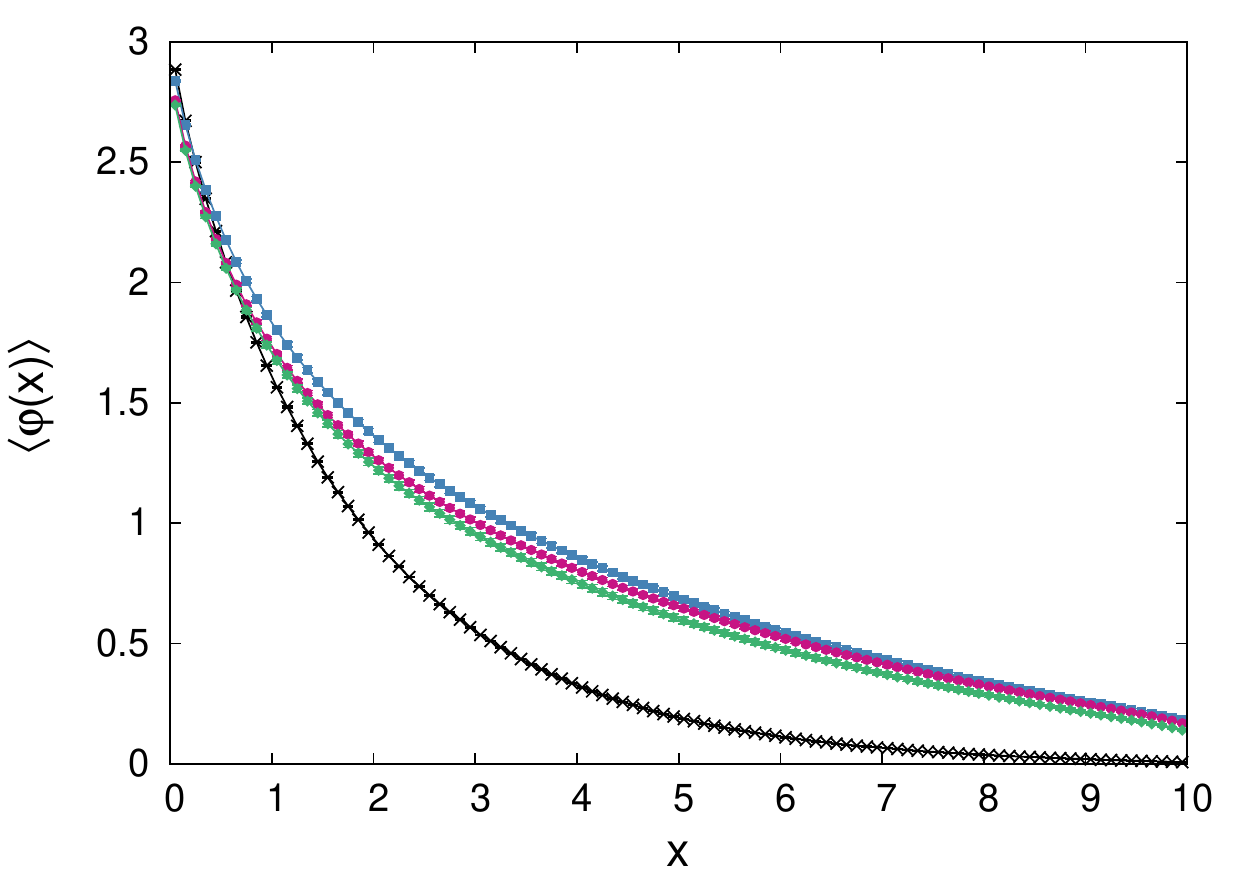}\,\,\,\,
\includegraphics[width=0.9\columnwidth]{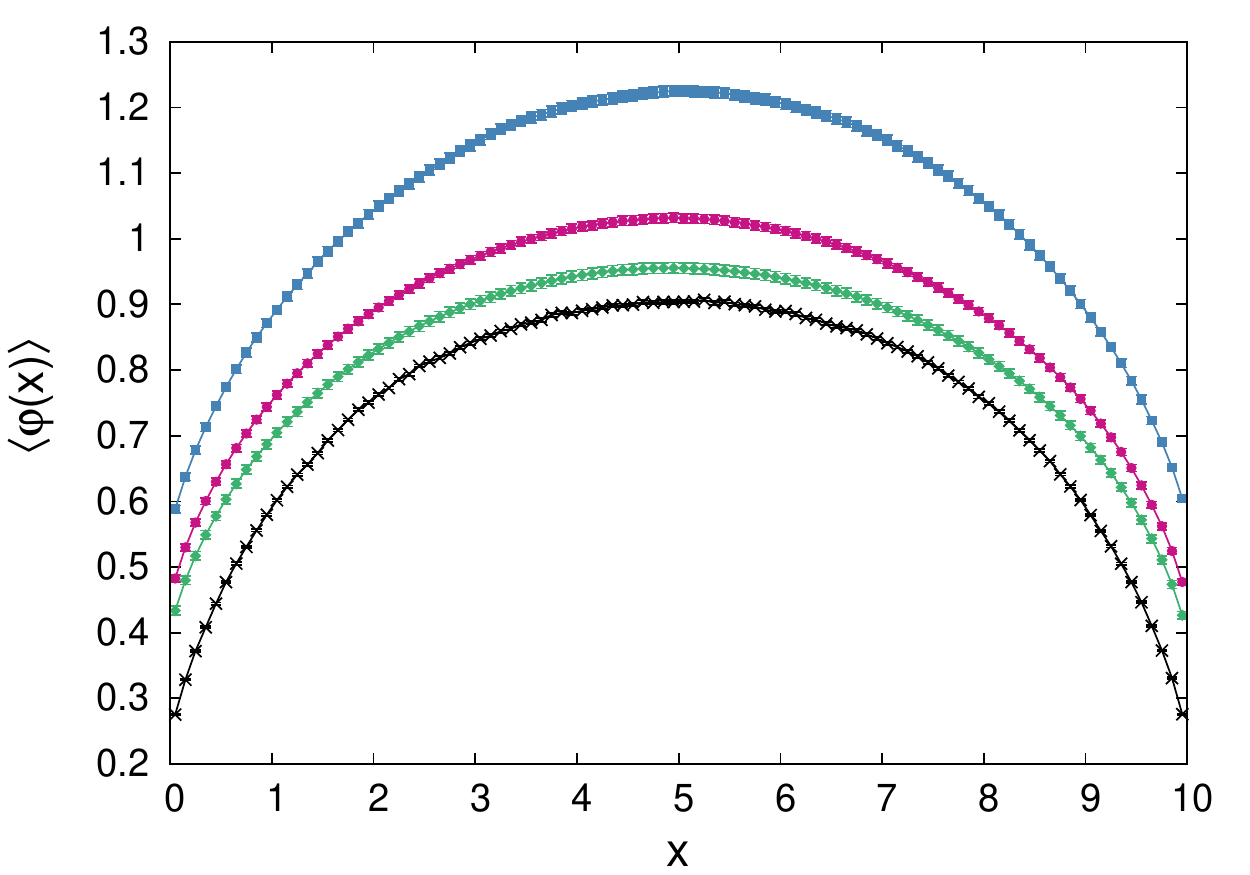}
\end{center}
\caption{Ensemble-averaged spatial scalar flux for the benchmark configurations: Case 3. Left column: {\em suite} I configurations; right column: {\em suite} II configurations. Black crosses represent the atomic mixing approximation, blue squares the $1d$ slab tessellations, red circles the $2d$ extruded tessellations, and green diamonds the $3d$ tessellations.}
\label{fig_space_3}
\end{figure*}

The spatial profiles of the ensemble-averaged scalar flux $\langle \varphi(x) \rangle= \langle \int \int \int\varphi({\bf r}, {\boldsymbol \omega}) d {\boldsymbol \omega} dy dz \rangle$ are reported in Figs.~\ref{fig_space_1} to \ref{fig_space_3}. In \tripoli{}, we estimate $\langle \varphi(x) \rangle$ by recording the flux within the spatial grid and by dividing the obtained result by the volume of each mesh. The corresponding data sets are available from the authors upon request. Consistently with the findings concerning the integral observables, the atomic mixing approximation usually leads to poor results as compared with the benchmark solutions for the spatial profiles. The spatial profiles for the atomic mixing approximation and for the $1d$ slab tessellations are in good agreement with those reported in~\citep{brantley_benchmark}. As a general remark, the scalar flux in $d=1$ is systematically larger than that in $d=2$, which in turn is larger than that in $d=3$. Some exceptions are nonetheless apparent, such as for instance in cases $1a$ for {\em suite} I configuration and $1c$ for {\em suite} II configuration. The impact of dimension on the benchmark solutions depends on the geometry and material configurations, and might vary between a few percent as in case $1a$ for both {\em suite} I and {\em suite} II, up to $100\%$ or more in case $3b$ for {\em suite} II.

\section{Conclusions}
\label{conclusions}

The key goal of this work was to compute reference solutions for linear transport in stochastic geometries. In order to establish a proper and easily reproducible framework, we have built our specifications upon the benchmark originally proposed by Adams, Larsen and Pomraning, and recently revisited by Brantley. We have thus considered a box of fixed side, with two free surfaces on opposite sides, and reflecting boundary conditions everywhere else. As a prototype example of stochastic media, we have adopted Markov geometries with binary mixing: such geometries have been numerically implemented by resorting to the algorithm for colored Poisson geometries.

Three kinds of Poisson tessellations of the box have been tested: $1d$ slab tessellations, $2d$ extruded tessellations, and full $3d$ tessellations. To the best of our knowledge, benchmark solutions for $2d$ and $3d$ tessellations with Markov mixing have never been studied before. Material compositions and correlation lengths, as well as initial and boundary conditions, have been assigned based on the benchmark specifications, amounting to a total of $18$ distinct cases for each dimension. A large number of random geometries and material compositions have been realized. For each realization, mono-energetic linear transport with isotropic scattering and absorption has been simulated by Monte Carlo method. The code \tripoli{} developed at CEA has been used for this purpose.

The physical observables that have been examined in this work are the reflection and transmission coefficients, and the scalar particle flux, averaged over the ensemble of available realizations. The full distributions of the reflection and transmission coefficients have been also examined, in order to evaluate the impact of correlation lengths and volumetric fractions on the dispersion of these observables around their average values.

The reference solutions presented in this work might be helpful for the validation of the fast but approximate methods that have been developed over the years so as to describe particle propagation in stochastic media with effective transport kernels, i.e., without having to average over medium realizations. In particular, benchmark solutions for the full $3d$ tessellations are essential to real-world applications. Effective models such as the Chord Length Sampling algorithm typically neglect the correlations on the particle paths induced by the interfaces between materials: future work will be thus aimed at assessing the impact of such approximations as a function of the system dimension, by comparison with benchmark solutions. Furthermore, the numerical tools that we have used in this paper in order to generate the colored Poisson geometries and to perform the stochastic transport are extremely flexible, and could thus easily accommodate several extensions or improvements of the current benchmark specifications. In particular, we might include anisotropy in the scattering kernels, energy-dependent cross sections and scattering distributions, particle production from fission, different geometrical shapes and arbitrary boundary conditions.

\section*{Acknowledgements}
TRIPOLI\textsuperscript{ \textregistered} and TRIPOLI-4\textsuperscript{ \textregistered} are registered trademarks of CEA. The authors wish to thank \'Electricit\'e de France (EDF) for partial financial support.

\clearpage

\begin{table*}
\begin{center}
\begin{tabular}{cccccccccccc}
\toprule
Case & $d$ & $\langle N_p | L \rangle$ & $\Lambda(L)$ & $\Lambda$ & $\Lambda_{\alpha}(L)$ & $\Lambda_{\alpha}$ & $\Lambda_{\beta}(L)$ & $\Lambda_{\beta}$ \\
 \midrule
& 1 & $102.1 \pm 0.1$ & $0.0990 \pm 0.0001$ & $0.099$ & $0.985 \pm 0.004$ & $0.99$ & $0.1098 \pm 0.0004$ & $0.11$ \\
1 & 2 & $8206 \pm 20$ & $0.0988 \pm 0.0003$ & $0.099$ & $0.944 \pm 0.007$ & $0.99$ & $0.1058 \pm 0.0009$ & $0.11$ \\
& 3 & $565646 \pm 3586$ & $0.0963 \pm 0.0006$ & $0.099$ & $0.90 \pm 0.02$ & $0.99$ & $0.098 \pm 0.002$ & $0.11$ \\
\midrule
& 1 & $11.14 \pm 0.03$ & $0.987 \pm 0.004$ & $0.99$ & $6.22 \pm 0.03$ & $9.9$ & $0.695 \pm 0.009$ & $1.1$ \\
2 & 2 & $101.4 \pm 0.7$ & $0.942 \pm 0.007$ & $0.99$ & $5.00 \pm 0.05$ & $9.9$ & $0.57 \pm 0.01$ & $1.1$ \\
& 3 & $817 \pm 15$ & $0.93 \pm 0.02$ & $0.99$ & $4.4 \pm 0.1$ & $9.9$ & $0.50 \pm 0.02$ & $1.1$ \\
\midrule
& 1 & $4.98 \pm 0.02$ & $2.48 \pm 0.02$ & $2.525$ & $3.63 \pm 0.03$ & $5.05$ & $3.70 \pm 0.03$ & $5.05$ \\
3 & 2 & $21.2 \pm 0.2$ & $2.22 \pm 0.02$ & $2.525$ & $2.96 \pm 0.04$ & $5.05$ & $3.03 \pm 0.04$ & $5.05$ \\
& 3 & $86 \pm 2$ & $2.07 \pm 0.05$ & $2.525$ & $2.68 \pm 0.09$ & $5.05$ & $2.53 \pm 0.08$ & $5.05$ \\
\bottomrule
\end{tabular}
\end{center}
\caption{Statistical properties of the Poisson tessellations used for the benchmark configurations, as a function of the dimension $d$. The quantity $\langle N_p | L \rangle$ denotes the average number of $d$-polyhedra composing the tessellation, $\Lambda(L)$ is the average correlation length measured by Monte Carlo ray tracing, and $\Lambda_{\alpha}(L)$ and $\Lambda_{\beta}(L)$ are the average chord lengths for material $\alpha$ and $\beta$, respectively, measured by Monte Carlo ray tracing. The corresponding expected theoretical values for infinite-size tessellations are denoted by $\Lambda$, $\Lambda_{\alpha}$ and $\Lambda_{\beta}$.}
\label{tab_geo}
\end{table*}

\begin{table*}
\begin{center}
\begin{tabular}{ccccccccccc}
\toprule
 & Case: & 1a & 1b & 1c & 2a & 2b & 2c & 3a & 3b & 3c \\
 \midrule
Atomic mixing & $\langle t \rangle$ & $122$ & $41$ & $65$ & $67$ & $40$ & $66$ & $117$ & $39$ & $66$ \\
\midrule
$d=1$ & $\langle t \rangle$ & $155$ & $63$ & $117$ & $94$ & $62$ & $75$ & $138$ & $45$ & $69$ \\
& $\sigma[t]$ & $48$ & $16$ & $25$ & $61$ & $14$ & $7$ & $53$ & $6$ & $6$ \\
\midrule
$d=2$ & $\langle t \rangle$ & $168$ & $77$ & $186$ & $91$ & $62$ & $82$ & $152$ & $46$ & $72$ \\
& $\sigma[t]$ & $9$ & $4$ & $50$ & $26$ & $7$ & $8$ & $54$ & $4$ & $5$ \\
\midrule
$d=3$ & $\langle t \rangle$ & $3962$ & $1711$ & $3582$ & $119$ & $63$ & $87$ & $144$ & $46$ & $75$ \\
& $\sigma[t]$ & $889$ & $364$ & $862$ & $36$ & $4$ & $4$ & $35$ & $3$ & $4$ \\
\bottomrule
\end{tabular}
\end{center}
\caption{Simulation times $t$ for the benchmark configurations, expressed in seconds. The cases of {\em suite} I.}
\label{tab_times1}
\end{table*}

\begin{table*}
\begin{center}
\begin{tabular}{ccccccccccc}
\toprule
 & Case: & 1a & 1b & 1c & 2a & 2b & 2c & 3a & 3b & 3c \\
 \midrule
Atomic mixing & $\langle t \rangle$ & $49$ & $11$ & $47$ & $46$ & $9$ & $43$ & $139$ & $8$ & $43$ \\
\midrule
$d=1$ & $\langle t \rangle$ & $101$ & $38$ & $102$ & $105$ & $27$ & $39$ & $195$ & $14$ & $42$ \\
& $\sigma[t]$ & $25$ & $9$ & $14$ & $255$ & $8$ & $8$ & $173$ & $4$ & $9$ \\
\midrule
$d=2$ & $\langle t \rangle$ & $530$ & $457$ & $630$ & $85$ & $31$ & $49$ & $172$ & $13$ & $43$ \\
& $\sigma[t]$ & $81$ & $84$ & $288$ & $80$ & $5$ & $6$ & $113$ & $3$ & $6$ \\
\midrule
$d=3$ & $\langle t \rangle$ & $63468$ & $61503$ & $62912$ & $102$ & $63$ & $84$ & $159$ & $16$ & $48$ \\
& $\sigma[t]$ & $16384$ & $13801$ & $16142$ & $34$ & $21$ & $24$ & $75$ & $3$ & $6$ \\
\bottomrule
\end{tabular}
\end{center}
\caption{Simulation times $t$ for the benchmark configurations, expressed in seconds. The cases of {\em suite} II.}
\label{tab_times2}
\end{table*}

\clearpage

\begin{table*}
\begin{center}
\begin{tabular}{ccccccccccc}
\toprule
Configuration & Observable & Atomic mixing & {\tt $1d$} & {\tt $2d$} & {\tt $3d$} \\
\midrule
& $\langle R \rangle$ & $0.4919 \pm 0.0004$ & $0.435 \pm 0.002$ & $0.4031 \pm 0.0006$ & $0.4065 \pm 0.0004$ \\
1a & $\langle T \rangle$ & $0.00484 \pm 7 \times 10^{-5}$ & $0.0147 \pm 0.0002$ & $0.0173 \pm 0.0001$ & $0.0162 \pm 0.0001$ \\
& $\langle \varphi \rangle$ & $5.499 \pm 0.007$ & $6.09 \pm 0.01$ & $6.356 \pm 0.008$ & $6.318 \pm 0.008$ \\
\midrule
& $\langle R \rangle$ & $0.0193 \pm 0.0001$ & $0.0841 \pm 0.0007$ & $0.0453 \pm 0.0002$ & $0.0376 \pm 0.0002$ \\
1b & $\langle T \rangle$ & $8 \times 10^{-6} \pm 3 \times 10^{-6}$ & $0.0017 \pm 0.0001$ & $0.00108 \pm 3 \times 10^{-5}$ & $0.00085 \pm 3 \times 10^{-5}$ \\
& $\langle \varphi \rangle$ & $1.077 \pm 0.001$ & $2.89 \pm 0.02$ & $2.165 \pm 0.005$ & $1.920 \pm 0.003$ \\
\midrule
& $\langle R \rangle$ & $0.4747 \pm 0.0004$ & $0.4743 \pm 0.0004$ & $0.4059 \pm 0.0004$ & $0.4036 \pm 0.0004$ \\
1c & $\langle T \rangle$ & $0.00384 \pm 6 \times 10^{-5}$ & $0.0159 \pm 0.0003$ & $0.0179 \pm 0.0001$ & $0.0164 \pm 0.0001$ \\
& $\langle \varphi \rangle$ & $5.172 \pm 0.0007$ & $6.95 \pm 0.03$ & $6.52 \pm 0.01$ & $6.296 \pm 0.0008$ \\
\bottomrule
\end{tabular}
\end{center}
\caption{Ensemble-averaged observables for the benchmark configurations: {\em suite} I - case $1$.}
\label{tab_suite1_case1}
\end{table*}

\begin{table*}
\begin{center}
\begin{tabular}{ccccccccccc}
\toprule
Configuration & Observable & Atomic mixing & {\tt $1d$} & {\tt $2d$} & {\tt $3d$} \\
\midrule
& $\langle R \rangle$ & $0.4919 \pm 0.0004$ & $0.235 \pm 0.003$ & $0.226 \pm 0.002$ & $0.223 \pm 0.002$ \\
2a & $\langle T \rangle$ & $0.00484 \pm 7 \times 10^{-5}$ & $0.0975 \pm 0.0009$ & $0.0955 \pm 0.0007$ & $0.0935 \pm 0.0008$ \\
& $\langle \varphi \rangle$ & $5.499 \pm 0.007$ & $7.63 \pm 0.02$ & $7.57 \pm 0.01$ & $7.55 \pm 0.02$ \\
\midrule
& $\langle R \rangle$ & $0.0193 \pm 0.0001$ & $0.285 \pm 0.0002$ & $0.196 \pm 0.001$ & $0.161 \pm 0.002$ \\
2b & $\langle T \rangle$ & $8 \times 10^{-6} \pm 3 \times 10^{-6}$ & $0.193 \pm 0.003$ & $0.143 \pm 0.002$ & $0.119 \pm 0.002$ \\
& $\langle \varphi \rangle$ & $1.077 \pm 0.001$ & $11.65 \pm 0.08$ & $9.00 \pm 0.06$ & $7.76 \pm 0.07$ \\
\midrule
& $\langle R \rangle$ & $0.4747 \pm 0.0004$ & $0.4304 \pm 0.0008$ & $0.3669 \pm 0.0006$ & $0.3438 \pm 0.0006$ \\
2c & $\langle T \rangle$ & $0.00384 \pm 6 \times 10^{-5}$ & $0.185 \pm 0.002$ & $0.176 \pm 0.002$ & $0.165 \pm 0.0002$ \\
& $\langle \varphi \rangle$ & $5.172 \pm 0.007$ & $12.50 \pm 0.06$ & $11.39 \pm 0.05$ & $10.76 \pm 0.06$ \\
\bottomrule
\end{tabular}
\end{center}
\caption{Ensemble-averaged observables for the benchmark configurations: {\em suite} I - case $2$.}
\label{tab_suite1_case2}
\end{table*}

\begin{table*}
\begin{center}
\begin{tabular}{ccccccccccc}
\toprule
Configuration & Observable & Atomic mixing & {\tt $1d$} & {\tt $2d$} & {\tt $3d$} \\
\midrule
& $\langle R \rangle$ & $0.7820 \pm 0.0004$ & $0.693 \pm 0.003$ & $0.672 \pm 0.003$ & $0.670 \pm 0.004$ \\
3a & $\langle T \rangle$ & $0.0667 \pm 0.0003$ & $0.161 \pm 0.002$ & $0.170 \pm 0.002$ & $0.169 \pm 0.003$ \\
& $\langle \varphi \rangle$ & $14.83 \pm 0.02$ & $16.35 \pm 0.05$ & $16.46 \pm 0.05$ & $16.35 \pm 0.08$ \\
\midrule
& $\langle R \rangle$ & $0.00202 \pm 4 \times 10^{-5}$ & $0.0349 \pm 0.0004$ & $0.0221 \pm 0.0004$ & $0.0167 \pm 0.0006$ \\
3b & $\langle T \rangle$ & $9 \times 10^{-6} \pm 3 \times 10^{-6}$ & $0.0740 \pm 0.002$ & $0.061 \pm 0.002$ & $0.045 \pm 0.003$ \\
& $\langle \varphi \rangle$ & $1.004 \pm 0.001$ & $5.01 \pm 0.06$ & $4.08 \pm 0.06$ & $3.49 \pm 0.08$ \\
\midrule
& $\langle R \rangle$ & $0.4747 \pm 0.0004$ & $0.443 \pm 0.001$ & $0.406 \pm 0.001$ & $0.395 \pm 0.001$ \\
3c & $\langle T \rangle$ & $0.00384 \pm 6 \times 10^{-5}$ & $0.101 \pm 0.002$ & $0.098 \pm 0.002$ & $0.085 \pm 0.003$ \\
& $\langle \varphi \rangle$ & $5.172 \pm 0.007$ & $8.80 \pm 0.07$ & $8.34 \pm 0.07$ & $7.9 \pm 0.1$ \\
\bottomrule
\end{tabular}
\end{center}
\caption{Ensemble-averaged observables for the benchmark configurations: {\em suite} I - Case $3$.}
\label{tab_suite1_case3}
\end{table*}

\begin{table*}
\begin{center}
\begin{tabular}{ccccccccccc}
\toprule
Configuration & Observable & Atomic mixing & {\tt $1d$} & {\tt $2d$} & {\tt $3d$} \\
\midrule
& $\langle R \rangle$ & $0.1374 \pm 0.0003$ & $0.1522 \pm 0.0004$ & $0.1590 \pm 0.0004$ & $0.1580 \pm 0.0004$ \\
1a & $\langle T \rangle$ & $0.1367 \pm 0.0004$ & $0.1521 \pm 0.0004$ & $0.1589 \pm 0.0004$ & $0.1580 \pm 0.0004$ \\
& $\langle \varphi \rangle$ & $7.978 \pm 0.008$ & $7.70 \pm 0.01$ & $7.511 \pm 0.008$ & $7.530 \pm 0.008$ \\
\midrule
& $\langle R \rangle$ & $0.0271 \pm 0.0002$ & $0.0723 \pm 0.0006$ & $0.0542 \pm 0.0003$ & $0.0481 \pm 0.0002$ \\
1b & $\langle T \rangle$ & $0.02685 \pm 0.0002$ & $0.0725 \pm 0.0006$ & $0.0541 \pm 0.0003$ & $0.0480 \pm 0.0002$ \\
& $\langle \varphi \rangle$ & $1.040 \pm 0.001$ & $3.73 \pm 0.01$ & $2.182 \pm 0.003$ & $1.809 \pm 0.003$ \\
\midrule
& $\langle R \rangle$ & $0.1290 \pm 0.0003$ & $0.1739 \pm 0.0009$ & $0.1630 \pm 0.0004$ & $0.1575 \pm 0.0004$ \\
1c & $\langle T \rangle$ & $0.1295 \pm 0.0003$ & $0.1738 \pm 0.0009$ & $0.1626 \pm 0.0004$ & $0.1575 \pm 0.0004$ \\
& $\langle \varphi \rangle$ & $7.406 \pm 0.008$ & $9.62 \pm 0.03$ & $7.77 \pm 0.01$ & $7.455 \pm 0.008$ \\
\bottomrule
\end{tabular}
\end{center}
\caption{Ensemble-averaged observables for the benchmark configurations: {\em suite} II - case $1$.}
\label{tab_suite2_case1}
\end{table*}

\begin{table*}
\begin{center}
\begin{tabular}{ccccccccccc}
\toprule
Configuration & Observable & Atomic mixing & {\tt $1d$} & {\tt $2d$} & {\tt $3d$} \\
\midrule
& $\langle R \rangle$ & $0.1374 \pm 0.0003$ & $0.1903 \pm 0.0004$ & $0.1894 \pm 0.0004$ & $0.1888 \pm 0.0004$ \\
2a & $\langle T \rangle$ & $0.1367 \pm 0.0004$ & $0.1898 \pm 0.0004$ & $0.1898 \pm 0.0004$ & $0.1885 \pm 0.0004$ \\
& $\langle \varphi \rangle$ & $7.978 \pm 0.008$ & $8.26 \pm 0.03$ & $7.45 \pm 0.03$ & $7.21 \pm 0.02$ \\
\midrule
& $\langle R \rangle$ & $0.0271 \pm 0.0002$ & $0.292 \pm 0.002$ & $0.225 \pm 0.002$ & $0.194 \pm 0.002$ \\
2b & $\langle T \rangle$ & $0.0268 \pm 0.0002$ & $0.290 \pm 0.002$ & $0.227 \pm 0.001$ & $0.193 \pm 0.002$ \\
& $\langle \varphi \rangle$ & $1.040 \pm 0.001$ & $10.70 \pm 0.05$ & $7.94 \pm 0.04$ & $6.54 \pm 0.05$ \\
\midrule
& $\langle R \rangle$ & $0.1290 \pm 0.0003$ & $0.313 \pm 0.002$ & $0.285 \pm 0.001$ & $0.269 \pm 0.001$ \\
2c & $\langle T \rangle$ & $0.1295 \pm 0.0003$ & $0.311 \pm 0.002$ & $0.287 \pm 0.001$ & $0.268 \pm 0.001$ \\
& $\langle \varphi \rangle$ & $7.406 \pm 0.008$ & $11.90 \pm 0.03$ & $10.38 \pm 0.03$ & $9.57 \pm 0.04$ \\
\bottomrule
\end{tabular}
\end{center}
\caption{Ensemble-averaged observables for the benchmark configurations: {\em suite} II - case $2$.}
\label{tab_suite2_case2}
\end{table*}

\begin{table*}
\begin{center}
\begin{tabular}{ccccccccccc}
\toprule
Configuration & Observable & Atomic mixing & {\tt $1d$} & {\tt $2d$} & {\tt $3d$} \\
\midrule
& $\langle R \rangle$ & $0.3710 \pm 0.0004$ & $0.409 \pm 0.001$ & $0.412 0.001$ & $0.409 \pm 0.002$ \\
3a & $\langle T \rangle$ & $0.3705 \pm 0.0004$ & $0.413 \pm 0.001$ & $0.409 0.001$ & $0.407 \pm 0.002$ \\
& $\langle \varphi \rangle$ & $25.93 \pm 0.03$ & $27.5 \pm 0.2$ & $24.2 \pm 0.2$ & $22.5 \pm 0.2$ \\
\midrule
& $\langle R \rangle$ & $0.0253 \pm 0.0002$ & $0.125 \pm 0.002$ & $0.102 \pm 0.002$ & $0.087 \pm 0.002$ \\
3b & $\langle T \rangle$ & $0.0250 \pm 0.0002$ & $0.130 \pm 0.002$ & $0.101 \pm 0.002$ & $0.085 \pm 0.002$ \\
& $\langle \varphi \rangle$ & $0.9584 \pm 0.0009$ & $5.84 \pm 0.05$ & $3.81 \pm 0.05$ & $2.95 \pm 0.06$ \\
\midrule
& $\langle R \rangle$ & $0.1290 \pm 0.0003$ & $0.220 \pm 0.002$ & $0.209 \pm 0.002$ & $0.199 \pm 0.003$ \\
3c & $\langle T \rangle$ & $0.1295 \pm 0.0003$ & $0.226 \pm 0.002$ & $0.206 \pm 0.002$ & $0.196 \pm 0.002$ \\
& $\langle \varphi \rangle$ & $7.406 \pm 0.008$ & $10.46 \pm 0.05$ & $8.81 \pm 0.05$ & $8.16 \pm 0.06$ \\
\bottomrule
\end{tabular}
\end{center}
\caption{Ensemble-averaged observables for the benchmark configurations: {\em suite} II - case $3$.}
\label{tab_suite2_case3}
\end{table*}

\clearpage

\clearpage

\end{document}